\documentclass[a4paper, 11pt, oneside]{Thesis}  % Use the "Thesis" style, based on the ECS Thesis style by Steve Gunn
\graphicspath{{Figures/}}  % Location of the graphics files (set up for graphics to be in PDF format)

\usepackage[round, comma, authoryear]{natbib}
\usepackage{verbatim}  % Needed for the "comment" environment to make LaTeX comments
\usepackage{vector}  % Allows "\bvec{}" and "\buvec{}" for "blackboard" style bold vectors in maths
\usepackage{amsmath}
\usepackage{parskip}
\usepackage{graphicx}
\hypersetup{
     colorlinks   = true,
     citecolor    = black,
     urlcolor     = black,
     linkcolor    = black
}

\usepackage{amssymb}
\usepackage{bm}

\usepackage{pdfpages}

\begin{document}
\thispagestyle{empty}
\begin{figure}[h]
\centering
\centerline{\includegraphics[width=7in]{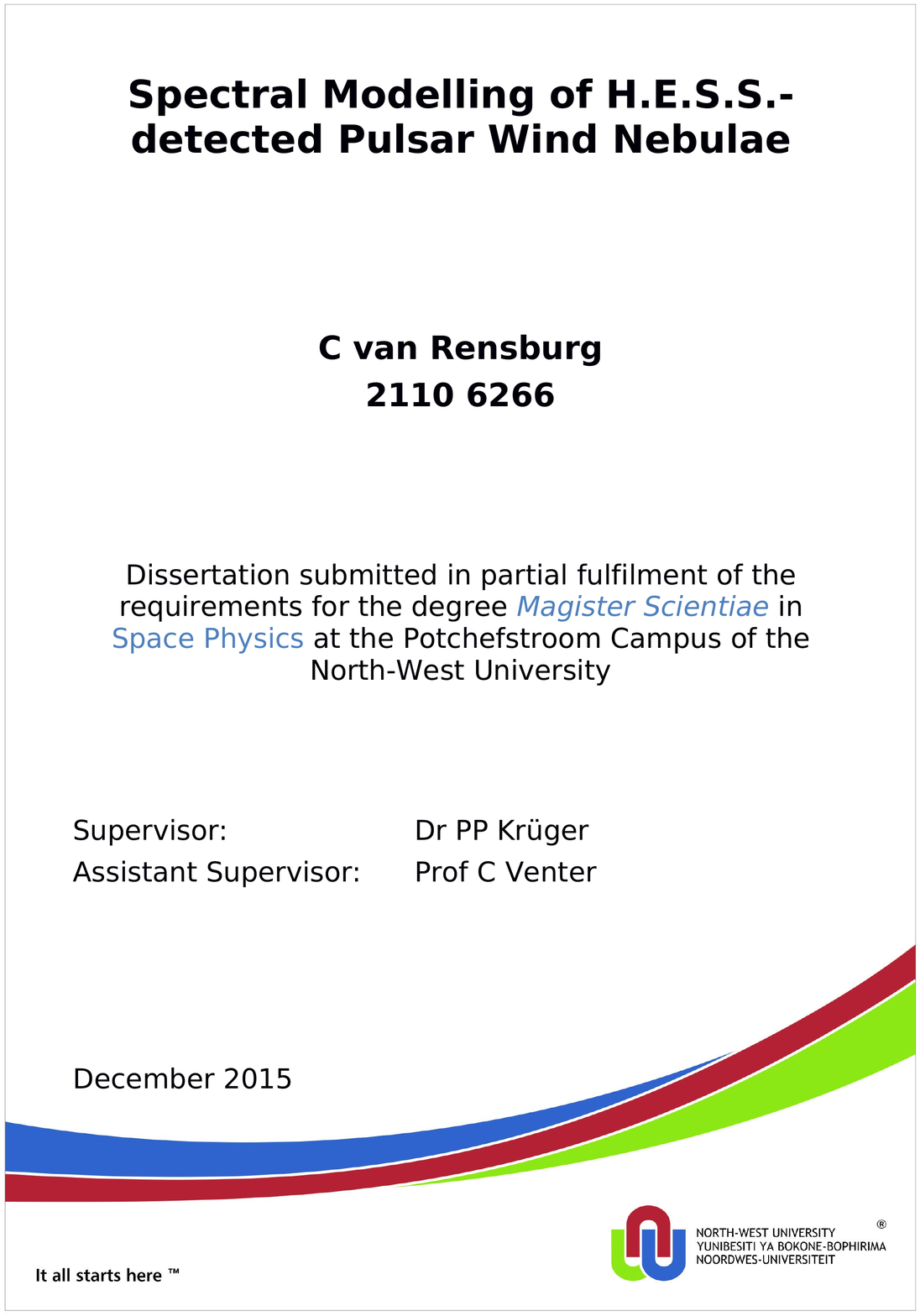}}
\end{figure}
\clearpage

\frontmatter	  % Begin Roman style (i, ii, iii, iv...) page numbering

% Set up the Title Page
\title  {Spectral Modelling of H.E.S.S.-detected Pulsar Wind Nebulae}
\authors  {\texorpdfstring
            {\href{your web site or email address}{Carlo van Rensburg}}
            {Carlo van Rensburg}
            }
\addresses  {\groupname\\\deptname\\\univname}  % Do not change this here, instead these must be set in the "Thesis.cls" file, please look through it instead
\date       {\today}
\subject    {}

%\maketitle
%% ----------------------------------------------------------------

\setstretch{1.3}  % It is better to have smaller font and larger line spacing than the other way round

% Define the page headers using the FancyHdr package and set up for one-sided printing
\fancyhead{}  % Clears all page headers and footers
\rhead{\thepage}  % Sets the right side header to show the page number
\lhead{}  % Clears the left side page header

\pagestyle{fancy}  % Finally, use the "fancy" page style to implement the FancyHdr headers

%% ----------------------------------------------------------------

%% ----------------------------------------------------------------
% The "Funny Quote Page"
\pagestyle{empty}  % No headers or footers for the following pages

\null\vfill
% Now comes the "Funny Quote", written in italics
\textit{``Not only is the Universe stranger than we think, it is stranger than we can think.”''}

\begin{flushright}
Werner Heisenberg
\end{flushright}

\vfill\vfill\vfill\vfill\vfill\vfill\null
\clearpage  % Funny Quote page ended, start a new page
%% ----------------------------------------------------------------

% The Abstract Page
\addtotoc{Abstract}  % Add the "Abstract" page entry to the Contents
\abstract{
\addtocontents{toc}{\vspace{1em}}  % Add a gap in the Contents, for aesthetics
In the last decade, ground-based Imaging Atmospheric Cherenkov Telescopes have discovered about 175 very-high-energy (VHE; $E >$ 100 GeV) gamma-ray sources, with more to follow with the development of H.E.S.S.\ II and CTA. Nearly 40 of these are confirmed pulsar wind nebulae (PWNe). We present results from a leptonic emission code that models the spectral energy density of a PWN by solving a Fokker-Planck-type transport equation and calculating inverse Compton and synchrotron emissivities. Although models such as these have been developed before, most of them model the geometry of a PWN as that of a single sphere. We have created a time-dependent, multi-zone model to investigate changes in the particle spectrum as the particles traverse through the PWN, by considering a time and spatially-dependent magnetic field, spatially-dependent bulk particle motion causing convection, diffusion, and energy losses (SR, IC and adiabatic). Our code predicts the radiation spectrum at different positions in the nebula, yielding novel results, e.g., the surface brightness versus the radius and the PWN size as function of energy. We calibrated our new model against more basic models using the observed spectrum of PWN G0.9+0.1, incorporating data from H.E.S.S.\ as well as radio and X-ray experiments. We fit our predicted radiation spectra to data from G21.5$-$0.9, G54.1+0.3, and HESS J1356$-$645 and found that our model yields reasonable results for young PWNe. We next performed a parameter study which gave significant insight into the behaviour of the PWN for different scenarios. Our model is now ready to be applied to a population of PWNe to probe possible trends such as the surface brightness as a function of spin-down of the pulsar.\\

}

\clearpage  % Abstract ended, start a new page
%% ----------------------------------------------------------------

\setstretch{1.3}  % Reset the line-spacing to 1.3 for body text (if it has changed)

\pagestyle{fancy}  %The page style headers have been "empty" all this time, now use the "fancy" headers as defined before to bring them back

%% ----------------------------------------------------------------
\lhead{\emph{Contents}}  % Set the left side page header to "Contents"
\tableofcontents  % Write out the Table of Contents

%% ----------------------------------------------------------------
\lhead{\emph{List of Figures}}  % Set the left side page header to "List if Figures"
\listoffigures  % Write out the List of Figures

%% ----------------------------------------------------------------
\lhead{\emph{List of Tables}}  % Set the left side page header to "List of Tables"
\listoftables  % Write out the List of Tables

%% ----------------------------------------------------------------
%\setstretch{1.5}  % Set the line spacing to 1.5, this makes the following tables easier to read
%\clearpage  % Start a new page
%\lhead{\emph{Abbreviations}}  % Set the left side page header to "Abbreviations"
%\listofsymbols{ll}  % Include a list of Abbreviations (a table of two columns)
%{
% \textbf{Acronym} & \textbf{W}hat (it) \textbf{S}tands \textbf{F}or \\
%\textbf{LAH} & \textbf{L}ist \textbf{A}bbreviations \textbf{H}ere \\

%}

%% ----------------------------------------------------------------
%\clearpage  % Start a new page
%\lhead{\emph{Physical Constants}}  % Set the left side page header to "Physical Constants"
%\listofconstants{lrcl}  % Include a list of Physical Constants (a four column table)
%{
% Constant Name & Symbol & = & Constant Value (with units) \\
%Speed of Light & $c$ & $=$ & $2.997\ 924\ 58\times10^{8}\ \mbox{ms}^{-\mbox{s}}$ (exact)\\

%}

%% ----------------------------------------------------------------
%\clearpage  %Start a new page
%\lhead{\emph{Symbols}}  % Set the left side page header to "Symbols"
%\listofnomenclature{lll}  % Include a list of Symbols (a three column table)
%{
% symbol & name & unit \\
%$a$ & distance & m \\
%$P$ & power & W (Js$^{-1}$) \\
%& & \\ % Gap to separate the Roman symbols from the Greek
%$\omega$ & angular frequency & rads$^{-1}$ \\
%}
%% ----------------------------------------------------------------
% End of the pre-able, contents and lists of things
% Begin the Dedication page

\addtocontents{toc}{\vspace{2em}}  % Add a gap in the Contents, for aesthetics

%% ----------------------------------------------------------------
\mainmatter	  % Begin normal, numeric (1,2,3...) page numbering
\pagestyle{fancy}  % Return the page headers back to the "fancy" style

% Include the chapters of the thesis, as separate files
% Just uncomment the lines as you write the chapters

% Chapter 1
	
\chapter{Introduction} % Write in your own chapter title
\label{Chapter1} 
\lhead{Chapter 1. \emph{Introduction}} % Write in your own chapter title to set the page header

\begin{figure}[h]
\centering
\begin{minipage}[b]{4in}
\centering
\includegraphics[width=3in]{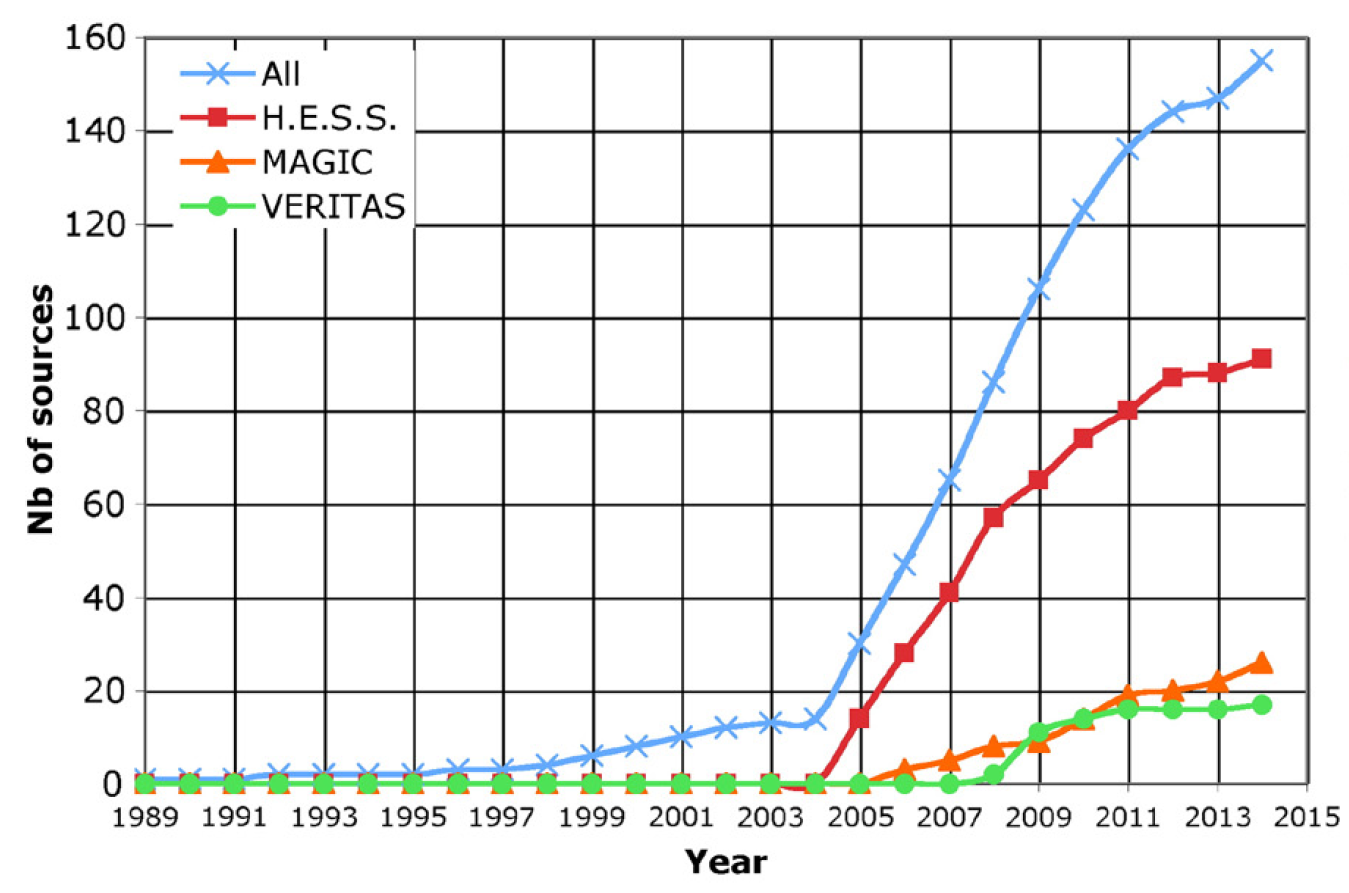}
\caption{\label{fig:degrange}Discovery of VHE gamma-ray sources, also indicating the contribution of H.E.S.S., MAGIC, and VERITAS \citep{Degrange2015}.}
\end{minipage} 
\end{figure}

In the last decade, ground-based Imaging Atmospheric Cherenkov Telescopes (IACTs) have discovered almost 175 very-high-energy (VHE, $E >$ 100 GeV) $\gamma$-ray sources. \cite{Hewitt2015} mention that, as of December 2014, nearly 40 of these are confirmed pulsar wind nebulae (PWNe). A systematic search with the \textit{Fermi}-LAT for GeV emission in the vicinity of TeV-detected sources yielded five high-energy gamma-ray PWNe and eleven PWN candidates. Other VHE source classes include supernova remnants, active galactic nuclei, or unidentified sources\footnote{tevcat.uchicago.edu}.  A subset of the unidentified sources may eventually turn out to be PWNe. Figure~\ref{fig:degrange} shows how the number of known VHE sources has increased over time, including the contribution of the three main ground-based gamma-ray telescopes, namely High Energy Stereoscopic System (H.E.S.S.), Very Energetic Radiation Imaging Telescope Array System (VERITAS), and Major Atmospheric Gamma Imaging Cherenkov Telescopes (MAGIC) \citep{Degrange2015}.

\begin{figure}[h]
\centering
\begin{minipage}[b]{5in}
\centering
\includegraphics[width=4in]{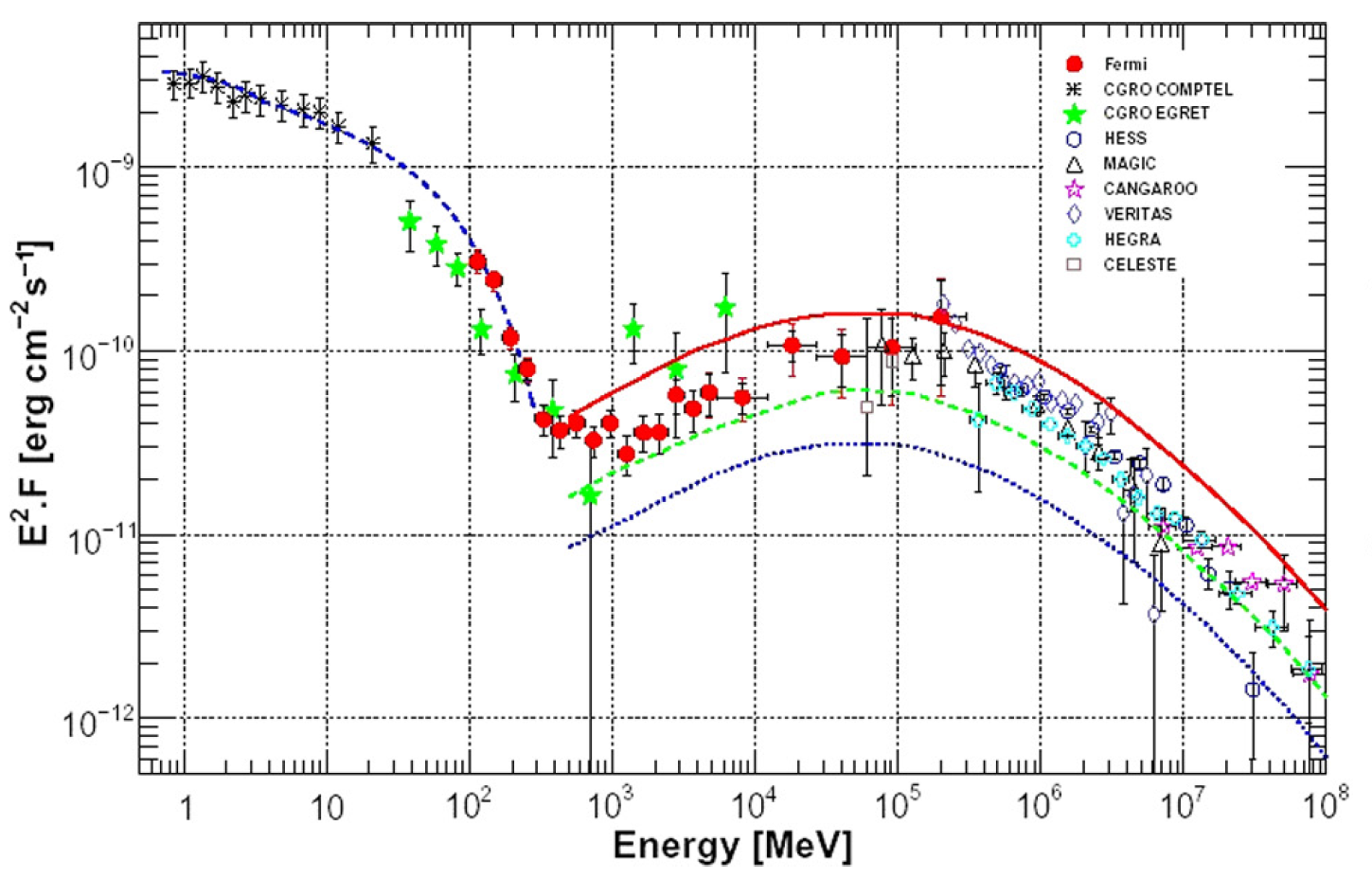}
\caption{\label{fig:degrange_2}Spectral energy distribution (SED) for the Crab Nebula showing emitted photons from radio to VHE gamma rays \citep{Degrange2015}.}
\end{minipage} 
\end{figure}

PWNe are associated with supernova remnants (SNRs). Historically they have been defined based on their observational properties, by having a centre-filled emission morphology, a flat spectrum at radio wavelengths, and a very broad spectrum of non-thermal emission ranging from the radio band all the way to high energy gamma rays \citep{Amato2014}. PWNe are visible through non-thermal emission from a magnetised plasma of relativistic particles fed by an energetic central pulsar. The non-thermal emission from the PWN is thought to result from two main processes: leptons in the plasma interacting with the magnetic field of the nebula, and producing synchrotron radiation (SR) up to several keV; secondly, low-energy photons, for example from the cosmic microwave background (CMB), can be upscattered to very high energies by energetic leptons via inverse Compton (IC) scattering.  Due to these two effects the radio, $X$-ray, and VHE $\gamma$-ray emissions are tightly linked, as all three emerge from the same lepton population. Figure~\ref{fig:degrange_2} shows the spectral energy distribution (SED) for the Crab Nebula to illustrate the two processes responsible for the non-thermal emission from the PWN, thus showing the SR bump on the left hand side and the IC bump on the right hand side.

Over the past decades theorists and observers alike have attempted to find and quantify relationships between the pulsar and the surrounding nebula. Although much progress has been made, with the young, nearby Crab Nebula being the archetypal source in this class, many unresolved issues remain. This interplay between theory and observations should also help us in identifying some unknown sources as being PWNe.

\section{Problem statement}
As mentioned above, there are many unanswered questions in PWN physics. For example, \cite{Gelfand2015} name a couple of these questions: How is the pulsar wind generated in the magnetosphere? What is responsible for converting the pulsar wind from a magnetically-dominated to a particle-dominated outflow? How are particles accelerated in these objects? \cite{Hewitt2015} add to these questions by stating that PWNe could be responsible for the so-called positron excess in the interstellar medium (ISM), where the ratio of positrons to electrons increase with energy. Experiments like \textit{Fermi}-LAT, \textit{PAMELA}, and \textit{AMS-02} have observed this increase in the positron-electron ratio for energies above 10 GeV, which is contrary to the standard theory that suggests that the ratio should simply decrease with energy. They state that as PWNe age, their magnetic field decreases which can cause particles that are trapped at the termination shock in the PWN to escape into the ISM. This may be the cause of the increased ratio of positrons to electrons. \cite{Kargaltsev2015} furthermore adds another question: the phenomena of `Crab flares'. It is currently known that the Crab Nebula exhibits a rapid variability in the GeV gamma-ray band. These rapid variabilities or flares cannot be predicted by current models and they do not fit into our current theory of PWNe and particle acceleration. This is also a challenge for observers, as the Crab Nebula is currently used as a standard candle for cross-calibrating X-ray and gamma-ray instruments. All these questions leave great room for research in this field.

\begin{figure}[htbp]
 \begin{minipage}{0.5\linewidth}
  \centering
  \includegraphics[width=1.0\linewidth]{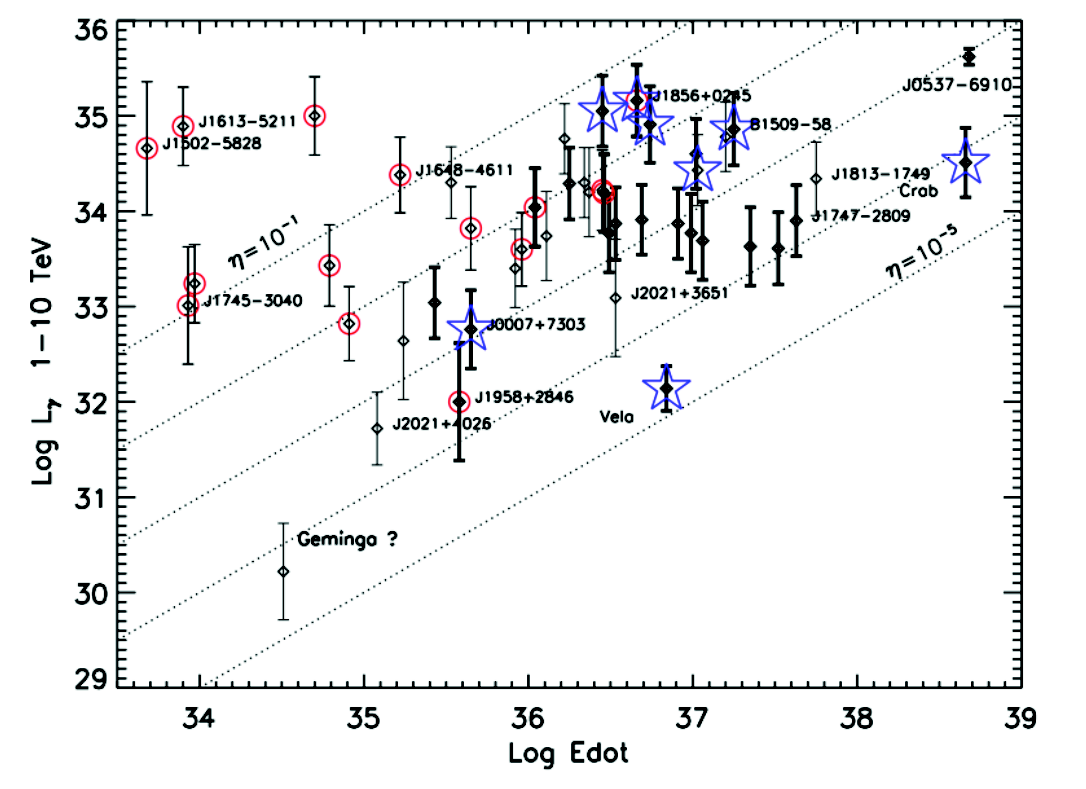}
  \caption{The non-correlation between the gamma-ray luminosity and the embedded pulsar's spin-down luminosity \citep{Kargaltsev2015}.}
  \label{fig:kargaltsev2015_2}
 \end{minipage}%
 \begin{minipage}{0.5\linewidth}
  \centering
  \includegraphics[width=1.0\linewidth]{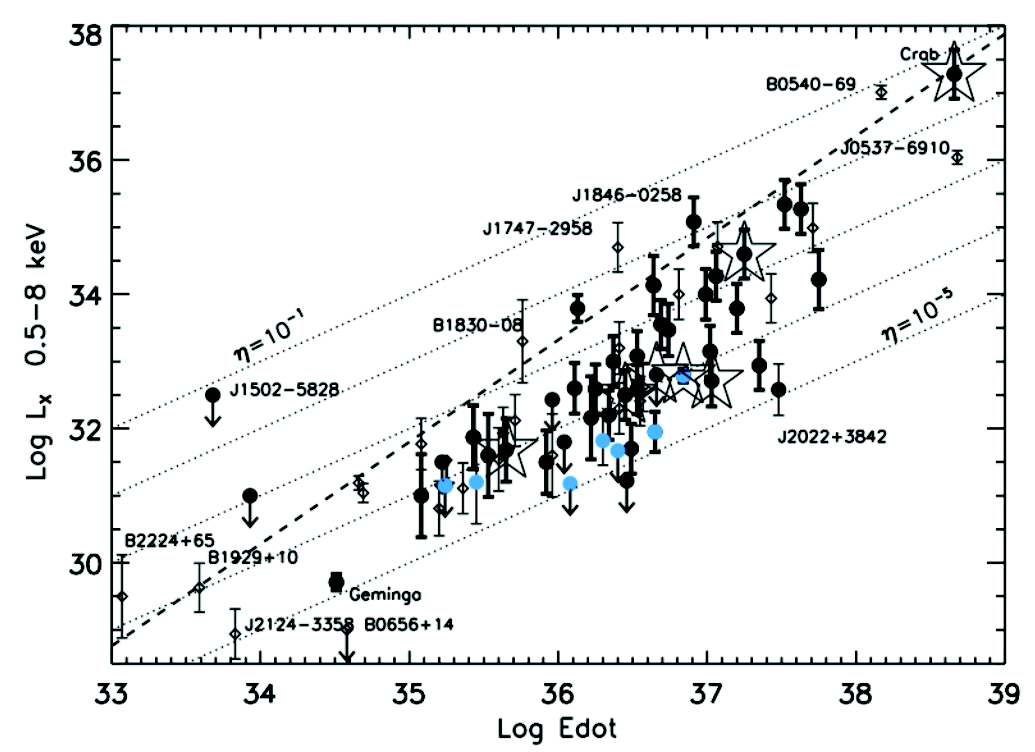}
  \caption{The correlation between the X-ray luminosity and the embedded pulsar's spin-down luminosity \citep{Kargaltsev2015}.}
  \label{fig:kargaltsev2015}
 \end{minipage}
\end{figure}

\cite{Kargaltsev2015} noted that the measured $\gamma$-ray luminosity (1$-$10 TeV) of the PWNe does not correlate with the spin-down luminosity of their embedded pulsars (Figure~\ref{fig:kargaltsev2015_2}). On the other hand, they found that the $X$-ray luminosity (0.5$-$8 keV) is correlated with the pulsar spin-down luminosity (Figure~\ref{fig:kargaltsev2015}). Furthermore, it is currently unknown whether there is any correlation between the TeV surface brightness of the PWNe and the spin-down luminosity of their embedded pulsars. Due to these reasons, it is necessary to create a spatially-dependent model to calculate the spectral energy density (SED) of the PWN. The spatial dependence will yield the flux as a function of the radius. This will allow us to model the surface brightness and thus probe this relationship between the TeV surface brightness of the PWNe and the spin-down energy of the embedded pulsar in future.

Currently there are too many free parameters in modelling PWNe and one zone models, although they can model the particle spectrum and SED from the PWN, can not constrain the magnetic field. We know that the magnetic field inside a PWN is not constant in space and one zone models use the average of the magnetic field over the entire PWN. This problem can solved with spatially-dependent modelling of PWNe and is addressed in this thesis.

\section{Research goal}
The main goal of this dissertation is to develop a time-dependent, multi-zone model of a PWN, including transport theory and pulsar physics. Such a code will allow us to model the evolving particle (lepton) population inside the PWN and thus also find the emitted SED. Similar models have been developed in the past by other researchers, but most of them model the PWN as a single sphere (no spatial dependence) and thus only model the average particle spectrum plus the radiation received from the PWN. With the development of new and improved telescopes, we are now able to view distant sources with a better angular resolution and better probe their detailed morphology. In Section~\ref{sec:HESS} a discussion on IACTs is given, describing the new developments of the current telescopes, e.g., the H.E.S.S. II telescope, and also the new Cherenkov Telescope Array (CTA) that will be built in the near future.

\begin{figure}[t]
\centering
\begin{minipage}[b]{5in}
\centering
\includegraphics[width=4in]{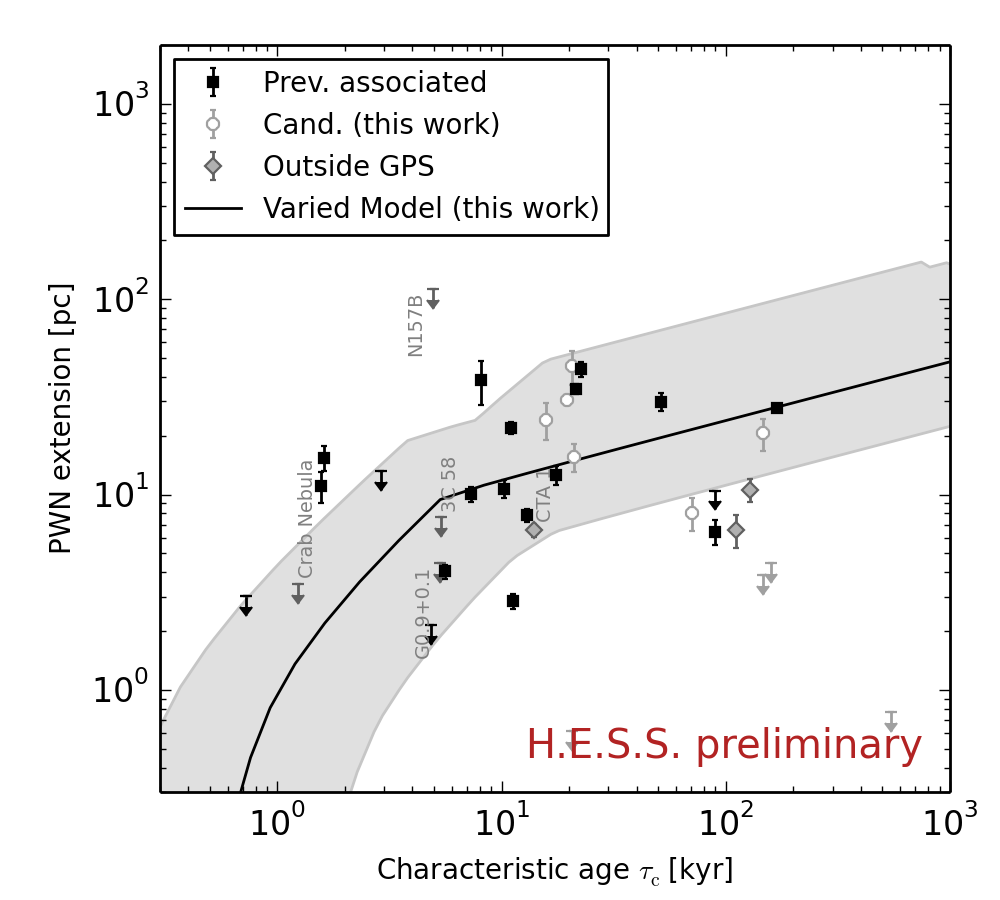}
\caption{\label{fig:klepser}PWN extension as a function of the characteristic age of the embedded pulsar, showing the increasing size of PWNe as they age \citep{Klepser2015}.}
\end{minipage} 
\end{figure}

Our model will allow us to calculate the evolution of the particle spectrum in the PWN, accompanied with the radiated SED, but most importantly it will allow us to calculate the surface brightness of the PWN, enabling us to make predictions regarding the size of the PWN. We will also be able to study how the size of the PWN changes with age and energy. This may be helpful to explain recent results by H.E.S.S. \cite[e.g.,][]{Klepser2015}. Figure~\ref{fig:klepser} shows the relationship between the extension of the PWN (PWN size) and the characteristic age of the embedded pulsar, indicating how PWNe increase in size as they age.

\section{Thesis outline}
\textbf{Chapter \ref{ch:ThBack}} is dedicated to giving the reader the necessary background to how a star transitions from a normal star to a PWN by undergoing a supernova explosion. Here I will explain the formation of a pulsar and some basic pulsar physics. I also summarised the characteristics and evolution of a PWN. I will discuss why a PWN is modelled with a two-component lepton injection spectrum. The two main processes that cause radiation from a PWN are SR and IC scattering. I discuss these two mechanisms in some detail and also describe the diffusion, convection, and adiabatic loss terms used to model the particle spectrum evolution.

In \textbf{Chapter \ref{ch:Model}} I discuss the development of our time-dependent, multi-zone model of a PWN. The geometry of the model is shown together with the form of the injected particle spectrum into the PWN. This is then used to show how the particle spectrum is calculated by taking into account the transport of particles, including effects that cause the particles to lose energy. The SED is calculated from the known particle spectrum and this SED is then projected onto a flat surface by doing a line-of-sight integration of the radiation to find the PWN image as viewed from Earth.

I will show the results from the PWN model in \textbf{Chapter \ref{ch:Results}} by first calibrating the model with other (spatially independent) models for PWN G0.9+0.1 and also a couple of other sources. I also show the results of a parameter study to investigate the effects of all the free parameters on the model predictions.

The conclusions and final remarks are given in \textbf{Chapter \ref{ch:concl}}.

Parts of this research have been published in \cite{vRensburg2014}.

\chapter{Theoretical background} % Write in your own chapter title
\label{ch:ThBack}
\lhead{Chapter \ref{ch:ThBack}. \emph{Theoretical Background}} % Write in your own chapter title to set the page header
In this chapter, I will discuss some background which will provide context for the modelling done in the next chapters of this dissertation. I will start by discussing what supernovae and pulsars are in Sections \ref{sec:Supernova} and \ref{sec:Pulsars}, give the definition of a PWN in Section~\ref{sec:PWN}, describe the relevant radiation processes for a PWN in Section~\ref{sec:rad_mech}, and discuss the diffusion, convection and other energy loss processes impacting the particle transport in Section~\ref{sec:Diff}. Lastly, I will discuss the H.E.S.S. telescope in Section~\ref{sec:HESS}, as well as the workings of Atmospheric Cherenkov telescopes (ACTs).

\section{Supernovae}
\label{sec:Supernova}
This dissertation is about the modelling of PWNe, which are directly related to pulsars as their name implies. Therefore, the first part of this chapter is dedicated to a short overview of the origin of pulsars and their link to supernovae.

\subsection{Thermonuclear supernovae (Type Ia)}
\begin{figure}[h]
\centering
\begin{minipage}[b]{5in}
\centering
\includegraphics[width=4in]{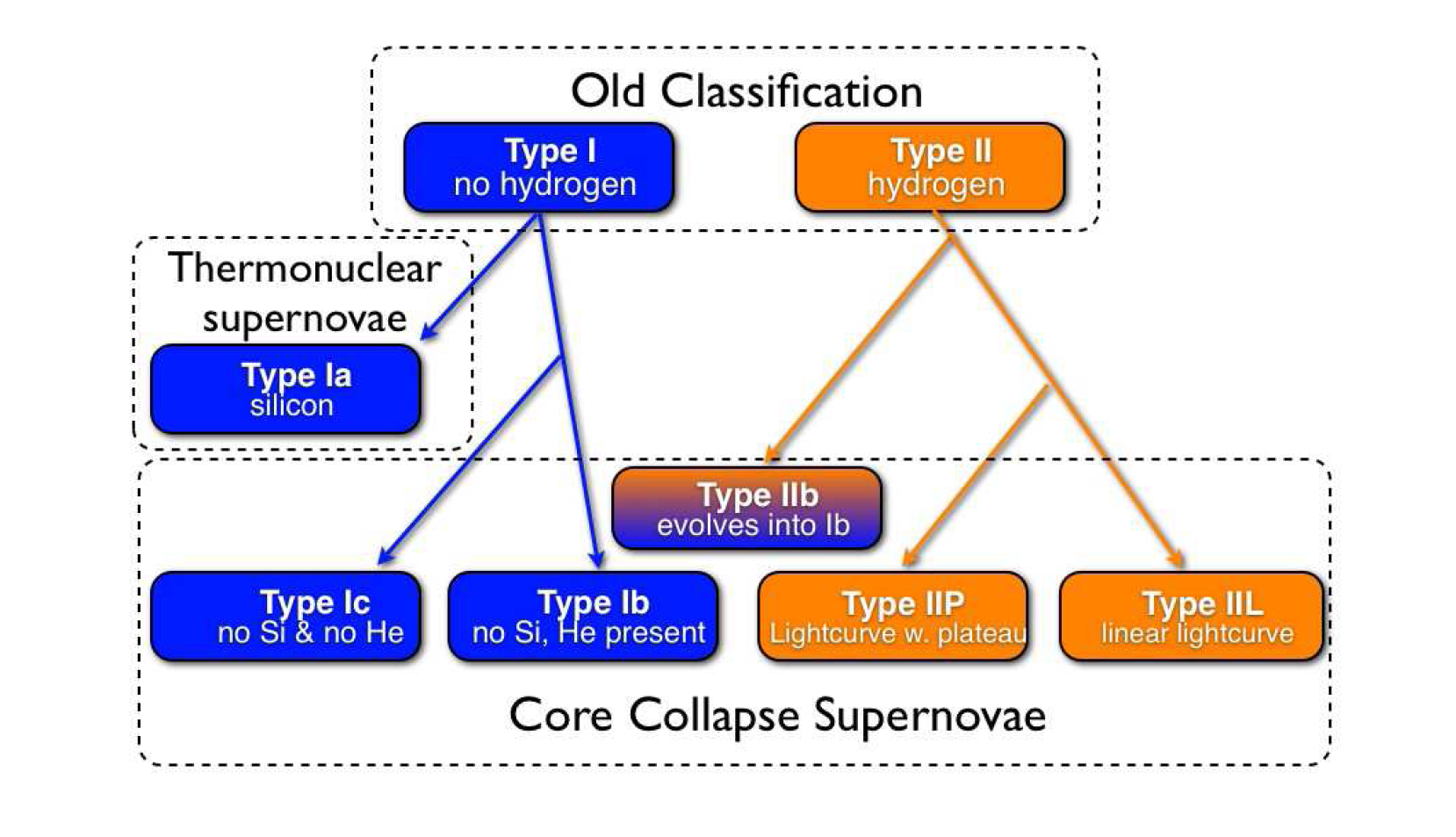}
\caption{\label{fig:SNR_class}The classification of supernovae, based on optical spectroscopy and light-curve shape \citep{Vink2012}.}
\end{minipage} 
\end{figure}

Supernova explosions are some of the most violent explosions in the universe, indicating the end of a stellar life cycle. There are two types of supernova explosions. The first type is a thermonuclear supernova (Type Ia) in which matter is accreted by a white dwarf from a companion star, or where a merger of two white dwarfs take place \citep{Schaefer2012}. According to \cite{Vink2012}, Type Ia supernovae do not result in the formation of a neutron star and are therefore not associated with PWNe. Therefore further detail relating to this type of supernova will not be discussed. The second type of supernova is associated with the gravitational core-collapse of a massive star (Type Ib, Ic, II). \cite{Vink2012} describes how these are categorized by the different optical spectra they produce. Figure~\ref{fig:SNR_class} shows what different line spectra are either present or not present in the different types of supernovae and also whether they are caused by thermonuclear reactions or a core collapse.

\subsection{Core-collapse supernovae}
\label{sec:SN_col}
According to \cite{Woosley2005} a massive star with a mass of $\gtrsim 8 \rm{M}_\odot$ will undergo fusion of hydrogen, helium, carbon, neon, oxygen, and silicon during its lifetime. After these fusion processes have been completed, an iron-rich core is left and this cannot supply energy through fusion to overcome the gravitational force acting on the star. The star will thus start to collapse.

Once the core collapse of the star has begun, two processes take over. First the electrons that are responsible for the thermal pressure inside the star are pushed into the iron core. Second, the radiation photo-disintegrates a fraction of the iron core into helium. Both of these processes will drain energy from the star, thereby accelerating the gravitational-collapse process. In the collapse process, a proto-neutron star is formed, where the short-range nuclear forces stop the collapse. This proto-neutron star will radiate approximately $10^{53}$ erg of energy in the form of neutrinos within a few seconds, the remnant being a neutron star with a radius of approximately $10$ km.
  
Approximately $10^{51}$ erg of kinetic energy is deposited into the stellar material surrounding the proto-neutron star, creating a bubble of radiation and electron-positron pairs. The expansion of the stellar material into the interstellar medium (ISM) is supersonic. This creates a forward shock wave that accelerates the ambient matter. The ambient matter collects in a thin shell behind the forward shock, creating a well-known shell-type supernova remnant (SNR).

According to \cite{McKee1974} the pressure inside the shell will drop due to the adiabatic losses suffered by the ejecta, so that the pressure inside the shell will be lower than the pressure behind the forward shock. This will result in the reverse shock being forced back to the centre of the shell. As the forward shock moves outward into the ejecta, the reverse shock heats, compresses, and decelerates the ejecta. The ejecta are separated from the shocked ISM by means of the creation of the reverse shock. The time needed for this reverse shock to propagate back to the centre was derived by \cite{FerreiraDeJ2008} as
\begin{equation}
t_{\rm{rs}} = 4 \times 10^3 \left(\frac{\rho_{\rm{ism}}}{10^{-24}\rm{g\ cm^{-3}}}\right)^{-1/3} \left(\frac{E_{\rm{snr}}}{10^{51}\rm{erg}}\right)^{-45/100} \left(\frac{M_{\rm{ej}}}{3M_{\odot}}\right)^{3/4} \left(\frac{\gamma_{\rm{ej}}}{5/3}\right)^{-3/2} \rm{yr},
\label{revSh}
\end{equation}
where $\rho_{\rm{ism}}$ is the density of the ISM, $E_{\rm{snr}}$ is the kinetic energy released in the supernova explosion, and $M_{\rm{ej}}$ and $\gamma_{\rm{ej}}$ are the mass and adiabatic indices of the ejecta, respectively. By inserting typical values of $E_{\rm{ej}} = 10^{51}{\rm \ erg}$, $\gamma_{\rm{ej}} = 1.67$, $M_{\rm{ej}} = 5 M_{\odot}$, and $\rho_{\rm{ism}} = 10^{-24}\rm{g\ cm^{-3}}$, we find $t_{\rm{rs}} \approx 6~000 \ \rm{yr}$.

\section{Pulsars}
\label{sec:Pulsars}
\cite{Lyne2006} mentions that in 1934 two astronomers, Walter Baade and Fritz Zwicky, proposed the existence of a new type of star called a neutron star. Such a neutron star represents one endpoint of a stellar life cycle. They wrote:

\begin{minipage}[b]{5.5in}
\center
\textit{...with all reserve we advance the view that a supernova represents the transition of an ordinary star into a neutron star, consisting mainly of neutrons. Such a star may posses a very small radius and an extremely high density.}
\end{minipage}

It took more than 30 years after this remark before pulsars were discovered. The realisation that a pulsar is a rapidly-rotating neutron star finally validated this proposal.  For a full discussion on the discovery of pulsars, see \cite{Lyne2006}.

\cite{Richards1969} studied the pulsar NP 0532 and found that the period of the pulsar was not constant, but instead it increased as time passed. The rate of this increase $\dot{P} = dP/dt$ can be related to the loss in rotational kinetic energy $E_{\rm{rot}}$ from the pulsar \citep{Lorimer2005}
\begin{equation}
L = -\frac{dE_{\rm{rot}}}{dt} = -\frac{d(I \Omega ^2/2)}{dt} = -I \Omega \dot{\Omega} = 4\pi^2I\dot{P}P^{-3},
\label{eq:lorimerLOSS}
\end{equation}
where $\Omega = 2\pi /P$ is the angular speed, $I$ the moment of inertia, and $L$ (also sometimes denoted by $\dot{E}_{\rm{rot}}$) the spin-down luminosity of the pulsar. A large fraction of the spin-down luminosity is carried away from the pulsar in the form of a pulsar wind. A fraction $\eta_{\rm{rad}}$ of the spin-down luminosity is, however, converted into pulsed emission. The value of $\eta_{\rm{rad}}$ is very difficult to calculate, but \cite{Abdo2010} found in their first \textit{Fermi}-LAT catalogue that $\eta_{\rm{rad}} \approx 1 \% - 10 \% $, with $\eta_{\rm{rad}}\approx 1 \% $ for the Crab pulsar. The largest fraction of $\dot{E}_{\rm{rot}}$ is therefore converted into particle acceleration. This gives birth to the pulsar wind, which forms the PWN.

When modelling a PWN, one needs to know how much energy is available from the pulsar, which acts as a central energy source. \cite{Pacini1973} noted that, while the electrodynamics involving pulsars remain controversial, the rotational energy loss of a pulsar may be written as
\begin{equation}
L(t) = L_0 \left(1+\frac{t}{\tau_c}\right)^{-(n+1)/(n-1)},
\label{eq:spinDown}
\end{equation} 
where $L_0$ is the luminosity at the birth of the pulsar, and $n$ is the braking index of the pulsar given by \citep{Lorimer2005} 
\begin{equation}
n = \frac{\Omega \ddot{\Omega}}{\dot{\Omega}^2},
\label{eq:braking}
\end{equation}
and $t$ is the time. For a dipolar magnetic field in vacuum, $n=3$. We will use this value later on. Another variable used in the modelling of a PWN is the characteristic spin-down timescale of the pulsar, defined as \cite{VdeJager2007}
\begin{equation}
\tau_c = \frac{P}{(n-1)\dot{P}} = \frac{4 \pi^2 I}{(n-1)P_0^2L_0},
\label{eq:timescale}
\end{equation}
with $P_0$ the birth period of the pulsar.

\section{Pulsar wind nebulae}
\label{sec:PWN}
The earliest recording of a supernova (SN) explosion was in 1 054 AD \citep{Stephenson2002}. This object is known today as the Crab Nebula. For many years it was presumed that a 16th magnitude star was embedded in the SNR and this was confirmed in the late 1960s with the discovery of a 33-ms pulsar. This pulsar has a spin-down rate of 36 ns per day. The kinetic energy dissipated from the pulsar, as discussed in Section~\ref{sec:Pulsars}, was similar to the energy that was presumed to be injected into the SNR at that time \citep{Gold1969}. After this discovery a theoretical understanding was developed where instead of a pulsar being completely isolated and its magnetised relativistic pulsar wind expanding indefinitely, the pulsar is surrounded by the SN ejecta (Section~\ref{sec:SN_col}). The surrounding SN ejecta will reach an equilibrium point where its pressure will be equal to the ram pressure from the pulsar wind and a termination shock will form. This termination shock can accelerate the leptons in the pulsar wind by interacting with the frozen-in magnetic field of the pulsar and causing SR with energies ranging from radio to X-rays. These leptons can also interact with the cosmic microwave background radiation (CMBR), as well as infrared radiation from dust and starlight, causing IC scattering that can scatter photons up to GeV and TeV energies.

In Section~\ref{sec:PWN_char}, I will discuss the characteristics of a PWN and its evolution in Section~\ref{sec:PWN_evo}. In the modelling of the PWN, a two-component lepton spectrum is used, the reason for this being discussed in Section~\ref{sec:PWN_lept}. For more details, see, e.g., the reviews by \cite{Gaensler06}, \cite{Kargaltsev2008} for PWN physics and X-ray observations, and \cite{Amato2014} and \cite{Bucciantini2014} for PWN theory.

\subsection{Characteristics of a PWN}
\label{sec:PWN_char}
According to \cite{Djannati2009}, a PWN has the following defining characteristics:
\begin{itemize}
  \item \cite{Weiler1978} coined the phrase `plerion', which in Greek means ``filled bag". This refers to a filled morphology, being brightest at the centre and dimming in all directions towards the edges. This is observed in all directions at all wavelengths due to the constant injection of energy by the embedded central pulsar, accompanied by the cooling of particles as they diffuse through the PWN;
  \item It has a structured magnetic field as inferred from polarisation measurements;
  \item A PWN has an unusually hard synchrotron radio spectrum. If $N_{\rm{e}}$ is the particle number density, then the particle spectrum producing the radio emission is described by $N_{\rm{e}} \propto E^{-\alpha}$, with $\alpha$ having an index of 1.0~$-$~1.6;
  \item Particle re-acceleration occurs at the termination shock and can be described by a power law (towards higher energies) as $N_{\rm{e}} \propto E^{-\alpha}$, with $N_{\rm{e}}$ the particle number density and $\alpha \sim 2-3$. This and the previous point imply a 2 component lepton injection spectrum.
  \item Some of the observed PWNe have a torus as well as a jet in the direction of the rotational axis of the embedded pulsar. In these cases the torus displays an under-luminous region at approximately $r_{\rm{ts}} = 0.03 - 0.3$ pc, with $r_{\rm{ts}}$ the radius of the termination shock.
  \item There is evidence of synchrotron cooling which means that the size of the X-ray PWN decreases with increasing energy.
\end{itemize}
The characteristics of a PWN can be expanded even further by using VHE gamma-ray observations \citep{Venter_Cherenkov05}:
\begin{itemize}
  \item The magnetisation parameter $\sigma$ (ratio of electromagnetic to particle energy flux, \citealt{Kennel1984}) of the pulsar wind is less than unity, with $\sigma \approx 0.003$ for the Crab Nebula and $0.01 \leq \sigma \leq 0.1$ for the Vela PWN. This is small when compared to the magnetisation parameter inside the magnetosphere of a pulsar where $\sigma \approx 10^3$. 
  \item The magnetic field of a PWN can be very weak in the early epochs due to the rapid expansion of the PWN. This can cause the VHE gamma-ray producing electrons to survive for a long time. If the magnetic field drops below a few $\mu G$ it can lead to a source that is undetectable at synchrotron frequencies but still detectable at TeV energies. This is a possible explanation for the number of unidentified TeV sources seen by H.E.S.S. Alternatively, `relic PWN' may form in late stages of the evolution, where the $B$-field has also dropped, leading to VHE sources with no low-energy counterparts.
\end{itemize}

\subsection{PWN evolution}
\label{sec:PWN_evo}
\begin{figure}[h]
\centering
\begin{minipage}[b]{5in}
\centering
\includegraphics[width=4in]{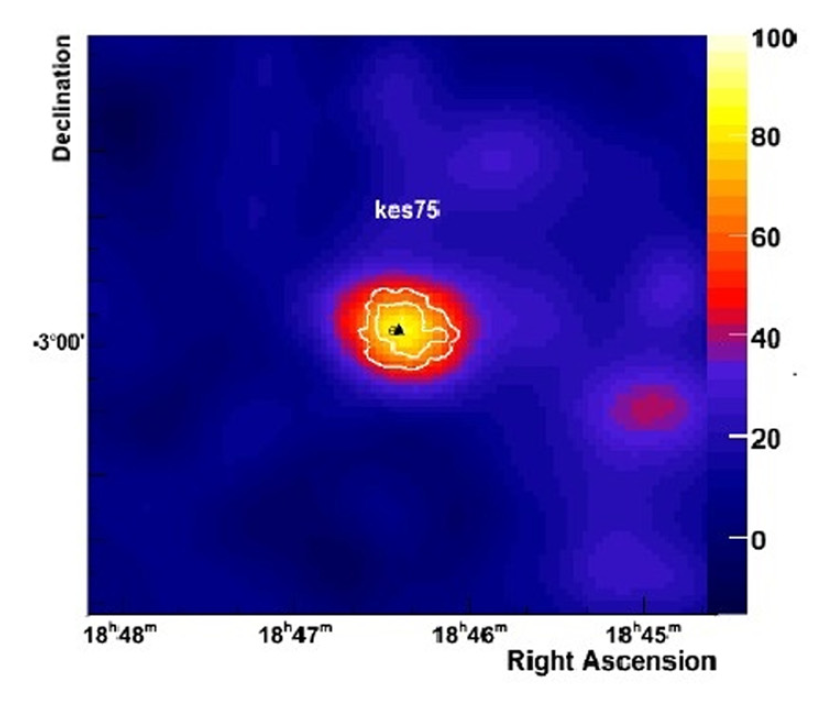}
\caption{\label{fig:PWN_young_1}PWN KES 75 showing a spherically symmetric PWN morphology usually associated with young pulsars \citep{Hewitt2015}.}
\end{minipage} 
\end{figure}

\begin{figure}[h]
\centering
\begin{minipage}[b]{5in}
\centering
\includegraphics[width=4in]{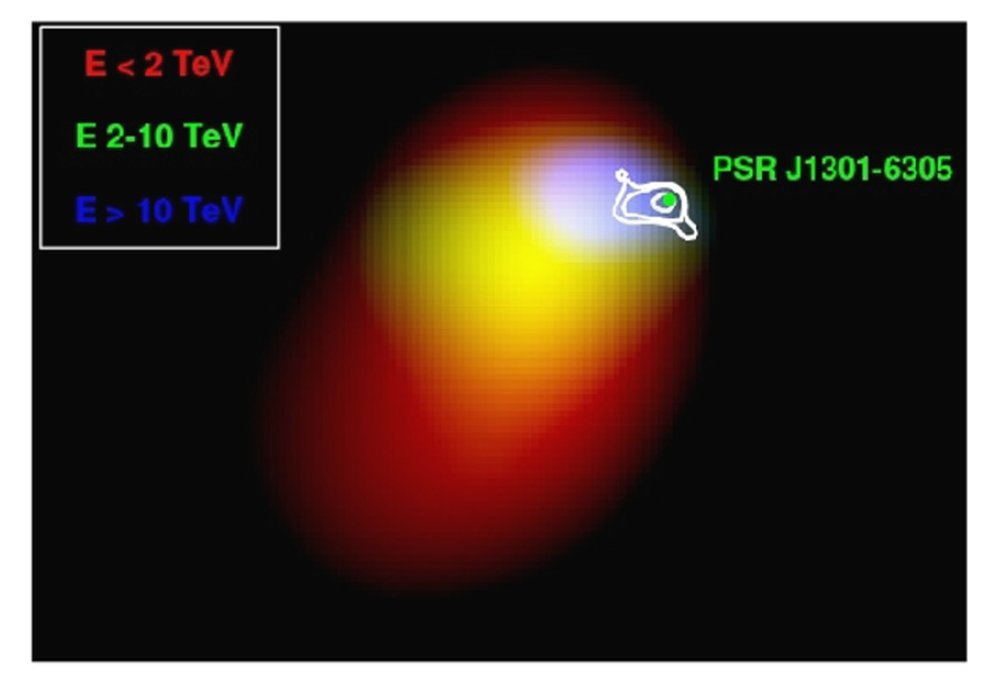}
\caption[HESS J1303$-$631 showing a `bullet-shaped', asymmetric PWN  \citep{Hewitt2015}.]{\label{fig:PWN_old_1}HESS J1303$-$631 showing a `bullet-shaped', asymmetric PWN. The red indicates photons below 2 TeV, yellow photons between 2 and 10 TeV, and blue photons above 10 TeV. \textit{XMM-Newton} X-ray contours are superimposed in white \citep{Hewitt2015}.}
\end{minipage} 
\end{figure}

The evolution of a PWN is tightly linked to the evolution of the pulsar's spin-down luminosity \citep{Gaensler06}. We consider two types of PWNe, namely young and old PWNe. Figure~\ref{fig:PWN_young_1} and \ref{fig:PWN_old_1} show a young and an old PWN.

At first the pulsar injects energy into the nebula, causing the PWN to expand supersonically into the slow-moving surrounding stellar ejecta. The rate at which this expansion occurs according to theoretical models is $R_{\rm{pwn}} \propto t^{\beta}$, where $R_{\rm{pwn}}$ is the outer boundary of the PWN, $t$ the age of the PWN, and $\beta \sim 1.1-1.2$ \citep{Reynolds1984}. We will consider PWNe in this first phase in our subsequent modelling. After the initial expansion phase, the reverse shock will propagate towards the centre of the SNR. When the reverse shock reaches $R_{\rm{pwn}}$ it initially compresses the PWN. This is followed by an unsteady contraction and expansion of $R_{\rm{pwn}}$, causing it to oscillate. After the oscillation phase of $R_{\rm{pwn}}$, the PWN enters another phase of steady expansion due to the ejecta being heated by the reverse shock. This second phase of steady expansion is characterised by the subsonic expansion of $R_{\rm{pwn}}$. According to \cite{Reynolds1984}, this expansion follows a power law given by $R_{\rm{pwn}} \propto t^{\beta}$, with $\beta \sim 0.3-0.7$.

As a first approach, it is commonly assumed that the PWN and the reverse shock are spherically symmetric. This is a good starting point but we know that this is not the full reality and in fact PWNe are much more complex. 

\cite{Blondin2001} performed simulations where the SNR is not expanding into a homogeneous ISM, but instead they added some inhomogeneity in the form of a pressure gradient to simulate the presence of, for example, a molecular cloud next to the SNR. As a result of the pressure inhomogeneity, the reverse shock will be asymmetric, causing the nebula to be displaced away from the pulsar. This causes the morphology of the PWN to have a `bullet' shape, with the pulsar located in the tip of the `bullet'. This is seen in many H.E.S.S. sources, so-called `offset-PWNe'. Figure~\ref{fig:PWN_old_1} shows such an example. Another cause for the PWN to exhibit a bullet shape can be due to the pulsar having some kick velocity with respect to the SNR, and thus it will also move away from the centre and form the bullet shape.

\subsection{Two-component lepton injection spectrum of the PWN}
\label{sec:PWN_lept}
In Section~\ref{sec:PWN_char}, I noted that a two-component lepton spectrum is required to explain the non-thermal emission from a PWN. Each of these components can be described by a power law given by $N_{\rm{e}} \propto E^{-\alpha}$, with $N_{\rm{e}}$ the particle number density. As mentioned, the first low-energy component responsible for the hard synchrotron radio spectrum and the GeV IC scattering has an index of $\alpha \sim 1-1.6$, while the second high-energy component responsible for the X-ray synchrotron and the TeV inverse Compton scattering has an index of $\alpha \sim 2-3$. 

Some PWN evolution models (see, e.g., \citealt{VdeJager2007}, \citealt{Zhang_2008}) use this broken-power-law distribution of the leptons as an injection spectrum into the PWN at the termination shock. They also assume that the transition from the one component to the other is a smooth one, thus having the same intensity at the transition. In contrast, \cite{Vorster2013} assumed that the transition from one component to the next is not necessarily smooth but that the injection spectrum can be modelled by a two-component particle spectrum that has a steep cutoff for the low-energy component in order to connect to the high-energy component, with each component characterised by a unique conversion efficiency. This causes a discontinuity in the particle spectrum but allows them to fit the steep slope of the X-ray data of many PWNe and is thus an observationally motivated injection spectrum. For the rest of my modelling however, I will use a broken power-law injection spectrum.

One can now ask about the origin of these two components as motivated by observations, and not simply a single power law spectrum. According to \cite{Axford_1977}, diffusive shock acceleration leads to a power-law spectrum with $N_{\rm{e}} \propto E^{-\alpha}$, with $\alpha = 2$ the maximum value. We can thus associate the high-energy component of the broken power law with this mechanism. It is however not so simple to explain the lower-energy component where $\alpha \sim 1.0-1.6$, as indicated by radio measurements. Relativistic MHD shock codes by \cite{Summerlin_2012} showed that it is possible for shocks to reproduce this hard spectrum if particles are subjected to shock drift acceleration. Particle-in-cell simulations by \cite{Spitkovsky_2008} also show that acceleration of particles at the termination shock leads to a Maxwellian spectrum with a non-thermal power-law tail. These ideas provide some basis for the assumption of a broken-power-law or two-component injection spectrum.

\section{Radiation mechanisms}
\label{sec:rad_mech}
Currently it is thought that IC scattering and SR are the two main mechanisms responsible for radiation from PWNe. These are the two processes invoked in Section~\ref{sec:RadSpec} where we calculate the SED. The SED consists of two components, where the high-energy component is due to the upscattering of photons to several TeV due to IC scattering (Section~\ref{sec:IC}) and the low-energy component spanning the radio and X-ray wavelengths is due to SR (Section~\ref{sec:SR}). 

\subsection{Inverse Compton scattering}
\label{sec:IC}
Here the upscattering of ``soft'' (low energy) background target photons to high energies when interacting with high-energy electrons is discussed. This process is called IC scattering.

The Thomson limit is valid when
\begin{equation}
\gamma \varepsilon \ll m_{\rm{e}}c^2, 
\label{eq:thom_lim}
\end{equation}
where $\gamma$ is the electron Lorentz factor, $m_{\rm{e}}$ is the mass of the electron, and $\varepsilon$ is the soft-photon energy. According to \cite{BlGould1970}, the mean energy of the Compton-scattered photon $\epsilon_1$ for an isotropic photon gas is given by
\begin{equation}
\langle \epsilon_1 \rangle = \frac{4}{3} \gamma^2 \langle \epsilon \rangle,
\end{equation}
where $\langle \epsilon \rangle$ is the mean energy of the soft photons. The total energy loss of a single electron is \citep{RybickiLightman1979}
\begin{equation}
-\frac{dE_{\rm{e}}}{dt} = \frac{4}{3} \sigma_{\rm{T}} c \gamma^2 U_{\rm{iso}},
\label{eq:IC_enrg_loss}
\end{equation}
where $\sigma_{\rm{T}} = 8 \pi r_0^2/3 = 6.65 \times 10^{-25} \rm{cm}^2$ is the Thomson cross section, with $r_0$ the Thompson scattering length (also known as the classical electron radius), and $U_{\rm{iso}}$ is the energy density of the isotropic photon field. The general IC scattered photon spectrum per electron is \citep{BlGould1970}
\begin{equation}
\frac{dN_{\gamma,\epsilon}}{dtd\epsilon_1}=\frac{\pi r_{0}^{2}c\:\:n(\epsilon)d\epsilon}{2\gamma^{4}\epsilon^{2}}\left(2\epsilon_{1}\:{\rm ln}\frac{\epsilon_{\gamma}}{4\gamma^{2}\epsilon}+\epsilon_{1}+4\gamma^{2}\epsilon-\frac{\epsilon_{1}^{2}}{2\gamma^{2}\epsilon}\right),
\end{equation}
where $n(\epsilon)$ is the photon number density associated with a blackbody distribution.

\begin{figure}[h]
\centering
\begin{minipage}[b]{5in}
\includegraphics[width=5in]{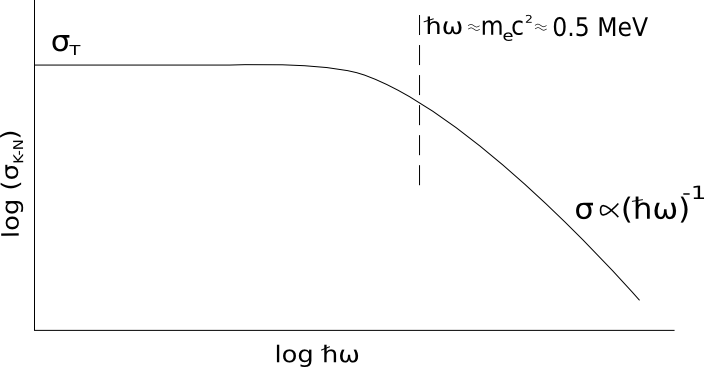}
\caption{\label{fig:Klein}A schematic diagram showing the dependence of the IC cross section on soft-photon energy. Arbitrary units are used. Adapted from \cite{Longair2011}.}
\end{minipage} 
\end{figure} 

The Klein-Nishina (K-N) limit is valid when
\begin{equation}
\gamma \varepsilon \gg m_{\rm{e}}c^2,
\label{eq:K-N_lim}
\end{equation}
and the scattered photon energy now becomes
\begin{equation}
\epsilon_1 \sim \gamma m_{\rm e}c^2.
\end{equation}
Figure~\ref{fig:Klein} shows how the Thomson cross section transitions to the (K-N) cross section as the soft-photon energy increases. This K-N cross section is given by \citep{RybickiLightman1979}
\begin{equation}
\sigma = \sigma_{\rm{T}}\frac{3}{4}\left[\frac{1+x}{x^3}\left\{\frac{2x(1+x)}{1+2x}-{\rm ln} (1+2x)\right\}+\frac{1}{2x}{\rm ln}(1+2x)-\frac{1+3x}{(1+2x)^2}\right],
\end{equation} 
with $x=\gamma \hbar \omega/m_{\rm e} c^2$. The single-electron energy loss rate in the extreme K-N for a blackbody photon distribution is given by \citep{BlGould1970}
\begin{equation}
-\frac{dE_{\rm e}}{dt} = \frac{1}{6}\pi r_{\rm e}^2\frac{(m_{\rm e}c\:kT)^2}{\hbar^3}\left[{\rm ln}\left(\frac{4\gamma\:kT}{m_{\rm e}c^2}\right)-1.98\right],
\end{equation}
where $k$ is the Boltzmann constant. The general equation for the upscattered photon spectrum per electron is given by \citep{Jones1968}
\begin{equation}
\frac{dN_{\gamma,\epsilon}}{dtdE_{\gamma}} = \frac{2\pi r_0^2m_{\rm e}c^3\:n(\epsilon)d\epsilon}{\gamma\epsilon}\left[2q\:{\rm ln}q + (1+2q)(1-q) + \frac{1}{2}\frac{(\Gamma_{\rm e}q)^2}{1+\Gamma_{\rm e}q}(1-q)\right],
\label{eq:IC_pord_rate}
\end{equation}
where $E_{\gamma} = \epsilon_{1}/\gamma m_{\rm e}c^2$ and $\Gamma_{\rm e}$ is the dimensionless parameter
\begin{equation}
\Gamma_{\rm e} = \frac{4\gamma}{m_{\rm e}c^2},
\end{equation}
and
\begin{equation}
q=\frac{E_{\gamma}}{\Gamma_{\rm e}(1-E_{\gamma})}.
\end{equation}
The total Compton spectrum can thus be calculated by integrating the production rate in Eq.~\eqref{eq:IC_pord_rate} over the soft-photon energy $\epsilon$ and Lorentz factor $\gamma$:
\begin{equation}
\left(\frac{dN}{d\epsilon_{1}}\right)_{\rm tot} = \int \!\!\!\!\! \int N_{\rm e}\left(\frac{dN_{\gamma,\epsilon}}{dtd\epsilon_{1}}\right)d\gamma d\epsilon,
\end{equation}
with $dN_{\rm e} = N_{\rm e}(\gamma)d\gamma$ the differential number of electrons per $\gamma$ interval. If we assume that the electron energy distribution is a power law, $N_{\rm e} \propto \gamma^{-p}$, interacting with a blackbody soft-photon distribution, then it follows that \citep{BlGould1970}
\begin{equation}
\left(\frac{dN}{d\epsilon_{1}}\right)_{\rm tot}\propto \epsilon_{1}^{-(p+1)/2} \rm \:\:\:\:- Thomson \:\:\:Limit,
\label{eq:thom_lim_spec}
\end{equation}
\begin{equation}
\left(\frac{dN}{d\epsilon_{1}}\right)_{\rm tot} \propto \epsilon_{1}^{-(p+1)} \rm - Extreme \:\:\:K-N \ Limit.
\end{equation}
Something to note is that the first expression in Eq.~(\ref{eq:thom_lim_spec}) is the same as for SR shown later in Eq.~(\ref{eq:SR_photon_spec}), but the spectrum is much softer in the extreme K-N regime.

\subsection{Synchrotron radiation}
\label{sec:SR}

\begin{figure}[h]
\centering
\begin{minipage}[b]{10cm}
\centering
\includegraphics[width=5cm]{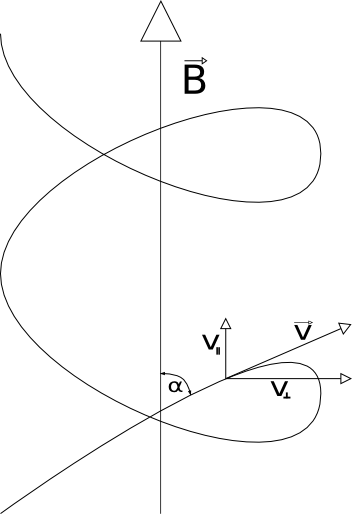}
\caption{\label{fig:Spiral}An electron spiralling around a magnetic field line, illustrating SR.}
\end{minipage} 
\end{figure} 

In this section, I discuss SR which is responsible for the low-energy component. SR occurs when charged particles (e.g., electrons) spiral around a magnetic field. Figure~\ref{fig:Spiral} is a schematic representation of an electron with a velocity $\vec{v}$ spiralling around a magnetic field $\vec{B}$ at a pitch angle $\alpha$. In the classical, non-relativistic case, a single particle gyrating in a magnetic field will radiate power according to the Larmor formula \citep{RybickiLightman1979}
\begin{equation}
P=\frac{2q^2a^2}{3c^3},
\end{equation}
where $a$ is the acceleration, and $q$ is the particle charge. If the relativistic case is considered and we assume that $dv_\parallel/dt = 0$, then the power radiated by an electron is given by
\begin{equation}
P=\frac{2q^2}{3c^3}\gamma^4\left(\frac{qB}{\gamma m_{\rm{e}} c}\right)^2 v_\perp^2,
\end{equation}
where $B$ is the magnetic field strength, and $v_{\perp}$ the the electron's speed perpendicular to the magnetic field. According to \cite{BlGould1970} we can also write the SR energy loss rate as 
\begin{equation}
\frac{dE_{\rm{SR}}}{dt} = -\left(\frac{2r_0^2}{3c}\right)\gamma^2 B^2 v_\perp^2,
\end{equation}
where $v_\perp^2 = v^2 \sin^2 \theta$, $r_0 = e^2/m_{\rm{e}}c^2$ for an electron, $e$ is the electron charge, and $\gamma$ is the electron's Lorentz factor. 

Next we need to calculate the radiative power from SR and to do this we rewrite the electron's speed as $\beta_\perp c$. Then by averaging over $\alpha$ for an isotropic distribution of velocities, we obtain $\langle \beta_{\perp} \rangle = \frac{2}{3}\beta^2$. Thus we find the total radiated power to be \citep{RybickiLightman1979}
\begin{equation}
P_{\rm{tot}} = \dot{E}_{\rm{SR}} = \frac{4}{3} \sigma_{\rm{T}} c \beta^2 \gamma^2 U_{\rm{B}} \propto E^2B^2,
\label{eq:SR_power_radiated}
\end{equation}
where $U_{\rm{B}} = \frac{B^2}{8\pi}$ is the magnetic energy density. The expression for $\dot{E}_{\rm{SR}}$ is similar to $\dot{E}_{\rm IC}$ (the the Thomson limit) in Eq.~\eqref{eq:IC_enrg_loss}. 

We can now calculate the single-particle spectrum. This spectrum is characterized by a critical frequency near which the spectrum reaches a maximum:
\begin{equation}
\omega_{\rm{c}} = \frac{3}{2} \gamma^3 \omega_{\rm{B}} \sin \alpha = \frac{3}{2}e\frac{B\sin \alpha}{m_{\rm{e}}c}\gamma^2,
\end{equation}
where $\omega_{\rm{B}}$ is the gyration frequency of rotation given by
\begin{equation}
\omega_{\rm{B}} = \frac{eB}{\gamma m_{\rm{e}}c}.
\end{equation}
The power emitted per frequency by a single electron is given by
\begin{equation}
P(\omega) = \frac{\sqrt{3}}{2\pi}\frac{e^3B\sin \alpha}{m_{\rm{e}}c}F(x),
\end{equation}
where
\begin{equation}
F(x) = x \int_x^{\infty}K_{\frac{5}{3}}(\xi)d\xi,
\end{equation}
with $x \equiv \omega/\omega_{\rm{c}}$, and $K_{\frac{5}{3}}$ a modified Bessel function of the second kind of order 5/3. The function $F(x)$ has different asymptotic forms for small and large values for $x$:
\begin{equation}
F(x) \sim \frac{4\pi}{\sqrt{3}\Gamma(1/3)}\left(\frac{x}{2}\right)^{1/3},\qquad x \ll 1,
\end{equation}
\begin{equation}
F(x) \sim \left(\frac{\pi}{2}\right)^{1/2}e^{-x}x^{1/2}, \qquad \quad x \gg 1.
\end{equation}
The spectral maximum occurs at $\omega_{\rm{max}} = 0.29\omega_{\rm{c}}$ \citep{Longair2011}.

If the number density $N_{\rm{e}}(E_{\rm{e}})$ of electrons in an energy range $(E_{\rm{e}},E_{\rm{e}}+dE_{\rm{e}})$, can be expressed as a power law
\begin{equation}
N_{\rm{e}}(E_{\rm{e}})dE_{\rm{e}} = CE_{\rm{e}}^{-p}dE_{\rm{e}}, \qquad E_1 < E_{\rm{e}} < E_2,
\end{equation}
one can show that the total SR power radiated by these particles is 
\begin{equation}
P_{\rm{tot}}(\omega) \propto \omega ^{-(p-1)/2} \propto \omega^s,
\end{equation}
where $s$ the index of the energy spectrum. Thus the photon spectrum is then similar to IC (in the Thomson limit) and is given by
\begin{equation}
\frac{dN_{\gamma}}{dE_{\gamma}} \propto \omega ^{-(p+1)/2}.
\label{eq:SR_photon_spec}
\end{equation}

\section{Diffusion, convection and adiabatic losses}
\label{sec:Diff}
According to \cite{Chen1984}, diffusion by means of Coulomb collisions has been understood for a long time. The diffusion coefficient $\kappa$ was thought to have a $1/B^2$ dependence but this result could not be verified in any of the experiments done. In 1946 Bohm gave an semi-empirical formula for the diffusion coefficient in their magnetic arc experiment. Their form of the diffusion coefficient was
\begin{equation}
\kappa = \frac{c}{3e}\frac{E}{B} = D_B.
\end{equation}
Any diffusion process following this law is thus called Bohm. 
%There are many different forms of diffusion. In the paper of \cite{Zheng2011} they study the solutions of the momentum diffusion equation of the particle distribution in a turbulent magnetic field. If the turbulence is given by a power spectrum of $W(k) \propto k^{-q}$, with $k$ the wave number, then the different types of diffusion that they consider is Bohm diffusion $(q=1)$, Kolmogorov diffusion $(q=5/3)$, Kreichnan diffusion $(q = 3/2)$, and the hard sphere approximation $(q = 2)$. These are different forms of diffusion and in their paper they show how the diffusion coefficient change for the different forms of the turbulent magnetic field.  

We currently don't have a very good idea of how turbulent the magnetic field is inside the PWN, although we have some idea from the polarized radio spectrum. Due to this uncertainty we do not know what form of diffusion coefficient we have to use and therefore we chose Bohm diffusion as a first approximation. To assume Bohm diffusion is a fairly common practice as it describes diffusion that is perpendicular to the magnetic field. In the modelling of the PWN, we use a axially-symmetric (azimuthal) magnetic field and thus we are only interested in radial diffusion perpendicular to the magnetic field, which will lead to particles moving from one zone to the next in the PWN. Due to this uncertainty in the form of the magnetic field we parametrised the magnetic field as
\begin{equation}
\kappa = \kappa_0\left(\frac{E}{E'_0}\right)^q
\end{equation}
adding two free parameters. The results from this is shown in Section~\ref{sec:Var_diff123}.

Convection is mass transfer due to the bulk motion of a fluid. We model the convection in the PWN by using a parametrized form for the velocity profile inside the PWN given by
\begin{equation}
V(r) = V_0\left(\frac{r}{r_0}\right)^{\alpha_V},
\label{V_profile_2}
\end{equation}
with $\alpha_V$ the velocity profile parameter and $r_0$ a reference radius (termination shock radius) where $V=V_0$. 

The particles will lose energy due to the PWN expansion in the form of adiabatic cooling, and the rate at which they lose energy is given by \citep[e.g.,][]{Zhang_2008}
\begin{equation}
\dot{E}_{\rm{ad}} = \frac{1}{3}(\nabla \cdot \mathbf{V})E_{\rm{e}}.
\label{eq:Edot_adiabatic}
\end{equation}
We assume the magnetic field in the PWN is azimuthal and may be parametrised by
\begin{equation}
B(r,t) = B_{\rm{age}}\left(\frac{r}{r_0}\right)^{\alpha_B}\left(\frac{t}{t_{\rm{age}}}\right)^{\beta_B},
\label{B_Field_1}
\end{equation}
with $B_{\rm{age}}$ the present-day magnetic field at $r=r_0$ and $t=t_{\rm{age}}$, with $t$ the time since the PWN's birth, and $\alpha_B$ and $\beta_B$ the magnetic field parameters. The magnetic field and bulk motion are linked together by Faraday's law of induction
\begin{equation}
\frac{\partial \mathbf{B}}{\partial t} = \nabla \times \left(\mathbf{V} \times \mathbf{B}  \right).
\label{eq:B_V}
\end{equation}
The Lorentz force $\mathbf{F} = q(\mathbf{E} + \mathbf{V} \times \mathbf{B})$ is set to zero, assuming that the plasma is a good conductor and thus a force-free environment. This assumption together with the Maxwell equation 
\begin{equation}
\frac{\partial \mathbf{B}}{\partial t} = -\nabla \times \mathbf{E},
\label{eq:maxwell}
\end{equation}
yields Eq \eqref{eq:B_V}. We assume that the timescale over which the magnetic field changes is much longer than the spatial scale of change for the velocity and magnetic field. Thus we set 
\begin{equation}
\frac{\partial \mathbf{B}}{\partial t} \simeq 0
\end{equation} 
so that
\begin{equation}
\nabla \times \left(\mathbf{V} \times \mathbf{B}  \right) \simeq 0.
\end{equation}
From this, and assuming spherical symmetry, Eq.~\eqref{eq:B_V} reduces to
\begin{equation}
VBr={\rm{constant}}=V_0B_0r_0.
\label{eq:vbr=c}
\end{equation}
It can now be shown that by placing Eq.~\eqref{V_profile_2} and Eq.~\eqref{B_Field_1} into Eq.~\eqref{eq:vbr=c}, the following relation holds:
\begin{equation}
\alpha_V+\alpha_B=-1.
\label{eq:a_v+a_b=-1}
\end{equation} 
This is a very important relationship and in the parameter study in Section~\ref{sec:alph_Valph_B} it will be shown what effect this has on the model. The ways the magnetic field and the bulk particle motion are implemented to the model are shown in Appendix~\ref{appen:disctr}, from Eq.~\eqref{ap_eq:qwerty} onward.

\section{Atmospheric Cherenkov telescopes (ACTs) and the High Energy Stereoscopic System (H.E.S.S.)}
\label{sec:HESS}
Our PWN model predicts a multi-wavelength radiation spectrum, ranging from the radio band to  the TeV band. In this section, however, I will discuss ACTs and the H.E.S.S.\ telescope in more detail. This is because we are members of the H.E.S.S.\ Collaboration as well as the South African Gamma-Ray Astronomy Programme (SA-GAMMA), and therefore our focus lies with radiation in the gamma-ray waveband in particular.

\subsection{Atmospheric Cherenkov telescopes (ACTs)}
There are currently three major ground-based gamma-ray telescopes in the world. These are H.E.S.S. in the Gamsberg mountain range in Namibia, the Very Energetic Radiation Imaging Telescope Array System (VERITAS) located at the basecamp of the Fred Lawrence Whipple Observatory in southern Arizona, and the Major Atmospheric Gamma Imaging Cherenkov Telescopes (MAGIC) located near the top of the Roque de los Muchachos on the Canary island of La Palma. These telescopes' predecessors were the High Energy Gamma Ray Astronomy (HEGRA) experiment that was located on La Palma in the Canary Islands and the CANGAROO telescope in Australia's Outback. The future of ACTs is the Cherenkov Telescope Array (CTA). This telescope array will have sites in both the northern and southern hemisphere and promises a factor of 5-10 improvement in sensitivity compared to current ground-based gamma-ray telescopes, as well as improved angular resolution. This telescope will have an energy range from well below 100 GeV to above 100 TeV\footnote{https://www.cta-observatory.org}.

\begin{figure}[h!]
\centering
\begin{minipage}[b]{5in}
\includegraphics[width=5in]{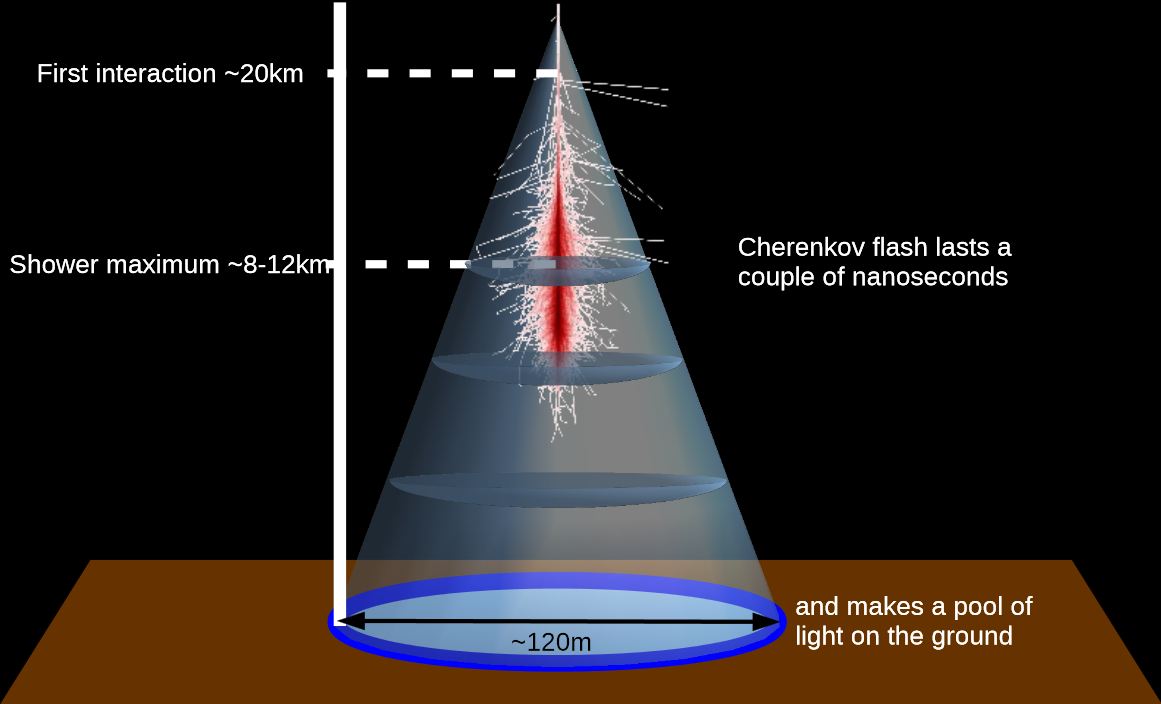}
\caption[Schematic view of a Cherenkov flash caused by a gamma ray]{\label{fig:act_1} Schematic view of a Cherenkov flash caused by a gamma ray (www.hermanusastronomy.co.za).}
\end{minipage} 
\end{figure} 

To view gamma rays with an ATC, the Cherenkov technique is used where an incident high-energy photon interacts with particles high up in the atmosphere and generates a shower of secondary particles. Figure~\ref{fig:act_1} is a schematic view of this process where the shower of particles reaches a maximum intensity at about 10~km and dies off deeper in the atmosphere. The particles essentially move at the speed of light in the atmosphere, emitting a faint blue light, \textit{Cherenkov radiation}, for a couple of nanoseconds. This blue flash of light illuminates the ground around the direction of the incident particle, creating a pool of light on the ground with a diameter of $\sim 120$ m. This is a very faint light flash, as a particle with an energy in the TeV range ($10^{12}$ eV) will only produce about 100 photons per m$^2$ at ground level. If a telescope is located within the light pool it will therefore ``see" the the air shower indirectly. The images seen by the telescope are the track of the air shower, which point back to the celestial body where the gamma ray originated. The intensity of the image can be used to calculate the energy of the incident gamma ray and the shape of the shower can be used to reject showers caused by other particles, e.g., cosmic rays.

\begin{figure}[t]
\centering
%\begin{minipage}[b]{5in}
\includegraphics[width=4in]{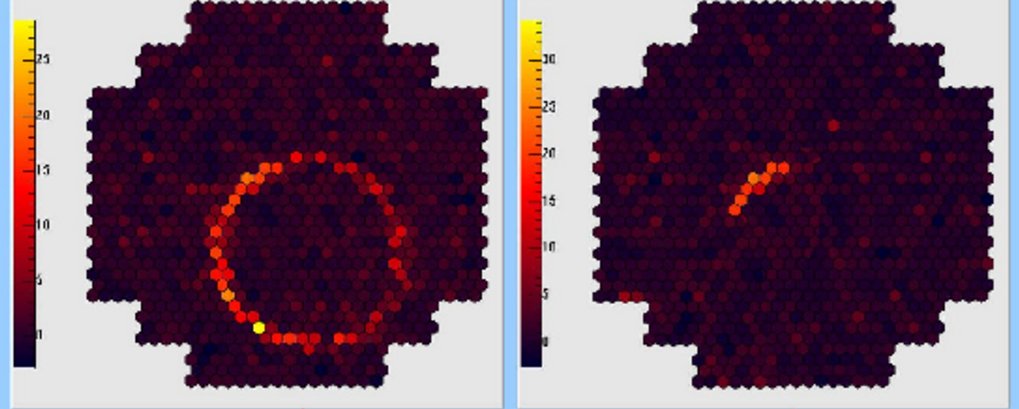}
\caption{\label{fig:act_2} Different shower patterns caused by high-energy muons. From \cite{Voelk2009}.}
%\end{minipage} 
\end{figure} 

Figure~\ref{fig:act_2} shows an example of the observed images caused by high-energy muons. The muon rings play a key part in the calibration of the photomultiplier tubes PMT of the cameras of the telescopes \citep{Chalme-Calvet_R_2014}

\begin{figure}[b]
\centering
%\begin{minipage}[b]{5in}
\includegraphics[width=5in]{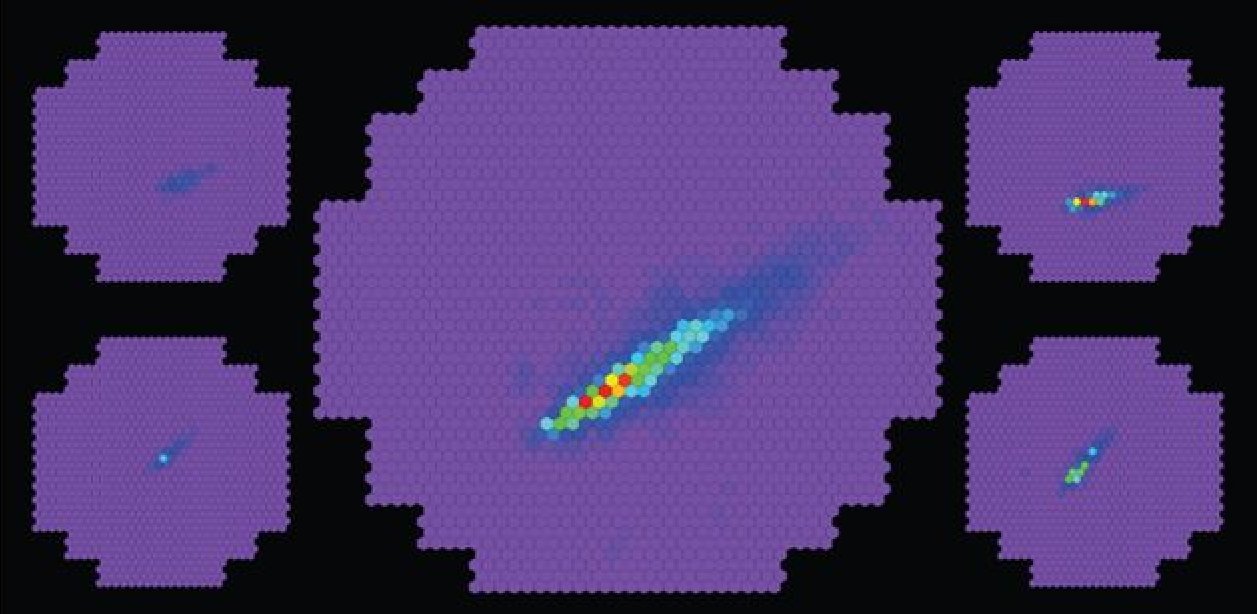}
\caption{\label{fig:act_3} Typical gamma-ray shower seen by the H.E.S.S. telescope array. From \cite{Hinton2013}.}
%\end{minipage} 
\end{figure} 

By using only one telescope it is difficult to reconstruct the geometry of the incident gamma ray and therefore multiple telescopes are used in an array to allow for a stereoscopic reconstruction of the direction of the incident gamma ray. Figure~\ref{fig:act_3} shows a typical gamma-ray shower as seen by the H.E.S.S.\ telescope array. 

\subsection{The H.E.S.S.\ array}
The review paper on the H.E.S.S.\ telescope by \cite{HESS_2013} will be used for this section (see also \citealt{deNaurois2015}). The H.E.S.S.\ experiment consists of an array of four 13-m (H.E.S.S.\ I) and one 28-m (H.E.S.S.\ II) ACTs located in the Khomas highland in Namibia. H.E.S.S.\ I started operations in 2003, with H.E.S.S.\ II seeing first light at 0:43 a.m.\ on 26 July 2012. In the recent past the four 13-m telescopes have undergone some maintenance where the 380 mirrors on each telescope have been recoated over a timespan of 2 years increasing the optical efficiency, which has decreased over the past ~8 years of operation. In future the Winston cones, phototubes and electronics will also be replaced. Another mirror upgrade is planned for 2016.

The addition of H.E.S.S.\ II to the H.E.S.S.\ array improves the sensitivity in the tens of GeV energy range and also decreases the energy threshold. This should allow for a more detailed search for pulsed emission from some Galactic sources, and improve the chances of viewing the VHE gamma-ray glow from gamma-ray bursts (GRBs), as well as the chance to detect new and more distant Galactic objects.

\subsection{VHE Galactic and extra-galactic sources}
\begin{figure}[t]
\centering
\begin{minipage}[b]{5in}
\includegraphics[width=5in]{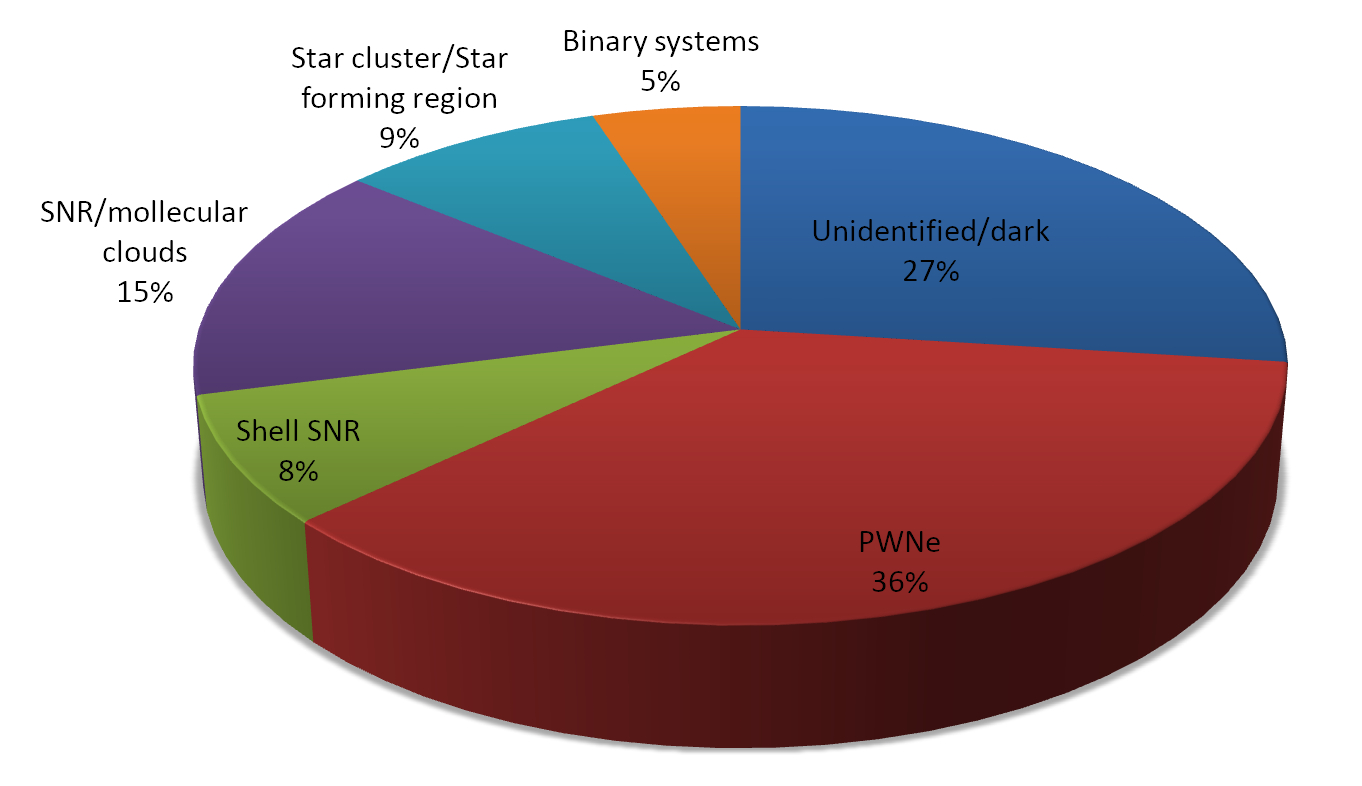}
\caption{\label{fig:hess} Fraction of different types of VHE gamma-ray emitters revealed by the H.E.S.S.\ Galactic Plane Survey.}
\end{minipage} 
\end{figure} 
The H.E.S.S.\ Galactic Plane Survey (GPS) revealed a large number of VHE sources in the Galactic Plane, with PWNe being the most abundant source type discovered. A thorough summary of all the known and unknown sources is given in TeVCat\footnote{tevcat.uchicago.edu} and Figure~\ref{fig:hess} shows the fraction of the different sources currently detected. A large number of the unknown sources may turn out to be PWNe, where the embedded pulsars have not (yet) been detected. 

Resolved supernova remnant shells, supernovae interacting with molecular clouds, binary systems, and stellar clusters are the next most abundant gamma-ray source classes in the Galactic Plane.

% Chapter 3

\chapter{Spatial-temporal-energetic modelling of a PWN} % Write in your own chapter title
\label{ch:Model}
\lhead{Chapter \ref{ch:Model}. \sc{Modelling the Radiation from a PWN}} % Write in your own chapter title to set the page header

In this chapter I describe the implementation of multi-zone, time-dependent code which will model the transport of particles through a PWN. The particles are injected by an embedded pulsar into a spherical shell and diffuse through space whilst undergoing energy losses. The geometry of the model is discussed in Section~\ref{sec:Mdimen}. The particles injected into the PWN are accelerated at the termination shock of the PWN, the form of this injected spectrum, and the transport equation used to model the particle spectral evolution are discussed in Section~\ref{sec:Injection}. The radiative energy losses that the particles undergo are discussed in Section~\ref{sec:rad_losses}. Diffusion and convection are dealt with in Section~\ref{sec:convDiff}. The transport of the particles is modelled by using a Fokker-Planck-type equation similar to the Parker equation \citep{Parker1965}. This equation is descretised and solved numerically as discussed in Section~\ref{sec:dnde}. Next, I discuss the calculation of the  broadband radiation spectrum in Section~\ref{sec:RadSpec} and the line-of-sight (LOS) calculation that projects the total radiation modelled from the PWN onto a flat surface in Section~\ref{sec:LOS}. This LOS calculation is done so that we can produce results as to the morphology of the PWN. Lastly, I will show some figures to prove that our model converges for a suitable number of bins in the different dimensions, and will also describe how the dynamical time step is calculated in Section~\ref{sec:Bins}.

\section{Model geometry}\label{sec:Mdimen}
\begin{figure}[t]
\centering
\begin{minipage}[b]{6in}
\includegraphics[width=6in]{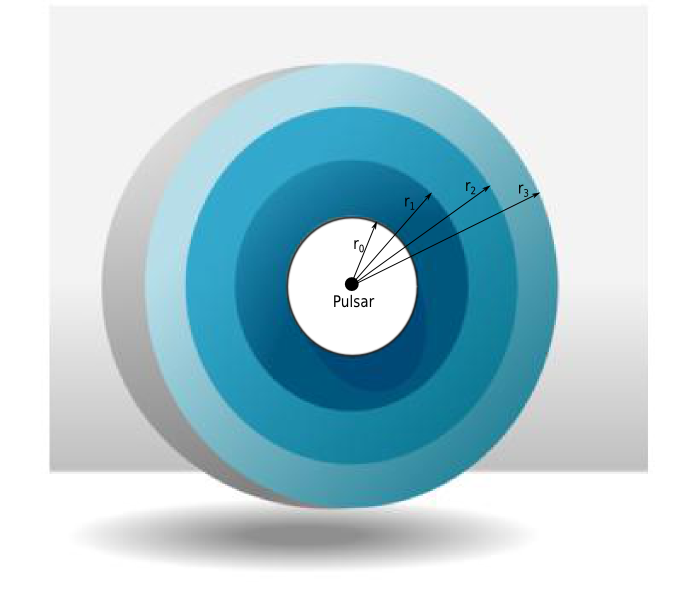}
\caption{\label{fig:PWN}Illustration of how the PWN model is set up.}
\end{minipage} 
\end{figure}

We make the simplified assumption that the geometrical structure of the PWN may be modelled as a sphere, as in Figure~\ref{fig:PWN}, into which particles are injected and allowed to diffuse and undergo energy losses. To simplify the model, we assumed spherical symmetry and that the only changes in the particle spectrum will be in the radial direction for a fixed particle energy. The model therefore has only one spatial dimension. 
In Figure~\ref{fig:PWN} it is shown how the model is set up with the pulsar in the middle and the different concentric zones (shells) of the PWN around the pulsar. The white region in the middle of the PWN is not modelled and the black circle separating the white and shaded regions (at radius $r_0$) is the termination shock where the particles are accelerated and injected into the PWN (this is the inner boundary). The model consists of three main dimensions in which the transport equation should be solved. The spatial, or radial dimension, the lepton energy dimension, and the time dimension. The radial dimension is divided into linear bins and is a static grid into which the PWN is allowed to expand. Therefore, there is a minimum radius $r_0$ at the termination shock, and a maximum radius $r_{\rm{max}}$ chosen to be much larger than the radius of the PWN ($R_{\rm{PWN}}(t)$). This radius $R_{\rm{PWN}}(t)$ will be calculated later from the predicted morphology of the PWN. The radial bin size is calculated using
\begin{equation}
\Delta r = \left(r_{\rm{max}} - r_{0} \right)/(N-1),
\end{equation}
with $N$ the number of bins and $dr$ the bin size in the radial dimension. Typical values used here are $r_{0} = 0.1~\rm{pc}$ and $r_{\rm{max}} = 16~\rm{pc}$, and the $k^{th}$ radius is given by $r_k = r_{\rm{min}} + (k-1)dr$, for $k = 1,2,\dots,N$.

The lepton energy dimension is divided into logarithmic bins. The way this is done is to choose a minimum ($E_{\rm{min}}=1.0 \times 10^{-7}~\rm{erg}$) and maximum ($E_{\rm{max}}=1.0 \times 10^{4}~\rm{erg}$) value for the energies, with the break in the spectrum at $E_{\rm{b}} \sim 0.1~\rm{erg}$, and then calculate the size of every energy bin. This is given by
\begin{equation}
\left(\Delta E\right)_i = \delta E_i,
\label{eq:delSR}
\end{equation}
with
\begin{equation}
\delta = \frac{1}{M-1} \ln \left(\frac{E_{\rm{max}}}{E_{\rm{min}}} \right),
\end{equation}
as discussed in Appendix~\ref{appen:delSR}. We can also calculate the $i^{th}$ energy bin using $E_i = E_{\rm{min}} e^{i\delta}$, for $i = 0,1,\dots,M-1$.

The time dimension is divided dynamically and starts at $t=0$, the time of birth of the PWN. It is allowed to reach the known age of the specific PWN modelled by incrementing the time by $dt$. The time step is calculated for each iteration of the code as discussed in Section~\ref{sec:Bins}.

\section{Transport equation and injection spectrum}\label{sec:Injection}
The transport of charged particles in a PWN is modelled by solving a Fokker-Planck-type equation similar to the Parker equation \citep{Parker1965} as mentioned earlier.  This equation includes diffusion, convection, energy losses (radiative and adiabatic), as well as a particle source. We start from the following form of the transport equation \citep{Moraal2013}
\begin{equation}
\frac{\partial f}{\partial t} = -\nabla \cdot \mathbf{S} + \frac{1}{p^2}\frac{\partial }{\partial p}\left(p^2\left\langle\dot{p}\right\rangle_{\rm{tot}} f\right) + Q(\mathbf{r},p,t),
\label{kopp1}
\end{equation}
with $Q(\mathbf{r},p,t)$ the particle injection spectrum, $f$ the distribution function, $r$ the spatial dimension, $p$ the momentum, and $\langle\dot{p}\rangle $ the total momentum rate of change. The term $\nabla \cdot \mathbf{S} = \nabla \cdot \left(\mathbf{V}f - \mathbf{\underline{K}} \nabla f\right)$ describes the general movement of particles in the PWN, with $\mathbf{V}$ the bulk motion of particles in the PWN and $\mathbf{\underline{K}}$ the diffusion tensor. However, we rewrite Eq.~\eqref{kopp1} in terms of energy and also transform the distribution function to a particle spectrum per unit volume. This is done by using the relations $U_p(\mathbf{r},p,t) = 4\pi p^2 f(\mathbf{r},p,t)$, to convert the distribution function to a particle spectrum, and $E^2 = p^2c^2 + E_0^2$ to convert the equation from momentum to energy space (Appendix~\ref{AppendixA}). We also assume that the diffusion is only energy dependent, $\mathbf{\underline{K}} = \kappa(E)$. Thus
\begin{equation}
\frac{\partial N_{\rm{e}}}{\partial t} = -\mathbf{V} \cdot (\nabla N_{\rm{e}}) +  \kappa \nabla^2 N_{\rm{e}} + \frac{1}{3}(\nabla \cdot \mathbf{V})\left( \left[\frac{\partial N_{\rm{e}}}{\partial \ln E} \right] - 2N_{\rm{e}} \right) +  \frac{\partial }{\partial E}(\dot{E}_{\rm{rad}}N_{\rm{e}}) +  Q(\mathbf{r},E,t).
\label{eq:transportFIN}
\end{equation} 
The derivation of this can be seen in Appendix~\ref{app:transport}. The units of $N_{\rm{e}}$ are the number of particles per unit energy and volume.

\begin{figure}[h]
\centering
\begin{minipage}[b]{5in}
\includegraphics[width=5in]{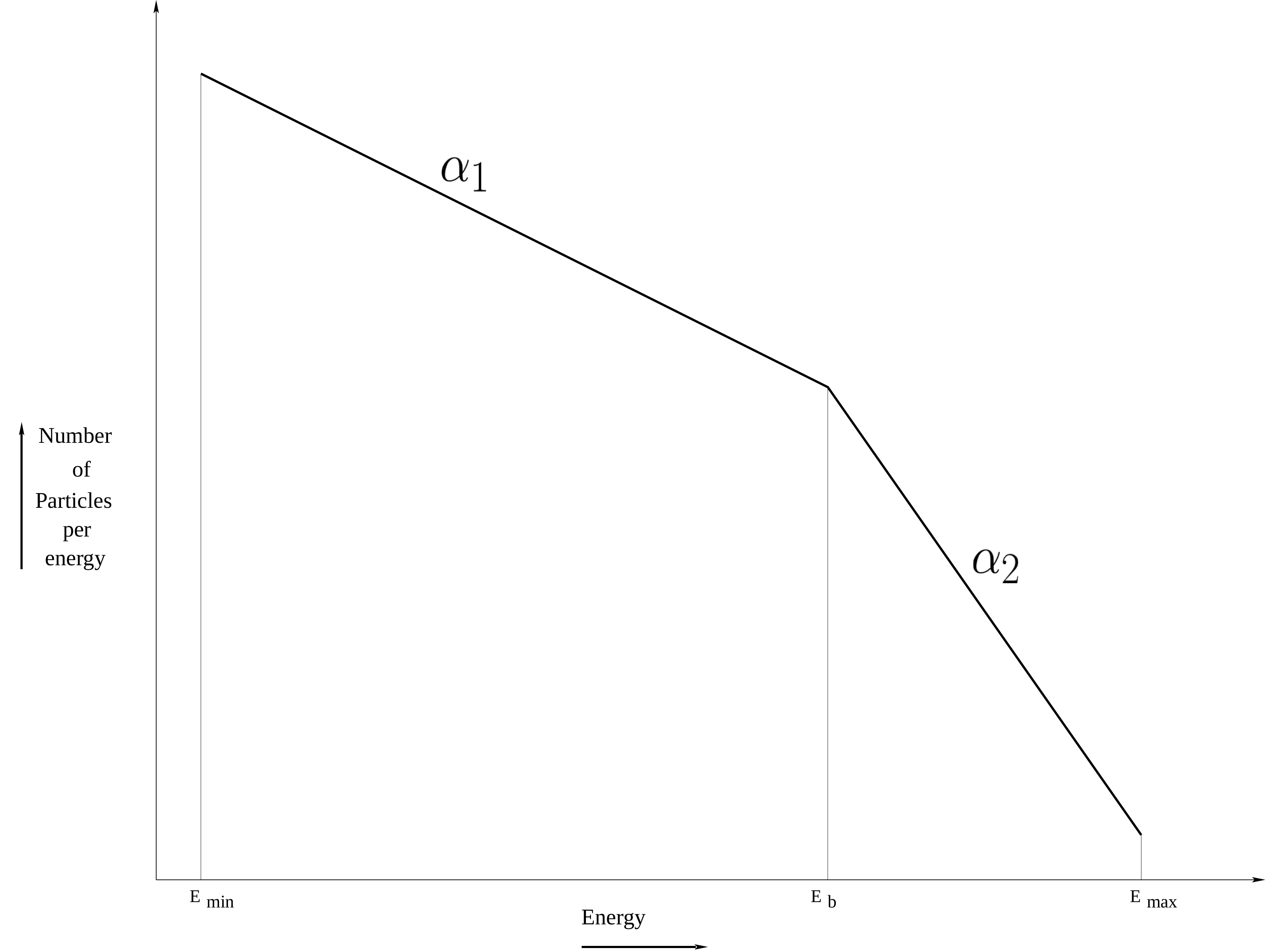}
\caption{\label{fig:injec}Schematic for the particle injection spectrum.}
\end{minipage} 
\end{figure} 
Following \cite{VdeJager2007}, we used a broken power law for the particle injection spectrum
\begin{equation}
Q(E_{\rm{e}},t) = \left\{\begin{matrix}
Q_0(t)\left(\frac{E_{\rm{e}}}{E_{\rm{b}}}\right)^{\alpha_1} E_{\rm{e}}<E_{\rm{b}}\\ 
Q_0(t)\left(\frac{E_{\rm{e}}}{E_{\rm{b}}}\right)^{\alpha_2} E_{\rm{e}}\geq E_{\rm{b}}.
\end{matrix}\right.
\label{brokenpowerlaw}
\end{equation}  
Here $Q_0$ is the time-dependent normalisation constant, $E_{\rm{b}}$ is the break energy, $E_{\rm{e}}$ is the lepton energy, and $\alpha_1$ and $\alpha_2$ are the spectral indices as shown in Figure~\ref{fig:injec}. To obtain $Q_0$ we use a spin-down luminosity $L(t) = L_0/\left(1+t/\tau_0\right)^2$ of the pulsar assuming $n=3$ \citep[e.g.,][]{Reynolds1984}, with  $\tau_0$ the characteristic spin-down timescale of the pulsar and $L_0$ the initial spin-down luminosity. Thus we set
\begin{equation}
  \epsilon L(t) = \int_{E_{\rm{min}}}^{E_{\rm{b}}}QE_{\rm{e}}dE_{\rm{e}} + \int_{E_{\rm{b}}}^{E_{\rm{max}}}QE_{\rm{e}}dE_{\rm{e}},
  \label{normQ}
\end{equation}
with $\epsilon$ the conversion efficiency of the spin-down luminosity to particle power. The way that this is descretised and used is discussed in Appendix~\ref{appen:normQ}.

\section{Radiative and adiabatic energy losses in the PWN}
\label{sec:rad_losses}
One way in which particle energy is dissipated from the system is due to radiation.  We incorporated SR and IC scattering, similar to calculations done by \cite{Kopp2013} in their globular cluster model. SR losses are given by \cite{BlGould1970}
\begin{equation}
  \left(\frac{dE}{dt}\right)_{\rm{SR}} = -\frac{\sigma_T c}{6\pi E_0^2} E_{\rm{e}}^2B^2,
  \label{SR}
\end{equation}
with $\sigma_T = \frac{8\pi}{3}r_{\rm{e}}^2 = 6.65 \times 10^{-25}\rm{cm}$ the Thompson cross section and $B$ the PWN magnetic field. 

The IC scattering energy loss rate is given by
\begin{equation}
  \left(\frac{dE}{dt}\right)_{\rm{IC}} = \frac{g_{\rm{IC}}}{E_{\rm{e}}^2} \sum_{l=1}^{3} \int \!\!\!\! \int n_{\varepsilon,l}(r,\varepsilon,T_l) \frac{E_\gamma}{\varepsilon} \zeta(E_{\rm{e}},E_{\gamma},\varepsilon) d \varepsilon dE_\gamma,
  \label{IC}
\end{equation}
with $n_{\varepsilon,l}(r,\varepsilon,T_l)$ the number density, $g_{\rm{IC}} = 2\pi e^4c$, $\varepsilon$ the soft-photon energy, $T_{l}$ the photon temperature of the $l^{th}$ blackbody component, $E_{\gamma}$ the TeV upscattered photon energy, and $\hat{\zeta}$ the collision rate
\begin{equation}
\zeta(E_{\rm{e}},E_{\gamma},\varepsilon) = \zeta_0 \hat{\zeta}(E_{\rm{e}},E_{\gamma},\varepsilon),
\end{equation} 
with $\zeta_0 = 2\pi e^4E_0c/\varepsilon E_{\rm{e}}^2$, and $\hat{\zeta}$ given by \citep{Jones1968}
\begin{equation}
\hat{\zeta}(E_{\rm{e}},E_{\gamma},\varepsilon) =\left\{\begin{matrix}
0 &\rm{if} &E_{\gamma} \leq \frac{\varepsilon E_0^2}{4 E_{\rm{e}}^2},\\ 
\frac{E_{\gamma}}{\varepsilon}-\frac{E_0^2}{4 E_{\rm{e}}^2} &\rm{if} &\frac{\varepsilon E_0^2}{4 E_{\rm{e}}^2} \leq E_{\gamma} \leq \varepsilon,\\ 
f(q,g_0) &\rm{if} &\varepsilon \leq E_{\gamma} \leq \frac{4\varepsilon E_{\rm{e}}^2}{E_0^2 + 4\varepsilon E_{\rm{e}}},\\ 
0 &\rm{if} &E_{\gamma} \geq \frac{4\varepsilon E_{\rm{e}}^2}{E_0^2 + 4\varepsilon E_{\rm{e}}}.
\end{matrix}\right.
\end{equation}
Here, $f(q,g_0) = 2q$ln$q+(1-q)(1+(2+g_0)q)$, $q=E_0^2E_{\gamma}/(4\varepsilon E_{\rm{e}}(E_{\rm{e}}-E_{\gamma}))$, and $g_0(\varepsilon,E_{\gamma}) = 2\varepsilon E_{\gamma}/E_0^2$. More details regarding the radiative energy losses are discussed in Section~\ref{sec:IC} and \ref{sec:SR}. 

The particles in the PWN also lose energy due to adiabatic processes caused by the bulk motion of the particles in the PWN as energy is expended to expand the PWN. The adiabatic energy losses are given by $\dot{E}_{\rm{ad}} = \frac{1}{3}(\nabla \cdot \mathbf{V})E_{\rm{e}}$ \citep{Zhang_2008}. The two radiation loss rates and the adiabatic energy loss rate can be added to find the total loss rate $\dot{E}_{\rm{tot}}$ used in Eq.~(\ref{eq:transportFIN}).

\section{Diffusion and convection}\label{sec:convDiff}
For the diffusion scalar coefficient $\kappa$, Bohm diffusion is assumed so that
\begin{equation}
  \kappa = \kappa_B\frac{E_e}{B},
\end{equation}
with $\kappa_B = c/3e$, and $c$ and $e$ denote the speed of light in vacuum and the elementary charge. The reason I choose Bohm diffusion is discussed in greater detail in Section~\ref{sec:Diff}. The bulk particle motion inside the PWN is parametrized by 
\begin{equation}
V(r) = V_0\left(\frac{r}{r_0}\right)^{\alpha_V},
\label{V_profile}
\end{equation}
with $\alpha_V$ the velocity profile parameter. Here $V_0$ is the velocity at $r_0$. When choosing a constant adiabatic timescale
\begin{equation}
\tau_{\rm{ad}} = \frac{E}{\dot{E}_{\rm{ad}}},
\label{eq:ad_parmtr}
\end{equation}
where $\dot{E}_{\rm{ad}} = 1/3(\nabla \cdot \mathbf{V})E_{\rm{e}}$, and by using the analytical solution for of $(\nabla \cdot \mathbf{V})$ in Eq.~\eqref{eq_ap:nablaDOTv}, we find that $V_0 = r_0/\tau_{\rm{ad}}$ and $\alpha_V = 1$.

\section{Calculation of the particle (lepton) spectrum}\label{sec:dnde}
\subsection{The discretised transport equation}
We assume spherical symmetry, thus $\frac{\partial}{\partial \theta} = 0$ and $\frac{\partial}{\partial \phi} = 0$, and that the only spatial direction in which $N_{\rm{e}}$ changes is the radial direction (i.e., $\nabla^2 N_{\rm{e}} = \frac{1}{r^2}\left[\frac{\partial}{\partial r}\left[r^2\frac{\partial N_{\rm{e}}}{\partial r}\right]\right]$). 

Eq.~(\ref{eq:transportFIN}) can now be discretised leaving us with
\begin{equation}
\begin{split}
(1-z+\beta)(N_{\rm{e}})_{i,j+1,k} &=2Q_{i,j,1} \triangle t\\
& + (1+z-\beta)(N_{\rm{e}})_{i,j-1,k} \\
& + (\beta+\gamma-\eta)(N_{\rm{e}})_{i,j,k+1} \\
& + (\beta-\gamma+\eta)(N_{\rm{e}})_{i,j,k-1} \\
& -2(\nabla \cdot \mathbf{V})_{i,j,k}\triangle t (N_{\rm{e}})_{i,j,k}\\
& + \frac{2}{(d E_{i+1,j,k}+dE_{i,j,k})} \times\\
& \left(r_{\rm{a}} \left(dE_{\rm{loss}}\right)_{i+1,j,k}(N_{\rm{e}})_{i+1,j,k} - \frac{1}{r_{\rm{a}}}\left(dE_{\rm{loss}}\right)_{i-1,j,k}(N_{\rm{e}})_{i-1,j,k}\right),
\end{split}
 \label{eq:fin_EQ}
\end{equation}
with $i$ the energy index, $j$ the time index, $k$ the radial index,  $\beta = 2\kappa \Delta t/ (\Delta r)^2$, $\gamma = 2\kappa \Delta t/(r\Delta r)$, $\eta = V_k\Delta t/\Delta r$, $\Delta r$ the bin size of the spatial dimension, $\Delta t$ the bin size of the time dimension, $dE_{\rm{loss}}=\dot{E}_{\rm{tot}}\Delta t$, and $V_k$ the bulk particle motion in the current radial bin. Also, $r_{a} = \Delta E_{i+1,j,k}/\Delta E_{i,j,k}$ with
\begin{equation}
z = \left(\frac{1}{\Delta E_{{i+1,j,k}}-\Delta E_{{i,j,k}}}\right)\left(\frac{1}{r_{\rm{a}}}-r_{\rm{a}}\right)\dot{E}_{{i,j,k}}.
\end{equation}
We first approached the discretisation process by using a simple Euler method. It soon became clear that this method was not stable. We then decided to use a DuFort-Frankel scheme to discretise Eq.\ (\ref{eq:transportFIN}). The details are given in Appendix~\ref{appen:disctr}. In solving this equation, we calculate the lepton spectrum of the PWN due to the injected particles from the embedded pulsar, taking into account their diffusion through the PWN and the IC scattering, SR, convection, and adiabatic energy losses. 
 
As mentioned previously, we use the parametrised form of the $B$-field given by 
\begin{equation}
B(r,t) = B_{\rm{age}}\left(\frac{r}{r_0}\right)^{\alpha_B}\left(\frac{t}{t_{\rm{age}}}\right)^{\beta_B},
\label{B_Field}
\end{equation}
with $B_{\rm{age}}$ the present day magnetic field at $r = r_0$ and $t = t_{\rm{age}}$, $t$ the time since the PWN's birth, and $\alpha_B$ and $\beta_B$ the magnetic field parameters. The magnetic field and the bulk particle motion in the PWN are linked, as noted in Eq.~\eqref{eq:B_V}. We can use this relationship to reduce the number of free parameters in the model as there are currently $3$ free parameters for the magnetic field and the bulk particle motion. Equation \eqref{eq:a_v+a_b=-1} shows that $\alpha_V + \alpha_B = -1$ thus reducing the number of free parameters by one. 

We limit the particle energy using $E_{\rm{max}} = \frac{e}{2}\sqrt{\frac{L(t)\sigma}{c(1+\sigma)}}$ \citep{VdeJager2007}, with $\sigma$ the ratio of electromagnetic to particle luminosity. Particles with $E_{\rm{e}}>E_{\rm{max}}$ are assumed to have escaped.

\subsection{Boundary conditions}\label{sec:boundary_conditions}
The multi-zone model divides the PWN into shells as seen in Figure~\ref{fig:PWN} to solve Eq.~\eqref{eq:fin_EQ} numerically. The particles are injected into zone one and allowed to propagate through the different zones, with the spectral evolution being governed by Eq.~\eqref{eq:fin_EQ}.  As the initial condition, all zones were assumed to be devoid of any particles, i.e., $N_{\rm{e}} = 0$ at $t = 0$, and a set of ``ghost points", that are also devoid of particles, were defined outside the boundaries in time, as the DuFort-Frankel scheme requires two previous time steps.

For the spatial dimension, the boundary conditions are reflective at the inner boundary to avoid losing particles towards the pulsar past the termination shock and at the outer boundary $r_{\rm{max}}$ the particles were allowed to escape. To model the escape of particles on the outer boundary, the particle spectrum was set to zero, and for the reflective boundary we needed zero flux at the innermost radial bin. Therefore we set
\begin{equation}
(N_{\rm{e}})_{i,j,0} = (N_{\rm{e}})_{i,j,1} \times \frac{\kappa/dr-V_{i,j,1}/2}{\kappa/dr+V_{i,j,1}/2},
\label{eq:reflective_boundary}
\end{equation}
which results in zero flux at the inner boundary. The energy boundary condition is governed by the minimum and maximum allowed particle energy given in Section~\ref{sec:Mdimen}.

The injection of particles into the PWN can also be seen as a boundary condition. We inject the particles at a certain rate and density $\dot{\rho}$ to be able to do the LOS calculation later. We assume the particle injection spectrum $Q^{\prime \prime}$ is uniformly distributed in the first zone and thus
\begin{equation}
\frac{Q^{\prime \prime}}{V^1_{\rm{shell}}} = Q,
\label{eq:injection_boundary}
\end{equation}
where $V^1_{\rm{shell}}$ is the volume of the first zone and $Q$ the injection spectrum per unit energy, time, and volume as used in Eq.~\eqref{eq:fin_EQ}.

\section{Calculation of radiation spectrum}\label{sec:RadSpec}
The time-dependent photon spectrum of each zone can now be calculated, using the electron spectrum solved for each zone. For IC we have \citep{Kopp2013}
\begin{equation}
\left(\frac{dN_\gamma}{dE_\gamma}\right)_{IC} = \frac{g_{IC}}{A} \sum_{j=1}^{3} \int \!\!\!\! \int n_{\varepsilon,j}(r,\varepsilon,T_j) \frac{\mathcal{N}_{\rm{e}}}{\varepsilon E_{\rm{e}}^2}\hat{\zeta}(E_{\rm{e}},E_\gamma,\varepsilon) d\varepsilon dE_{\rm{e}} , 
\label{eq:ICrad}
\end{equation}
where $A=4\pi d^2$, $d$ the distance to the source, and $\mathcal{N}_{\rm{e}} = N_{\rm{e}}V_{\rm{shell}}$ is the number of electrons per energy in a spherical shell around $r$. We consider multiple blackbody components of target photons, for example cosmic background radiation (CMB) with a temperature of $T_1 = $ 2.76 K and an average energy density of $u_1 = $ 0.23 eV/cm$^3$, Galactic background infrared photons, $T_2 = $ 35 K and $u_2 = 0.5$ eV/cm$^3$, and starlight with $T_3 = $ 4 500 K and $u_3 = $ 50 eV/cm$^3$.

For SR we have
\begin{equation}
\left(\frac{dN_\gamma}{dE_\gamma}\right)_{\rm{SR}} = \frac{1}{A}\frac{1}{hE_\gamma}\frac{\sqrt{3}e^3B(r,t)}{E_0} \int \!\!\!\! \int_0^{\pi/2} \mathcal{N}_{\rm{e}}(E_{\rm{e}},r)F \left(\frac{\nu}{\nu_{\rm{cr}}(E_{\rm{e}},\alpha,r)}\right)\sin^2 \alpha d\alpha dE_{\rm{e}},
\label{eq:SRrad}
\end{equation}
with $\nu_{\rm{cr}}$ the critical frequency (with pitch angle $\alpha$, which we assume to be $\pi /2$ so that $\sin^2 \alpha = 1$) given by
\begin{equation}
\nu_{\rm{cr}}(E_{\rm{e}},\alpha,r) = \frac{3ec}{4\pi E_0^3}E_{\rm{e}}^2 B(r,t),
\end{equation}
and
\begin{equation}
F(x) = x\int_x^\infty K_{5/3}(y)dy,
\end{equation}
where $K_{5/3}$ the modified Bessel function of order $5/3$.

The calculation of the radiation spectrum is done by using the code of \cite{Kopp2013} and is not done in this thesis. The total radiation spectrum at Earth is found by calculating Eq.~(\ref{eq:ICrad}) and Eq.~(\ref{eq:SRrad}) for each zone in the model and adding them. Additionally, the radiation per unit volume can also be calculated by dividing the radiation by the volume of the zone where the radiation originated from. Examples of this will be shown in Chapter \ref{ch:Results}.

\section{Calculation of the line-of-sight flux}\label{sec:LOS}
In this section I discuss how the line-of-sight (LOS) integration is done. We do the LOS integration to project the total flux from the PWN onto a flat surface to find the surface brightness and to thus find the flux as a function of radius. This will allow us to estimate the size of the PWN and also to study the size of the PWN as a function of energy. In order to do the LOS integration of the radiation from the PWN, we need to use the radiation per unit volume (as explained in the previous section) and multiply it with the volume in a particular LOS as viewed from Earth. 
\begin{figure}[t]
\centering
\begin{minipage}[b]{6in}
\includegraphics[width=6in]{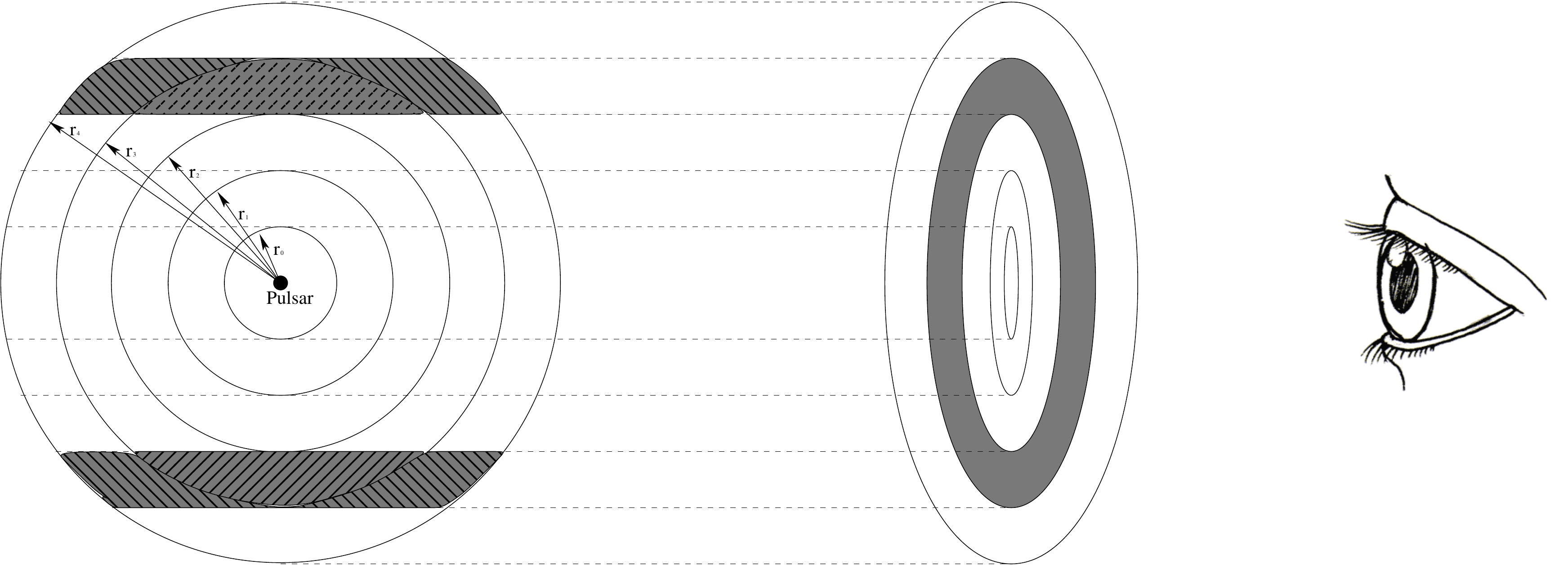}
\caption{\label{fig:LOS}Schematic for the geometry of the LOS calculation.}
\end{minipage} 
\end{figure}
Figure~\ref{fig:LOS} is a schematic representation of how this is done. The pulsar plus the multi-zone model of the surrounding PWN are on the left hand side of Figure~\ref{fig:LOS} and the right hand side shows how LOS cylinders are chosen through the PWN, with the observer looking on from the right. The source is very far from Earth and cylinders instead of cones are chosen as a good first approximation. Cylinders intersecting the spherical zones are used, both having the same radii. This results in the observer viewing the projected PWN as several $2D$ ``annuli", for example the shaded region in Figure~\ref{fig:LOS}, all with different radii. The radiation in a certain annulus can thus be calculated if the volume of the intersection between a particular cylinder and the spheres is known.
\begin{figure}[htbp]
 \begin{minipage}{0.5\linewidth}
  \centering
  \includegraphics[width=0.9\linewidth]{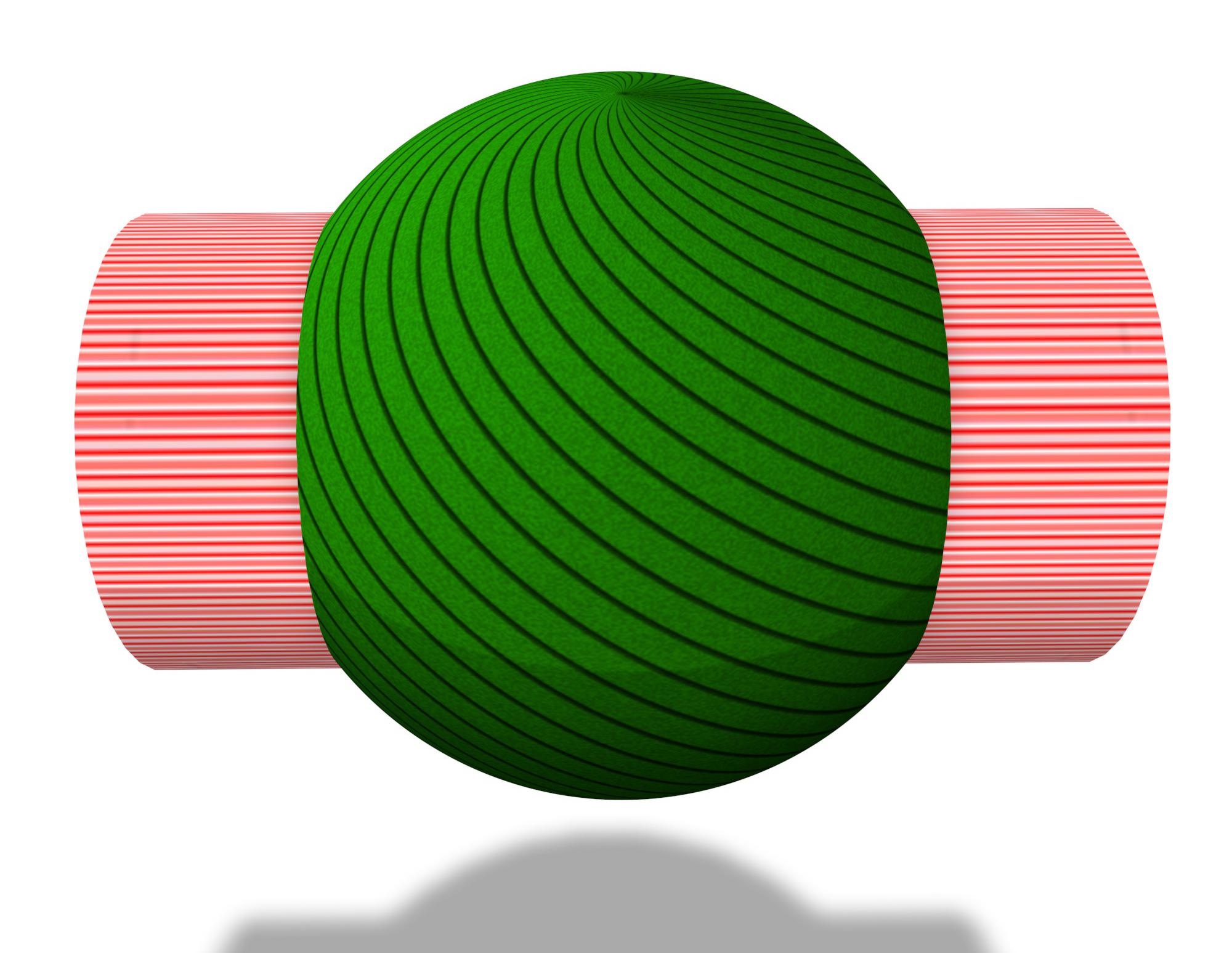}
  \caption{Intersection between a sphere and a cylinder.}
  \label{fig:cylshp}
 \end{minipage}%
 \begin{minipage}{0.5\linewidth}
  \centering
  \includegraphics[width=1.0\linewidth]{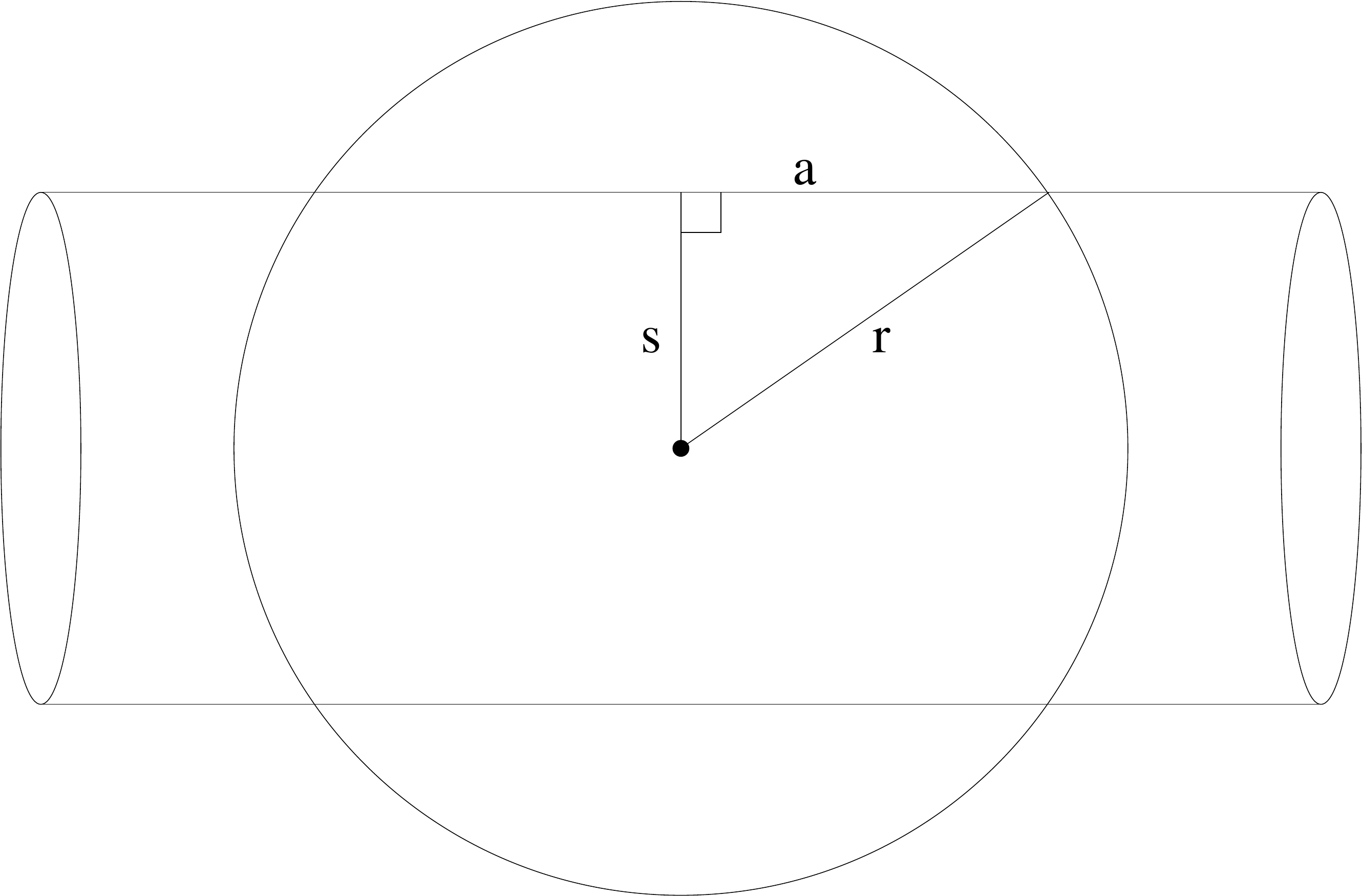}
  \caption{Schematic of the intersection between a sphere and a cylinder.}
  \label{fig:cylshematic}
 \end{minipage}
\end{figure} 

The intersection between a solid cylinder and a sphere can be seen in Figure~\ref{fig:cylshp}, with the schematic representation in Figure~\ref{fig:cylshematic}. If $V_{\rm{int}}$ is the volume of the intersecting part, then by noting that $a=\sqrt{r^2-s^2}$, where $s$ is the radius of the cylinder and $r$ the radius of the sphere, it is possible to calculate the intersection volume $V_{\rm{int}}$ by using cylindrical coordinates as
\begin{equation}
\begin{split}
  V_{\rm{int}}(s)& =  \int_0^{2 \pi} d \phi \int_0^s 2asds\\
   & = 4\pi \int_0^s s\sqrt{r^2-s^2} ds\\
   & = \frac{4\pi}{3}\left[-\left(r^2-s^2\right)^{\frac{3}{2}} + r^3\right],
\end{split}
\label{eq:intersection}
\end{equation}
where $\phi$ is the azimuthal angle. The volume in Eq.~(\ref{eq:intersection}) is not the volume required, as the volume for a specific annulus is needed. This can be calculated by subtracting the correct volumes from one another. For example, if the volume is required for a particular annulus with radius $s_k$ and sphere with $r_i$, then a single intersection volume is given by 
\begin{equation}
  V_{\rm{annuli}} = \left(V_{i,k} - V_{i,k-1}\right) - \left(V_{i-1,k} - V_{i,k-1}\right).
\label{eq:washer}
\end{equation}
The radiation at Earth can thus be calculated for a specific LOS by using the volume for all intersections of cylinders and spheres calculated in Eq.~(\ref{eq:washer}) and multiplying it by the radiation per unit volume for the specific zone found in Section~\ref{sec:RadSpec}. The total radiation for the specific LOS, or annulus, can be calculated by adding the radiation for a specific LOS together for all the different zones. To find the total radiation at Earth from the PWN, the radiation from all the different LOSs (annuli) are added together. 

\begin{figure}[h]
\centering
\begin{minipage}[b]{5in}
\centering
\includegraphics[width=4in]{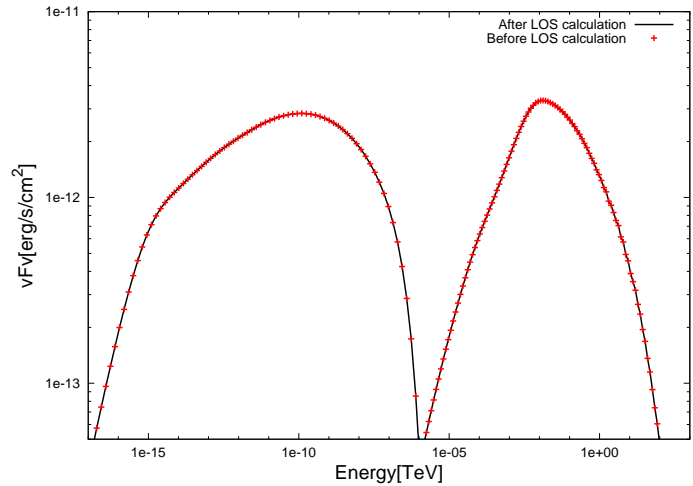}
\caption[SED comparison between the total flux before and after the LOS calculation.]{\label{fig:LOS_com}SED comparison between the total flux before and after the LOS calculation. The solid line is the total flux after the LOS calculation (sum of the flux in all annuli) and the red crosses are the total flux before the LOS calculation (sum of flux in all shells).}
\end{minipage} 
\end{figure}

The total flux form the PWN, as calculated in Section~\ref{sec:RadSpec}, should be the same as the total radiation after the LOS calculation, as nothing is changed except that the flux is now projected onto a flat surface. To see this the total flux before (as in Section~\ref{sec:RadSpec}) and after the LOS calculation (as mentioned in the previous paragraph) are compared and can be seen in Figure~\ref{fig:LOS_com}.
From Figure~\ref{fig:LOS_com} it is clear that the LOS calculation is functioning correctly as the total flux before (red crosses) and after the LOS calculation (black line) are exactly the same. We can now use this projected flux to calculate the surface brightness for the PWN at different viewing angles and thus use this to calculate the size of the PWN, as will be shown later in Section~\ref{sec:Space}.

\section{The effect of using a different number of bins}\label{sec:Bins}
When modelling the PWN we have to choose the number of bins in the 3 dimensions that the PWN is modelled. These are the spatial, temporal, and energy dimensions. In this section I will show that the model output converges when choosing a suitable number of bins.

Here and in the next chapter I will show particle spectrum figures. In these figures the particle spectrum, as calculated, is a number density. Therefore the units of the particle spectrum are particles/erg/cm$^3$. In the figures however, we have integrated over all space and multiplied with square of the electron energy $E^2$ thus the units of [$E^2 dN_{\rm{e}}/dE$] are erg.

\begin{figure}[b]
\centering
\begin{minipage}[b]{5in}
\centering
\includegraphics[width=4.4in]{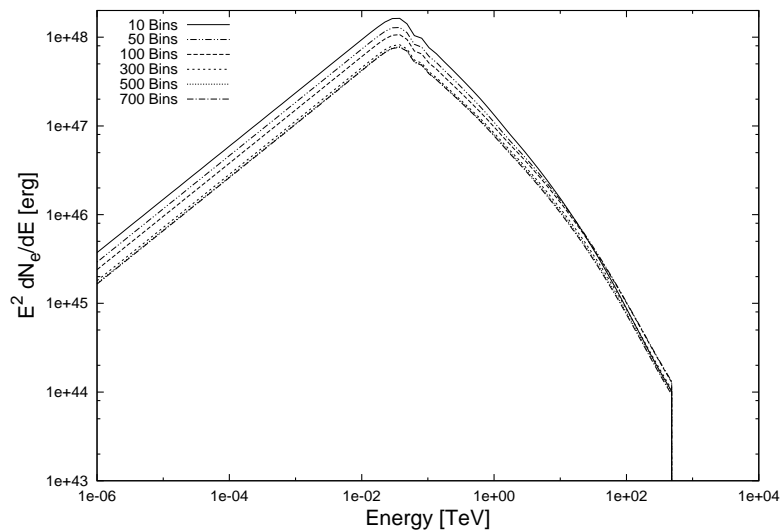}
\caption{\label{fig:Bins_conv_Par}Particle spectrum for PWN G0.9+0.1 showing that an increased number of spatial bins resulting in model convergence.}
\end{minipage} 
\end{figure}

\begin{figure}[t]
\centering
\begin{minipage}[b]{5in}
\centering
\includegraphics[width=4.4in]{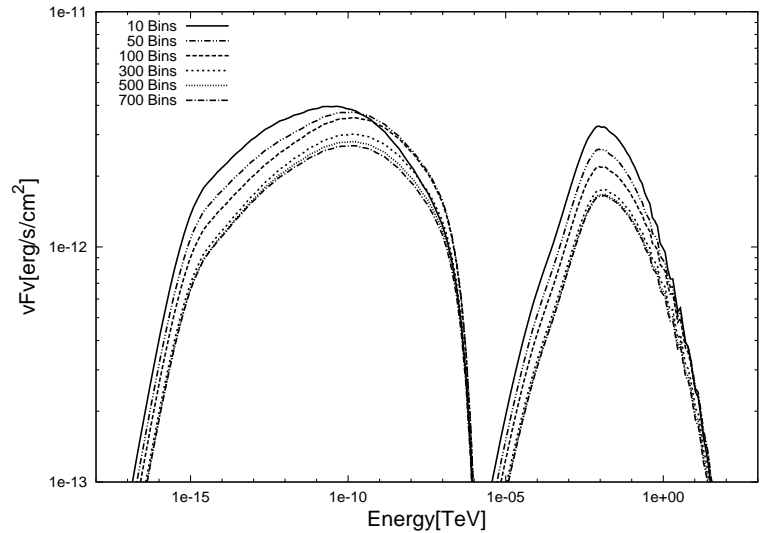}
\caption{\label{fig:Bins_conv}SED for the PWN G0.9+0.1 showing that an increased number of spatial bins resulting in model convergence.}
\end{minipage} 
\end{figure}

Figures \ref{fig:Bins_conv_Par} and \ref{fig:Bins_conv} show the particle spectrum and radiation spectrum for PWN G0.9+0.1 where the number of bins in the spatial dimension is increased from 10 up to 700 bins. We can see that at approximately 300 bins, the model starts to converge and therefore 300 bins were used throughout the rest of the modelling, as more bins do not necessarily increase the accuracy of the model, do increase the run time substantially.

The time step in the code is designed to be dynamic to improve the run time of the code. For the code to produce consistent results, the time step has to be much smaller than the energy-loss timescale and the diffusion timescale. Therefore, for each time iteration in the code, the energy-loss timescale and the diffusion timescale are calculated and the time step is then set to a small fraction of the smallest of the two timescales.

\begin{figure}[h]
\centering
\begin{minipage}[b]{5in}
\centering
\includegraphics[width=5in]{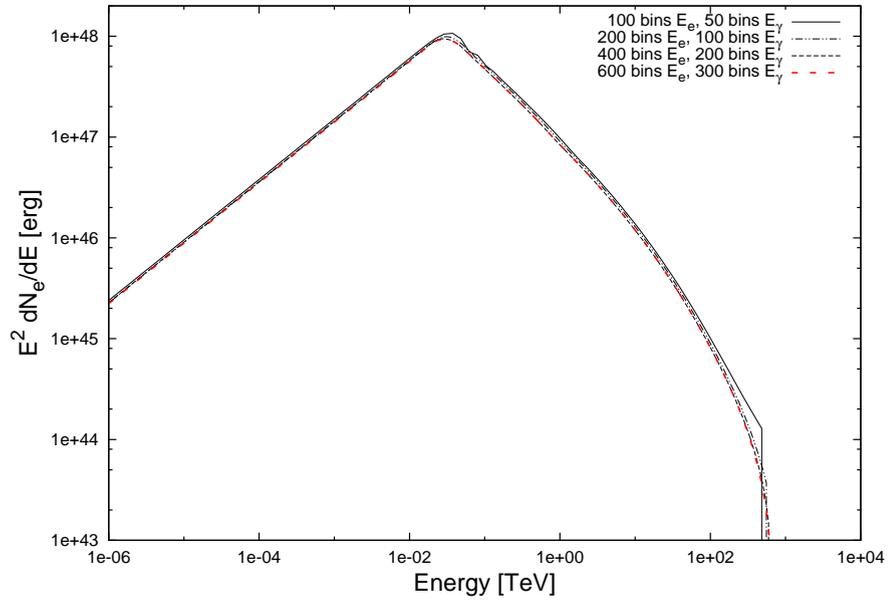}
\caption{\label{fig:Bins_conv_Par_ener}Particle spectrum for PWN G0.9+0.1 showing the effect of a change in the number of energy bins.}
\end{minipage} 
\end{figure}

\begin{figure}[h]
\centering
\begin{minipage}[b]{5in}
\centering
\includegraphics[width=5in]{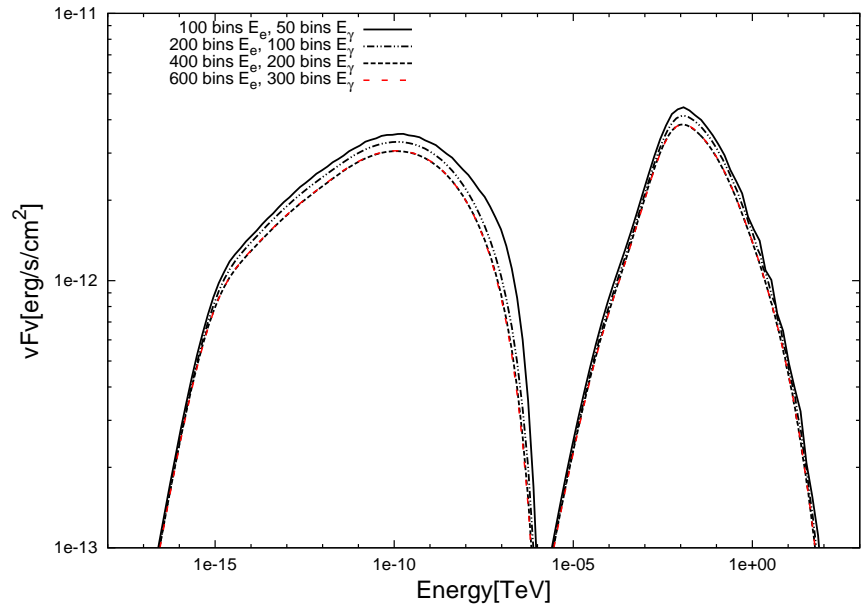}
\caption{\label{fig:Bins_conv_ener}SED for the PWN G0.9+0.1 showing the effect of a change in the number of energy bins.}
\end{minipage} 
\end{figure}

For the energy bin sizes, the number of bins was halved, doubled and multiplied by three to test the convergence of the code. This can be seen in Figures \ref{fig:Bins_conv_Par_ener} and \ref{fig:Bins_conv_ener}. Here we can see that if the number of energy bins are doubled (dashed line) or multiplied by three (red dashed line), the solution does not change at all. Therefore we selected 200 bins in the electron energies $E_{\rm{e}}$ and 100 bins in the photon energy $E_{\rm{\gamma}}$.

% Chapter 4

\chapter{Code calibration, parameter study, and SED fits} % Write in your own chapter title
\label{ch:Results}
\lhead{Chapter \ref{ch:Results}. \sc{Results and parameter study}} % Write in your own chapter title to set the page header
In this chapter the particle spectrum resulting from the solution of the Fokker-Planck-type transport equation will be shown together with the predicted radiation spectrum from the modelled PWN. Firstly, I will calibrate the newly developed code against results from other authors (Section~\ref{sec:G0.9intro}). Secondly, I will perform a parameter study to investigate the model behaviour when values of the different parameters are changed (Section~\ref{sec:parameterST}). Lastly, I will show the spatially-dependent results now possible with our new code (Section~\ref{sec:Space}).

\section{Calibration of the code}\label{sec:G0.9intro}
In this section I will use PWN G0.9+0.1 as a case study to calibrate the newly developed code. Following a short summary of the multi-wavelength properties of G0.9+0.1, I will compare my code's SED predictions with the results of two independent studies.

\subsection{Multi-wavelength observations of G0.9+0.1}
\cite{HelfandB1987} observed G0.9+0.1 for 45 minute integrations at 20~cm and 6~cm which led to the discovery of the composite nature of this bright, extended source near the the Galactic centre (GC) in the radio band. SNR G0.9+0.1 has therefore become a well-known supernova remnant, which is estimated to have an age of a few thousand years, and recognised as such from its radio morphology. This source exhibits a flat-spectrum radio core ($\sim2'$ across), corresponding to the PWN, and also clearly shows steeper shell components ($\sim 8'$ diameter shell).

While performing a survey on the GC, \cite{Sidoli2004} serendipitously observed SNR G0.9+0.1 using the \textit{XMM-Newton} telescope. Their observations provided the first evidence of X-ray emission from G0.9+0.1. \cite{Sidoli2004} fit an absorbed power-law spectrum that yields a spectral index of $\Gamma \sim 1.9$ with a flux of $F = 4.8 \times 10^{-12}$ erg cm$^{-2}$ s$^{-1}$ in the energy band 2$-$10 keV. This translates to a luminosity of $L_{\rm X} \sim 5 \times 10^{34}$ erg s$^{-1}$ for a distance of 10 kpc.

\begin{figure}[t]
\centering
\begin{minipage}[b]{5in}
\centering
\includegraphics[width=5in]{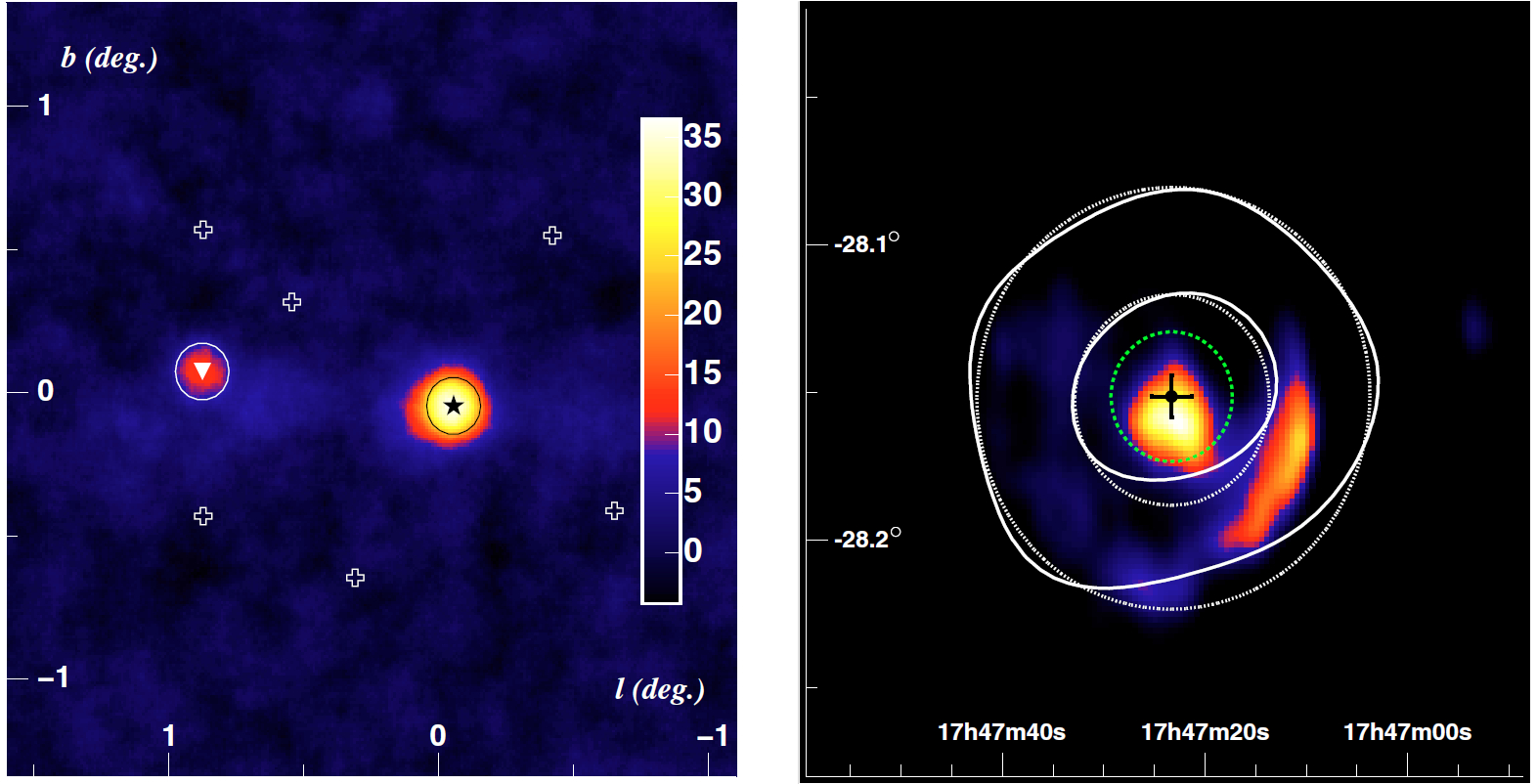}
\caption[Significance sky map for the field of view of the H.E.S.S.\ Galactic centre observations.]{\label{fig:G0.9+0.1_HESS}Left: Significance sky map for the field of view of the H.E.S.S.\ GC observations. G0.9+0.1 is marked with the triangle and Sgr A* is marked with the star. Right: 90~cm radio flux map of G0.9+0.1 from \cite{LaRosa2000} (colour scale), overlayed with the H.E.S.S.\ contours at 40$\%$ and 80$\%$ peak brightness (solid lines). The simulated point spread function of the H.E.S.S.\ telescope is shown by the dotted lines (also at 40$\%$ and 80$\%$ peak brightness). The innermost green dotted line illustrates the 95$\%$ confidence limit on the size of the emission region \citep{G0.9+0.1_HESS}.}
\end{minipage} 
\end{figure}

\cite{G0.9+0.1_HESS} studied VHE gamma rays from the GC with the H.E.S.S.\ telescope. The cameras on the H.E.S.S.\ telescope have a large field of view ($\sim 5^{\circ}$) and point sources at an angular distance of up to $\sim 2^{\circ}$ from the camera centre can be observed with good sensitivity. Thus during the observation of Sgr A$^*$ two sources of VHE gamma rays were clearly visible. These can be seen on the significance sky map of the H.E.S.S.\ telescope in Figure~\ref{fig:G0.9+0.1_HESS}.

The panel on the left shows the position of G0.9+0.1, marked with the triangle, with respect to Sgr A$^*$ which is marked with a star. The six telescope pointings are shown as crosses. The panel on the right shows the 90 cm radio flux map of G0.9+0.1 from \cite{LaRosa2000} overlayed with H.E.S.S.\ contours at $40\%$ and $80\%$ peak brightness (solid lines). The simulated point-spread function of the HESS telescope, also at $40\%$ and $80\%$ peak brightness, is indicated by the dotted lines. The innermost green dashed line illustrates the $95\%$ confidence limit on the size of the emission region. \cite{G0.9+0.1_HESS} performed a power-law fit to the spectrum and found a photon index of $2.29 \pm 0.14_{\rm{stat}}$ with a flux of $(5.5 \pm 0.8_{\rm{stat}})\times 10^{-12}$~cm$^{-2}$~s$^{-1}$ for energies above 200 GeV. This flux is only $\sim 2 \%$ of the flux from the Crab Nebula, making G0.9+0.1 one of the weakest sources ever detected at TeV energies.

Some years later, pulsar PSR J1747$-$2809 was discovered in PWN G0.9+0.1 with $P = 52$~ms and $\dot{P} = 1.85\times10^{-13}$ \citep{G0.9+0.1data}. We used these values to calculate $\dot{P_0}$ (time derivative of birth period) and $\tau_0 = P_0/2\dot{P_0}$ (assuming a birth period of $P_0 = 43$ ms and no decay of the pulsar $B$-field) allowing us to add some constraints on some of the parameters that were previously free.

\subsection{Calibration with the model of \cite{VdeJager2007}}
In this section we will use PWN G0.9+0.1 as a calibration source to test our new model against a previous more recent model and then also against a more modern model in Section~\ref{sec:Cal_Torres}. The assumed model parameters used to calibrate our model against that of \cite{VdeJager2007} are listed in Table~\ref{tbl:G0.9}. The latter is a one-zone model (no spatial dependence). 

\begin{figure}[h]
\begin{minipage}[b]{5in}
\includegraphics[width=5in]{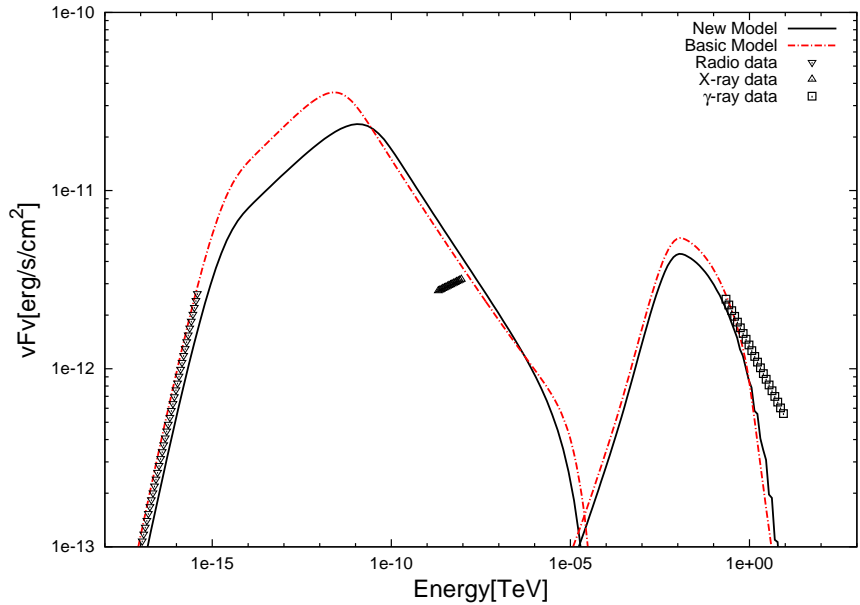}
\caption{\label{fig:Calibrate}Calibration model against the model of \cite{VdeJager2007} for PWN G0.9+0.1.}
\end{minipage} 
\end{figure}

In Table~\ref{tbl:G0.9}, $n$ is the braking index as in Eq.~\eqref{eq:braking}, $\beta_{\rm{VdJ}}$ is the magnetic field parameter as in Eq.~\eqref{eq:fieldCV}, $B(t_{\rm{age}})$ is the present-day magnetic field and in this first calibration with \cite{VdeJager2007} $B(t_{\rm{age}})=40.0~\mu$G is used, noting that this model was developed before the discovery of PSR J1747$-$2809 associated with PWN G0.9+0.1. The more accurate value for the present-day magnetic field, 14.0 $\mu$G, is used in the calibration against \cite{Torres2014} in Section~\ref{sec:Cal_Torres} as we now know $P$ and $\dot{P}$ for the embedded pulsar, as mentioned in the previous section. Also, $\epsilon$ is the conversion efficiency as mentioned in Eq.~\eqref{normQ}, $t_{\rm{age}}$ is the age of the PWN, $\tau_0$ is the characteristic spin-down timescale of the pulsar, $d$ is the distance to the PWN, $\alpha_1$ and $\alpha_2$ are the power law indices of the broken power law injection spectrum as in Eq.~\eqref{brokenpowerlaw}, and $L_0$ the birth spin-down luminosity. The sigma parameter ($\sigma$) is the ratio of the electromagnetic to particle luminosity and is used to calculate the maximum particle energy as discussed in the paragraph after Eq.~\eqref{B_Field}. We chose three soft-photon components: the CMB with a temperature of $T_1 = $ 2.76~K and an average energy density of $u_1 = $ 0.23~eV/cm$^3$, Galactic background infrared photons as component 2, and optical starlight as component 3. For these assumed model parameters we find the SED as shown in Figure~\ref{fig:Calibrate}. The radio data are from \cite{HelfandB1987}, the $X$-ray data from \cite{Sidoli2004} and \cite{Porquet2003}, and the gamma-ray data from \cite{G0.9+0.1_HESS}. The solid line represents our predicted SED while the dashed line shows the output from the model of \cite{VdeJager2007}.

To fit the new model to the model of \cite{VdeJager2007} we had to remove the effects of the bulk particle motion as their model did not incorporate such bulk motion of particles and only considered spatial diffusion in the particle transport. Thus their model did not include adiabatic energy losses nor convection. The way the effect of these processes are removed from the new model is by simply setting the bulk motion inside the PWN to zero. \cite{VdeJager2007} also modelled the magnetic field by parametrising it as
\begin{equation}
B(t) = \frac{B_0}{1+\left( \frac{t}{\tau_0} \right)^{\beta_{\rm{VdJ}}}}.
\label{eq:fieldCV}
\end{equation}
Our model was adapted to also parametrize the magnetic field using the same time-dependent form. These two simple changes to the model allowed us to calibrate our model against theirs as seen in Figure~\ref{fig:Calibrate}.

\begin{table}[t]
\centering
\begin{minipage}[b]{5in}
\small{
\begin{tabular}{|lll|}
\hline
Model Parameter & Symbol &  Value\\  \hline \hline
Braking index & $n$ & 3  \\
$B$-field parameter & $\beta_{\rm{VdJ}}$ & 0.5 \\
Present-day $B$-field & $B(t_{\rm{age}})$ & 40.0 $\mu \rm{G}$ \\
Conversion efficiency  &    $\epsilon$    &  0.6\\
Age & $t_{\rm{age}}$ & 1~900 yr\\
Characteristic timescale & $\tau_0$ & 3~681 yr\\
Distance & $d$ & 8.5 kpc  \\
$Q$ index 1& $\alpha_1$& -1.0 \\
$Q$ index 2& $\alpha_2$& -2.6 \\
Initial spin-down power($10^{38}\rm{erg}$ $\rm{s^{-1}}$)& $L_0$ & 0.99  \\ 
Sigma parameter & $\sigma$ & 0.2 \\
Soft-photon component 1 & $T_1$ and $u_1$ & $2.76$ K, $0.23~\rm{eV/cm^3}$ \\
Soft-photon component 2 & $T_2$ and $u_2$ & $35$ K, $0.5~\rm{eV/cm^3}$ \\  
Soft-photon component 3 & $T_3$ and $u_3$ & $4500$ K, $50~\rm{eV/cm^3}$ \\ \hline
\end{tabular}
}
\caption{Values of model parameters as used in the calibration against the model of \cite{VdeJager2007} for PWN G0.9+0.1.}
\label{tbl:G0.9}
\end{minipage}
\end{table}

Our time-dependent, multi-zone PWN model does not reproduce the results of \cite{VdeJager2007} exactly, but the SEDs are quite close. The reason for this is the fact that the older model did not take into account IC losses in the particle transport, since it assumed SR losses to dominate. This led to losses being underestimated, leaving an excess of high-energy particles. Their IC radiation is therefore slightly higher than our new model prediction. Other differences may result from our very different treatment of the particle transport, as they included no diffusion in their model. 

One thing to note here is that in Table~\ref{tbl:G0.9} the two variables, $\epsilon$ and $\sigma$, are independent. They are, however, related by
\begin{equation}
\epsilon = \frac{1}{1+\sigma}.
\end{equation}
This inconsistency is only present in the calibration with \cite{VdeJager2007} and is correctly implemented in the rest of the thesis.

Our model fits the data well, but still has trouble to fitting the slope of the X-ray spectrum. \cite{Vorster2013} modelled PWN G21.5$-$0.9 where they showed that they also encountered the problem of fitting the slope of the X-ray data when using a broken-power-law injection spectrum. They mention that most models use a broken power law that connects smoothly, i.e., having the same intensity at the transition, as we assumed in our model. They next show that by using a two-component particle injection spectrum that does not transition smoothly (instead the low-energy component cuts off steeply in order to connect to the lower-intensity, high-energy component) allows them to fit both the radio and X-ray spectral slopes. This is something worth noting for future development of our current code. 

\subsection{Calibration with the model of \cite{Torres2014}}\label{sec:Cal_Torres}
As a second calibration we used results from a more recent study by \cite{Torres2014}, where they created a time-dependent model of young PWNe by modelling them with a single-sphere model. We also use PWN G0.9+0.1 as the calibration source. The assumed model parameters for this second calibration are given in Table~\ref{tbl:G0.9_Torres}. The magnetic field is now modelled according to Eq.~\eqref{B_Field}, hence the values of $\alpha_{\rm{B}}$ and $\beta_{\rm{B}}$ in Table~\ref{tbl:G0.9_Torres}. The bulk motion of the particles is parametrised by Eq.~\eqref{V_profile} using model parameters $\alpha_{\rm{V}}$, $V_0$, and $r_0$.
\begin{table}[t]
\centering
\begin{minipage}[b]{5in}
\small{
\begin{tabular}{|lll|}
\hline
Model Parameter & Symbol &  Value\\  \hline \hline
Braking index & $n$ & 3  \\
$B$-field parameter & $\alpha_B$ & 0.0 \\
$B$-field parameter & $\beta_B$ & -1.3 \\
$V$-field parameter & $\alpha_V$ & 1.0 \\
Present-day $B$-field & $B(t_{\rm{age}})$ & 14.0 $\mu \rm{G}$ \\
Conversion efficiency  &    $\epsilon$    &  0.99\\
Age & $t_{\rm{age}}$ & 2~000 yr\\
Characteristic timescale & $\tau_0$ & 3~305 yr\\
Distance & $d$ & 8.5 kpc  \\
$Q$ index 1& $\alpha_1$& -1.4 \\
$Q$ index 2& $\alpha_2$& -2.7 \\
Initial spin-down power($10^{38}\rm{erg}$ $\rm{s^{-1}}$)& $L_0$ & 1.1  \\ 
Sigma parameter & $\sigma$ & 0.01 \\
Soft-photon component 1 & $T_1$ and $u_1$ & $2.76$ K, $0.23~\rm{eV/cm^3}$ \\
Soft-photon component 2 & $T_2$ and $u_2$ & $30$ K, $2.5~\rm{eV/cm^3}$ \\  
Soft-photon component 3 & $T_3$ and $u_3$ & $3000$ K, $25~\rm{eV/cm^3}$ \\ \hline
\end{tabular}
}
\caption{Values of model parameters as used in the calibration against the model of \cite{Torres2014} for PWN G0.9+0.1.}
\label{tbl:G0.9_Torres}
\end{minipage}
\end{table}
\begin{figure}[b]
\begin{minipage}[b]{5in}
\includegraphics[width=5in]{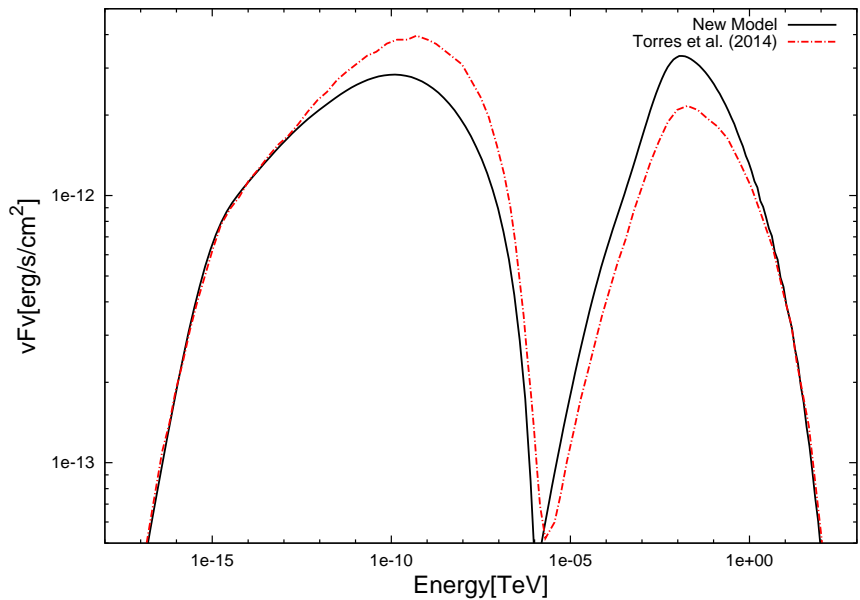}
\caption{\label{fig:Calibrate_Torres}Calibration model against the model of \cite{Torres2014} for PWN G0.9+0.1.}
\end{minipage} 
\end{figure}
Some of the parameters are different from those used during the calibration with the model of \cite{VdeJager2007}. One of these changes is the present-day magnetic field that is now set to $14~\mu G$, versus the previous value of $40~\mu G$. The reason for this is the discovery of pulsar J1747$-$2809 in the PWN G0.9+0.1. Thus $P$ and $\dot{P}$ are now known, so that a more accurate calculation of the present-day magnetic field ($B \propto \sqrt{P \dot{P}}$) can be made. The magnetic field is parametrised using $\alpha_B = 0$ and $\beta_B = -1.3$ which, from Eq.~\eqref{B_Field}, indicates that the magnetic field is constant in the spatial dimension. This is consistent with what \cite{Torres2014} assumed in their one-zone model. They model the time dependence of the magnetic field using
\begin{equation}
\int_0^t (1-\epsilon) L(t')R_{\rm{PWN}}(t')dt' = W_{B}R_{\rm{PWN}}
\label{eq:Torres_B}
\end{equation}
where
\begin{equation}
W_{B}=\frac{4\pi}{3}R^3_{PWN}(t)\frac{B^2(t)}{8\pi}
\end{equation}
and mention that if the age of the PWN is less than the characteristic age ($t_{\rm{age}} < \tau_0$), then $B(t) \propto t^{-1.3}$. Therefore we set the value of $\beta_B = -1.3$. One thing to note here is the usage of $R_{\rm{PWN}}$. \cite{Torres2014} explicitly uses a time-dependent PWN radius for G0.9+0.1, setting $R_{\rm{PWN}}(t_{\rm{age}}) = 2.5$ pc. We however do not. Instead we choose an $r_{\rm{max}}$ that is larger than $R_{\rm{PWN}}$ and then later calculate the size of the PWN by noting where the surface brightness has decreased by two thirds. This is possible for us since we have information about the morphology of the PWN. These results are shown in Section~\ref{sec:Space}. 

The way the velocity is parametrised is by setting $\alpha_V = 1.0$. This is done so that our model can have the same adiabatic energy loss rate as assumed by \cite{Torres2014}. They have a constant adiabatic energy loss timescale and to reproduce this in our model, we have to set $\alpha_V = 1$ (see Eq.~[\ref{eq:Edot_adiabatic}] and [\ref{eq_ap:nablaDOTv}]). This is, however, not physical, in view of the relationship between $V(r)$ and $B(r,t)$ in Eq.~\eqref{eq:a_v+a_b=-1}. From these equations it is clear that $\alpha_V = -1$ when $\alpha_B = 0$. The changes in $B(r,t)$ and $V(r)$ are the only substantial difference. The rest of the parameters are very similar to the previous case, e.g., the indices of the injection spectrum and the soft-photon components used in the calculation of the IC spectrum.

Figure~\ref{fig:Calibrate_Torres} compares our predicted SED with the model predictions of \cite{Torres2014}, with their results shown by the dashed-dotted line and our model SED shown as the solid line. The differences in the two models stem from the different way in which the transport of particles is handled. In our code we incorporated a Fokker-Planck-type transport equation and \cite{Torres2014} modelled the transport by using average timescales.

\subsection{Calibration using other sources also modelled by \cite{Torres2014}}\label{sec:fits}
\begin{figure}[t]
 \begin{minipage}{0.5\linewidth}
  \centering
  \includegraphics[width=1.0\linewidth]{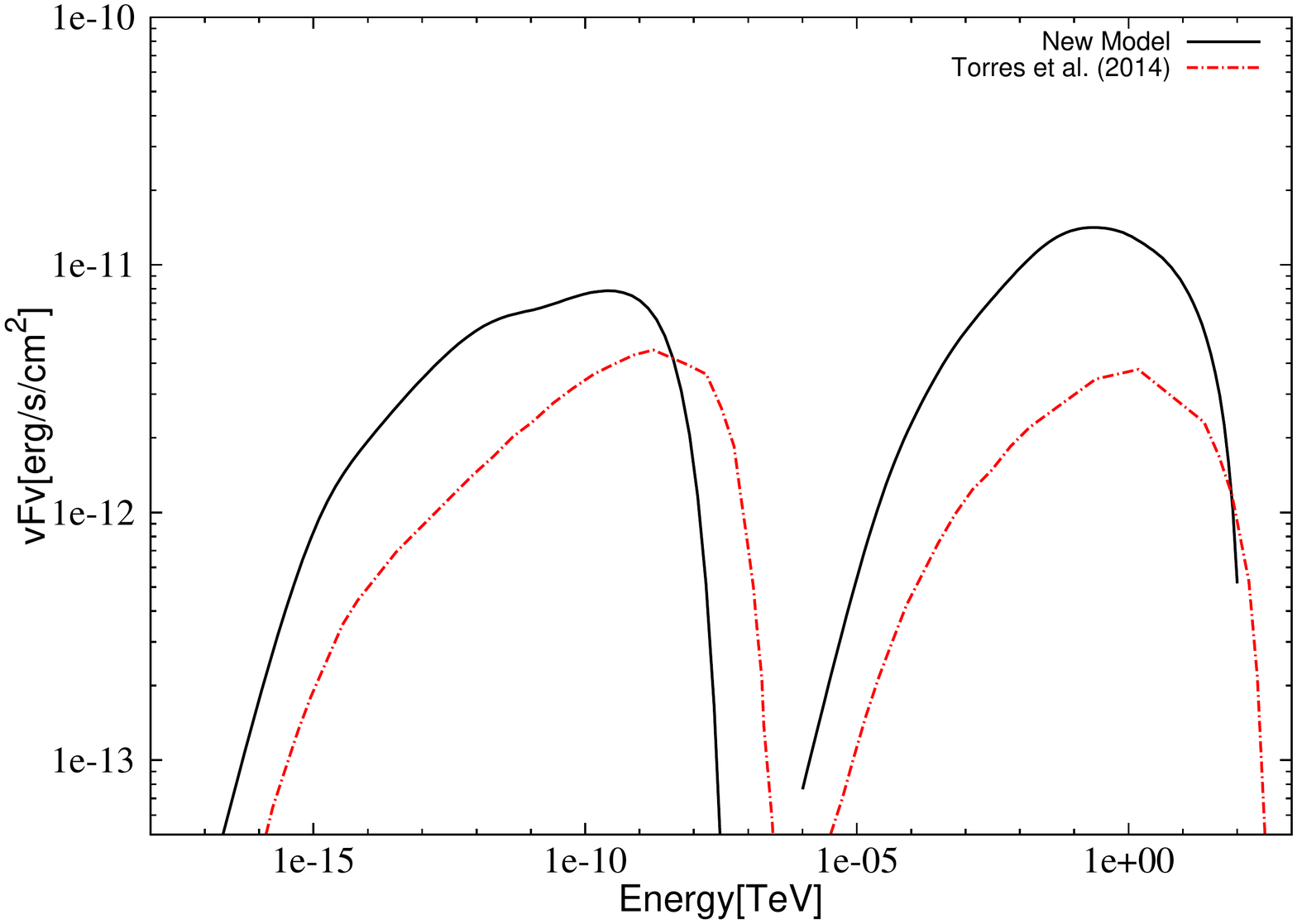}
  \caption{Our model against the model of \cite{Torres2014} for HESS J1356$-$645.}
  \label{fig:HESS_J1356a}
 \end{minipage}%
 \begin{minipage}{0.5\linewidth}
  \centering
  \includegraphics[width=1.0\linewidth]{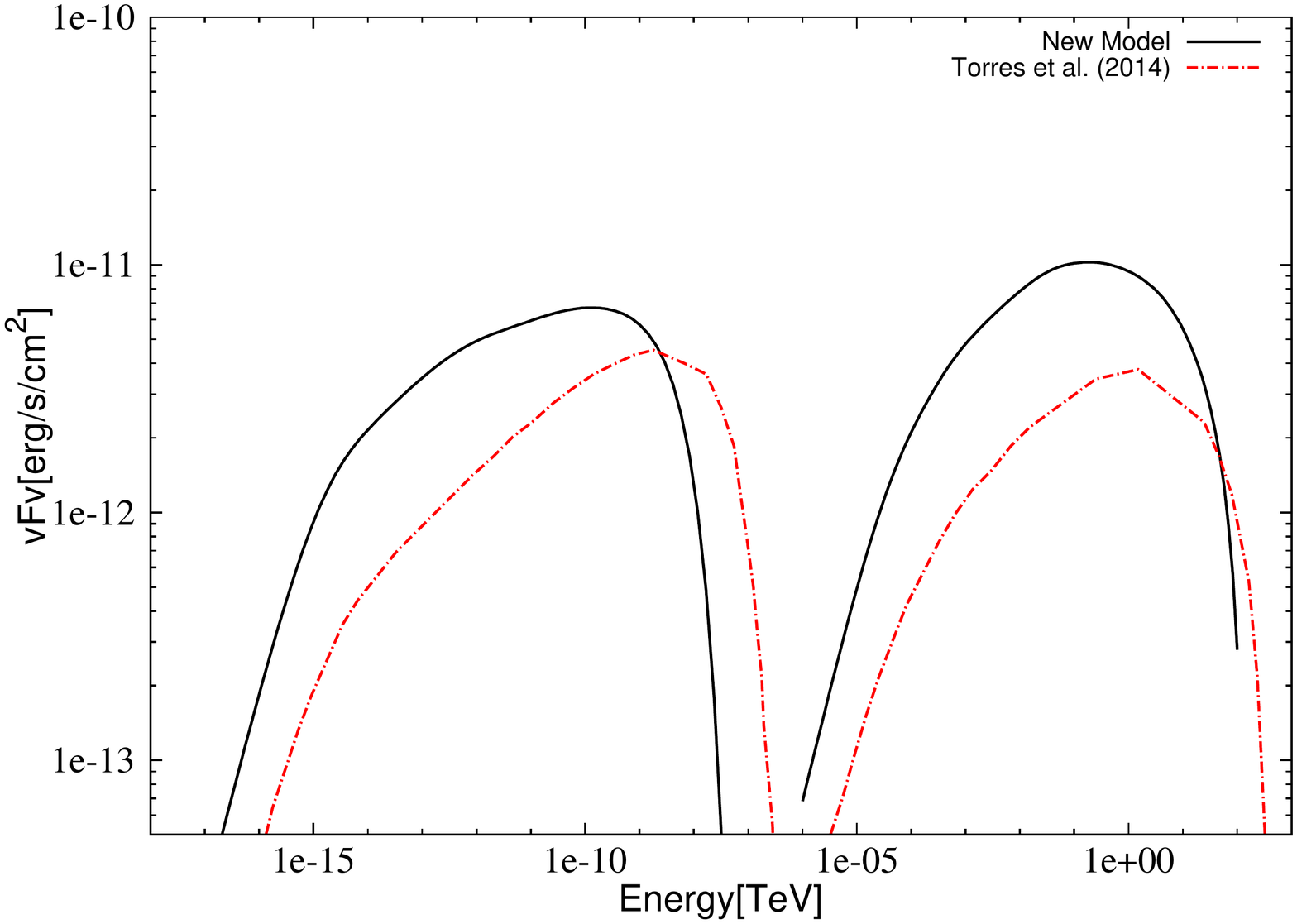}
  \caption{Our model against the second model of \cite{Torres2014} for HESS J1356$-$645.}
  \label{fig:HESS_J1356b}
 \end{minipage}
\end{figure}

From the previous sections it is clear that our new model provides a good fit to the SED of G0.9+0.1, but we are also interested in other young PWNe. As a further quick test of the code, we chose three other sources, G21.5-0.9, G54.1+0.3, and HESS J1356$-$645, and compared our model predictions with those of \cite{Torres2014};
\begin{figure}[t]
 \begin{minipage}{0.5\linewidth}
  \centering
  \includegraphics[width=1.0\linewidth]{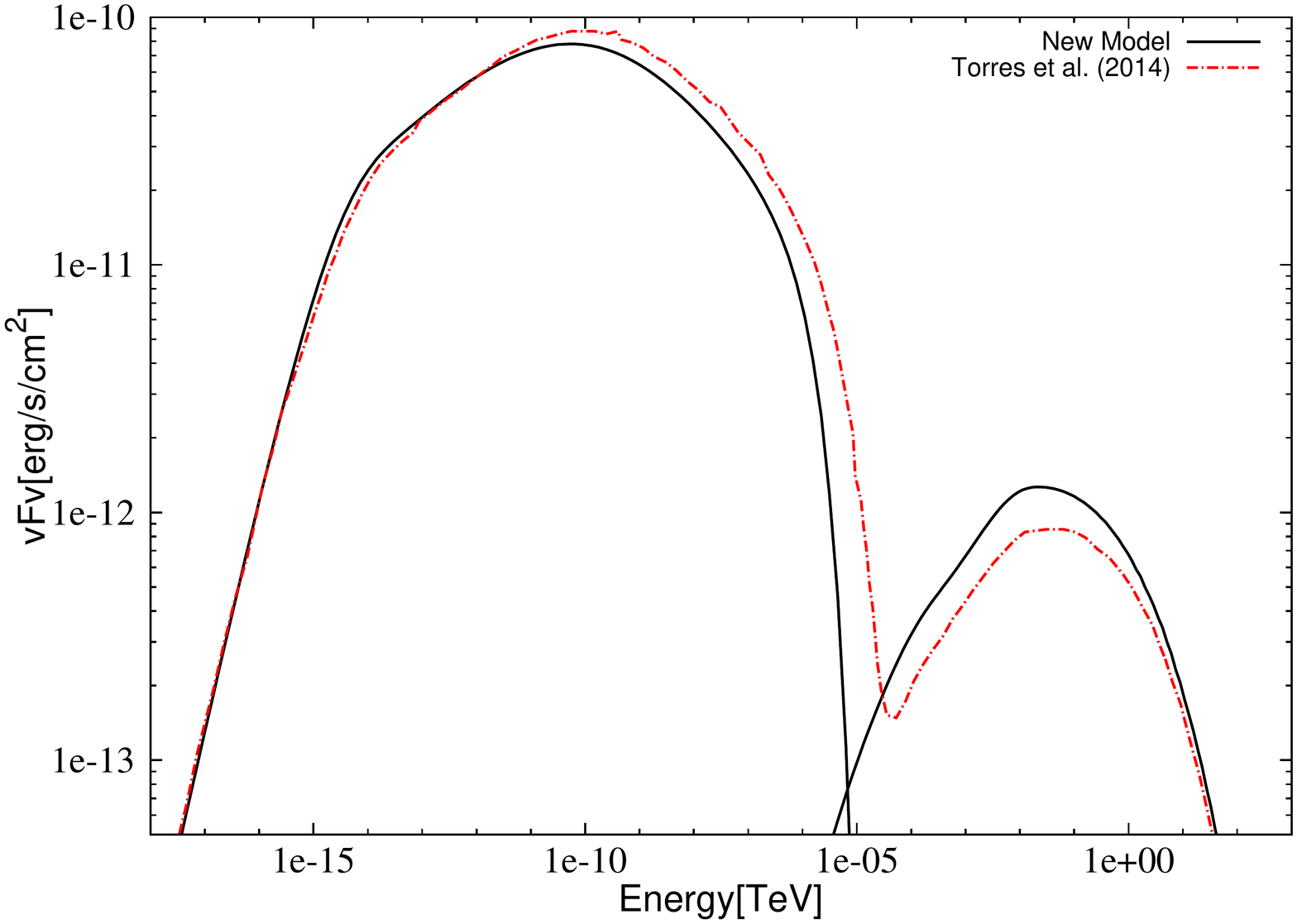}
  \caption{Our model against the model of \cite{Torres2014} for G21.5-0.9.}
  \label{fig:G21}
 \end{minipage}%
 \begin{minipage}{0.5\linewidth}
  \centering
  \includegraphics[width=1.0\linewidth]{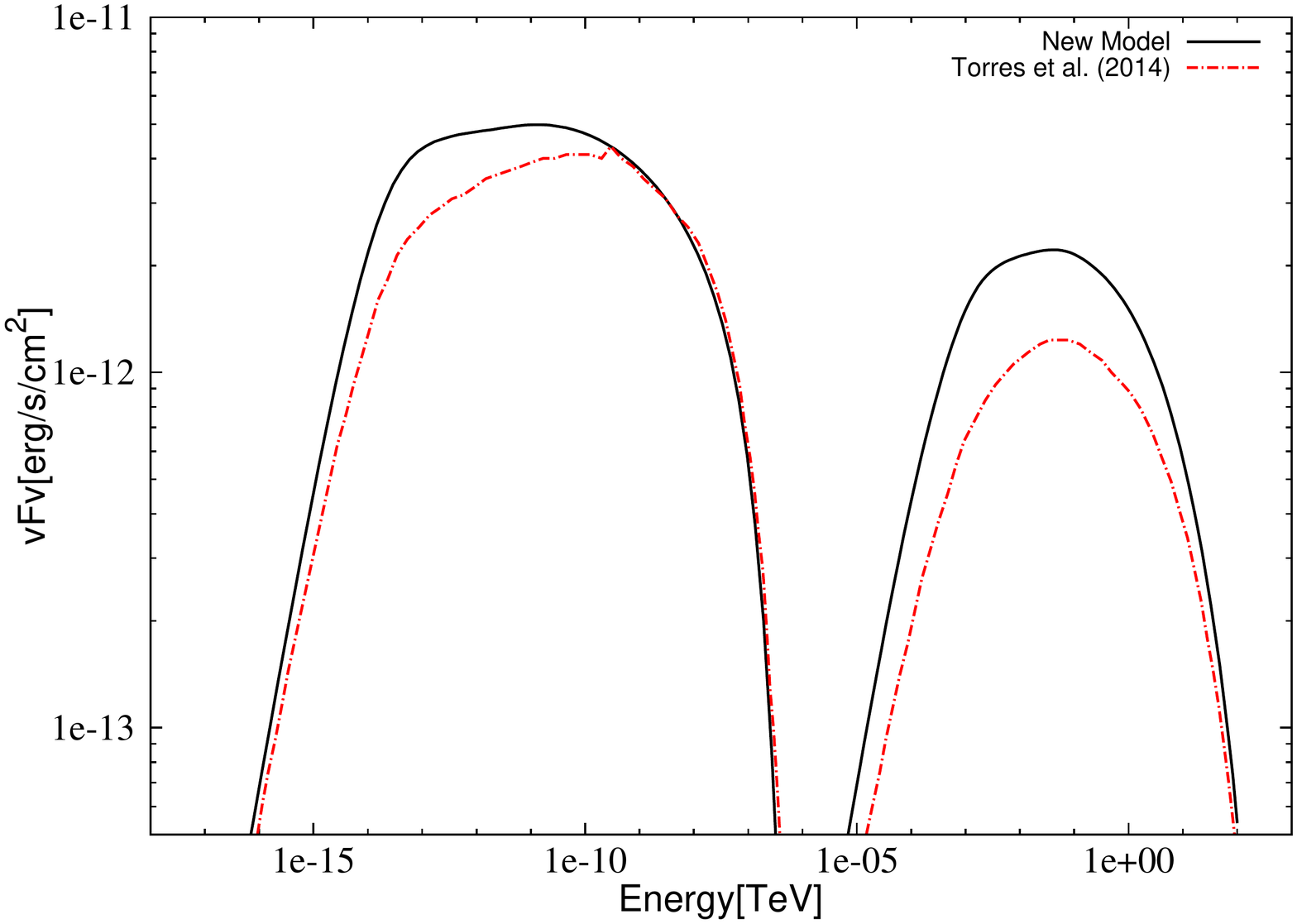}
  \caption{Our model against the model of \cite{Torres2014} for G54.1+0.3.}
  \label{fig:G54}
 \end{minipage}
\end{figure}
see Figures \ref{fig:G21}, \ref{fig:G54}, \ref{fig:HESS_J1356a}, and \ref{fig:HESS_J1356b}. For the first two sources, our model seems to also provide a good fit. The main reason for the slight differences is the fact that the magnetic field is not modelled in exactly the same way, as they use Eq.~\eqref{eq:Torres_B} and we use the parametrised form for the magnetic field in the PWN as in Eq.~\eqref{B_Field}. This once more shows that our approximation of $B(t) \propto t^{-1.3}$ for $t_{\rm{age}} < \tau_0$ is a good one, but for HESS J1356$-$645 this is no longer the case, since the age of the PWN exceeds the characteristic timescale $\tau_0$. This can be seen in Figure~\ref{fig:HESS_J1356a} and \ref{fig:HESS_J1356b} where \cite{Torres2014} used two different models to model this source. Both show substantially different results from those of our model. This shows that our model is currently only suitable for young PWNe and needs further development in future. 

\section{Parameter study}\label{sec:parameterST}
In the previous section I showed that our model calibrates well with two independent PWN models. We can now investigate the effects of all the different free parameters in the model on the particle spectrum and the SED. As a reference model for this section, we use the same parameters that were used in the calibration against \cite{Torres2014} for G0.9+0.1, as in Figure~\ref{fig:Calibrate_Torres}. The SED of the PWN is calculated at Earth for each zone and then these are added to find the total flux from the PWN.

\subsection{Evolution of the PWN}
\label{sec:time_evol}
In Section~\ref{sec:dnde}, I showed how the lepton spectrum evolves as the particles are injected into the first zone of the PWN, with an injection spectrum as in Figure~\ref{fig:injec}, and are then allowed to radiate, as discussed in Section~\ref{sec:RadSpec}. The PWN can be modelled for different ages and this causes the particle spectrum, and thus the radiation spectrum, to change as the PWN ages. Here the present-day magnetic field is kept constant and parametrised as in Eq.~\eqref{B_Field}.

\begin{figure}[b]
\centering
\begin{minipage}[b]{5in}
\centering
\includegraphics[width=5in]{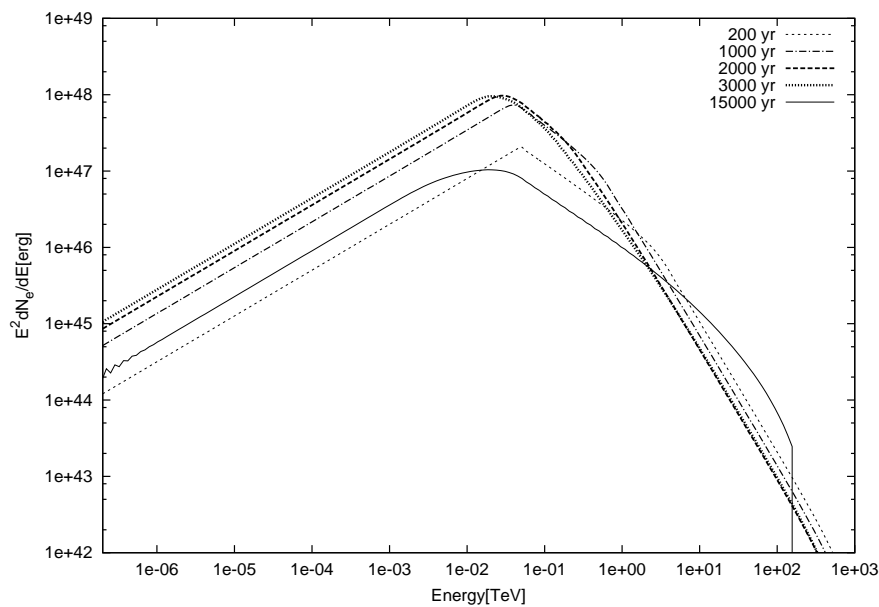}
\caption[Evolution of the lepton spectrum versus age.]{\label{fig:dnde_time}Evolution of the lepton spectrum. The different line types indicate the age of the PWN with the dashed line at 2 000~yr (current age of the PWN), and the rest of the lines indicating the evolution of the PWN.}
\end{minipage} 
\end{figure}

\begin{figure}[t]
\centering
\begin{minipage}[b]{5in}
\centering
\includegraphics[width=5in]{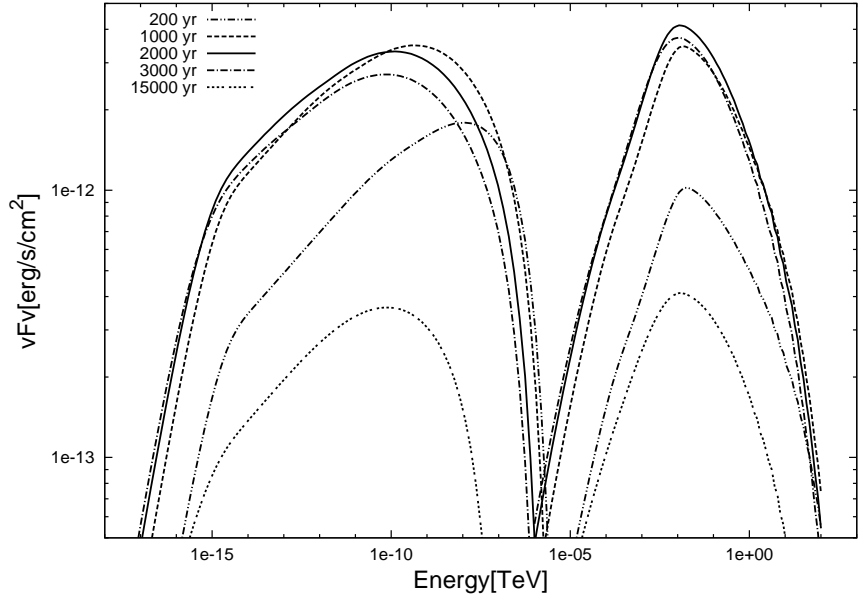}
\caption{\label{fig:SED_age}SED for PWN~G0.9+0.1 with a change in the age of the PWN. The solid line shows SED for $2~000$~yr (current age of the PWN). The other lines show the time progression of the SED from the PWN.}
\end{minipage} 
\end{figure}

In Figure~\ref{fig:dnde_time} the time evolution of the lepton spectrum is shown with the radiation spectrum shown in Figure~\ref{fig:SED_age}. From these two figures it can be seen that when the PWN is still very young ($t_{\rm{age}}\sim 200$ yr) the particle spectrum closely resembles the shape of the injection spectrum. As the PWN ages, however, it starts to fill up with particles (giving an increased $N_{\rm{e}}$) and at some stage the PWN is totally filled, at $T$ in the order of a few thousand years. After this the particle spectrum decreases. This is due to the particles losing energy over time due to SR, IC and adiabatic energy losses, and also due to the fact that the embedded pulsar is spinning down, resulting in fewer particles being injected into the PWN. If the particle spectrum for an age of 15 000 yr is observed, then the effect of the spun-down pulsar can be clearly seen in Figure~\ref{fig:dnde_time}. By this time the embedded pulsar has spun down so that the total particle spectrum is lower than it is at $\approx 200$ yr due to the fact that now more particles are escaping from the PWN than are being injected by the pulsar. Also note the leftward shift of $E_{\rm{b}}$ due to radiative losses. The bump at high energies for 15 000 yr is due to a pile-up of particles. This occurs due to the decreased magnetic field, resulting in an increased diffusion and also decreased SR energy losses. These losses are energy-dependent and therefore the high-energy particles will be affected most. The increased diffusion will cause the particles to build up as they do not escape, since our chosen $r_{\rm{max}} \gg R_{\rm{PWN}}$. This will be discussed in more detail in Section~\ref{sec:alph_Valph_B}.

The particle spectrum in Figure~\ref{fig:dnde_time} not only goes up and down as the PWN ages, but the whole spectrum shifts to lower energies. This can be seen by looking at where the spectrum peaks and also at the tails at high and low energies. This is due to the fact that the particles lose energy through different mechanisms, as discussed in Section~\ref{sec:rad_losses}. Due to the SR energy losses, the particle spectrum will develop a break at some break energy. The SR loss scale is given by
\begin{equation}
  \dot{E}_{\rm{SR}} = -\frac{\sigma_T}{6\pi m_{\rm{e}}^2 c^3} E_{\rm{e}}^2B^2.
  \label{SR123}
\end{equation}
By using $\dot{E}_{\rm{SR}}$ to calculate the timescale for synchrotron losses ($\tau_{\rm{SR}}$) and setting it equal to the age of the PWN ($t_{\rm{age}}$), one may estimate where the break is expected in the spectrum:
\begin{equation}
  \tau_{\rm{SR}} = \frac{E_{\rm{e}}}{\dot{E}_{\rm{SR}}} = t_{\rm{age}} \Rightarrow E_{\rm{e}} \propto \frac{1}{t_{\rm{age}} \langle B \rangle^2}.
  \label{eq:SR_timescale}
\end{equation}
Thus from Eq.~\eqref{eq:SR_timescale} we can see that the break should move to lower energies as the PWN ages. In Eq.~\eqref{eq:SR_timescale} we have to use the average magnetic field $\langle B \rangle$ over the lifetime of the PWN as the present-day magnetic field is too small. This is visible in Figure~\ref{fig:dnde_time} where the break for 200 yr is at $\approx$ 2 TeV, for 1 000 yr $\approx$ 0.6 TeV, for 2 000 yr $\approx$ 0.2 TeV, and for 5 000 yr $\approx$ 0.15 TeV. We can check this by comparing it to the predicted break energy as in Eq.~\eqref{eq:SR_timescale}.

The particle spectrum is reflected by the radiation spectrum (Figure~\ref{fig:SED_age}). Here the radiation spectrum also increases as the PWN ages, up to a maximum at a few thousand years and then starts to decrease and die down. Over the majority of the PWN lifetime, the spectral peak shifts towards lower energies due to accumulating energy losses.

\subsection{Magnetic field}\label{sec:B_field_par}
The magnetic field $B(r,t)$ inside the PWN plays a large role in determining the shape of the SED, and is characterised by the free parameters $B_{\rm{age}}$, $\alpha_B$ and $\beta_B$ (Table~\ref{tbl:G0.9_Torres}). The energy losses due to SR, the diffusion as well as the SR spectral shape (and peak energy) are dependent on the magnetic field strength (and indirectly, the IC spectrum). Due to this fact it is important to investigate what effects a change in the magnetic field strength will have on the particle and radiation spectrum of the PWN. As default parameter the present-day magnetic field is set to $14~\mu G$ and the present-day magnetic field is then changed to $10~\mu G$, $20~\mu G$ and to $40~\mu G$ to see what effect this will have. 

\begin{figure}[h!]
\centering
\begin{minipage}[b]{5in}
\centering
\includegraphics[width=5in]{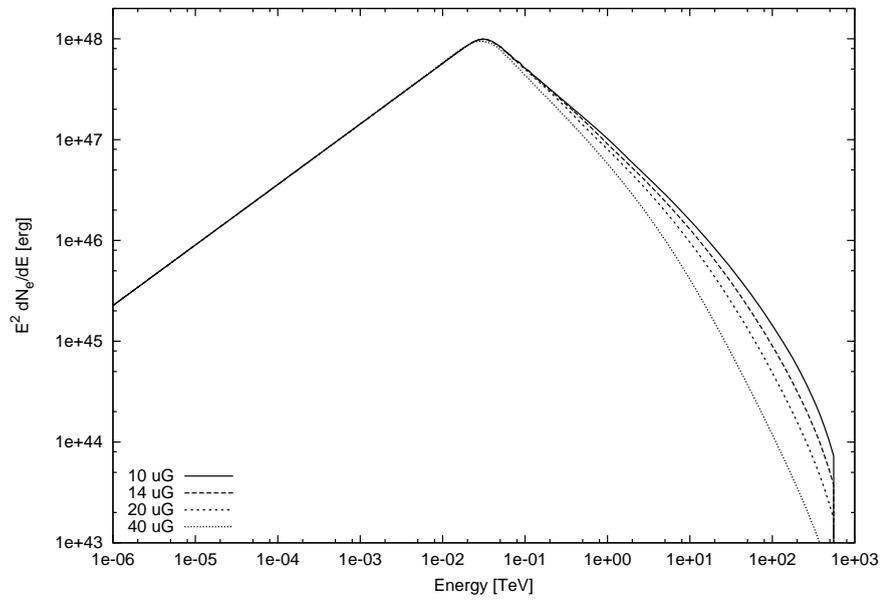}
\caption{\label{fig:Par_Change_B}Particle spectrum for PWN~G0.9+0.1 with a change in the present-day magnetic field.}
\end{minipage} 
\end{figure}

\begin{figure}[h!]
\centering
\begin{minipage}[b]{5in}
\centering
\includegraphics[width=5in]{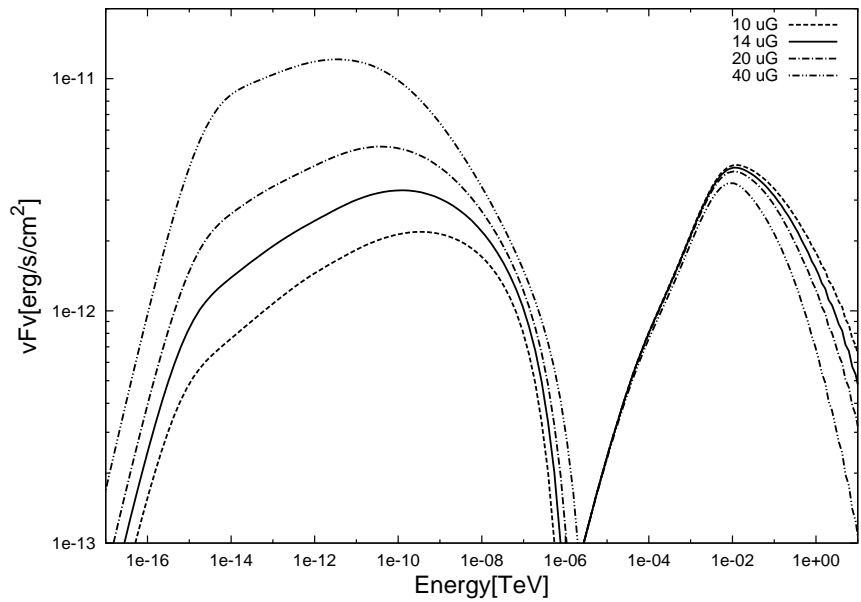}
\caption{\label{fig:SED_Change_B}SED for PWN~G0.9+0.1 with a change in the present-day magnetic field.}
\end{minipage} 
\end{figure}

The magnetic field inside the PWN is modelled by 
\begin{equation}
B(r,t) = B_{\rm{age}}\left(\frac{r}{r_0}\right)^{\alpha_B}\left(\frac{t}{t_{\rm{age}}}\right)^{\beta_B},
\label{B_Field123}
\end{equation}
where $r_0$ is the radius of the termination shock, and for this section, the values for $\alpha_B$ and $\beta_B$ are fixed to 0.0 and -1.3, respectively, as mentioned earlier, so only the value of $B_{\rm{age}}$ was changed. As the magnetic field in the PWN increases from 10 $\mu G$ to 40 $\mu G$ the particle spectrum becomes softer at high energies, since $\dot{E}_{\rm{SR}} \propto E_{\rm{e}}^2B^2$. Thus higher-energy particles lose more energy so that there are fewer particles at high energies left to radiate. The IC spectrum in Figure~\ref{fig:SED_Change_B} is therefore lower for a larger magnetic field. The SR power is directly proportional to the magnetic field strength squared and thus as the magnetic field increases, the SR also increases. The diffusion is modelled by Bohm diffusion 
\begin{equation}
\kappa = \frac{c}{3e}\left(\frac{E_{\rm{e}}}{B}\right),
\end{equation}
which is inversely proportional to the magnetic field. Therefore, an increased magnetic field will result in a decrease in diffusion as well as a smaller source (see Section~\ref{sec:Space}). 

Changes to $\alpha_B$ will be discussed in Section~\ref{sec:Space} as it is a spatial parameter.

\subsection{Bulk particle motion}\label{sec:V_field_par}
\begin{figure}[h!]
\centering
\begin{minipage}[b]{5in}
\centering
\includegraphics[width=5in]{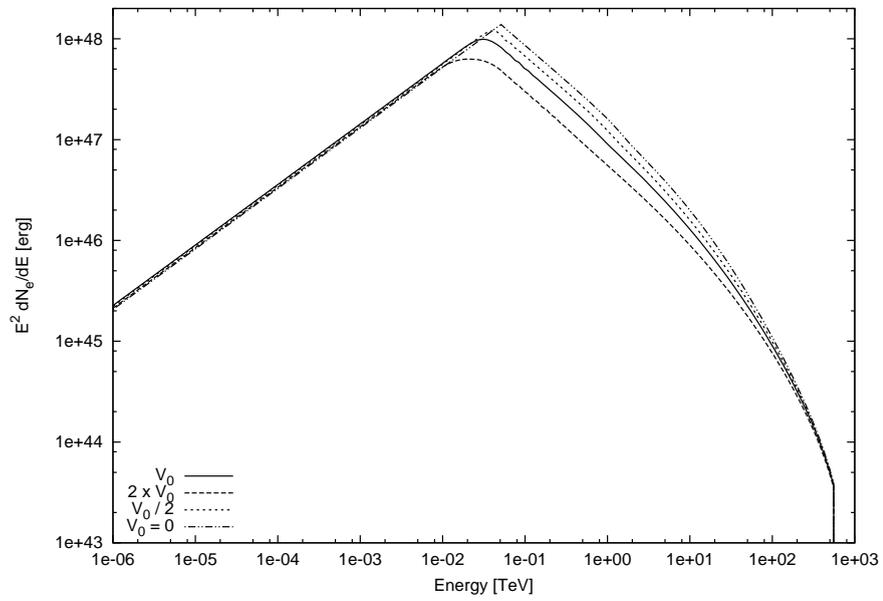}
\caption{\label{fig:Par_Change_V}Particle spectrum for PWN~G0.9+0.1 with a change in the bulk speed of the particles.}
\end{minipage} 
\end{figure}

\begin{figure}[h!]
\centering
\begin{minipage}[b]{5in}
\centering
\includegraphics[width=5in]{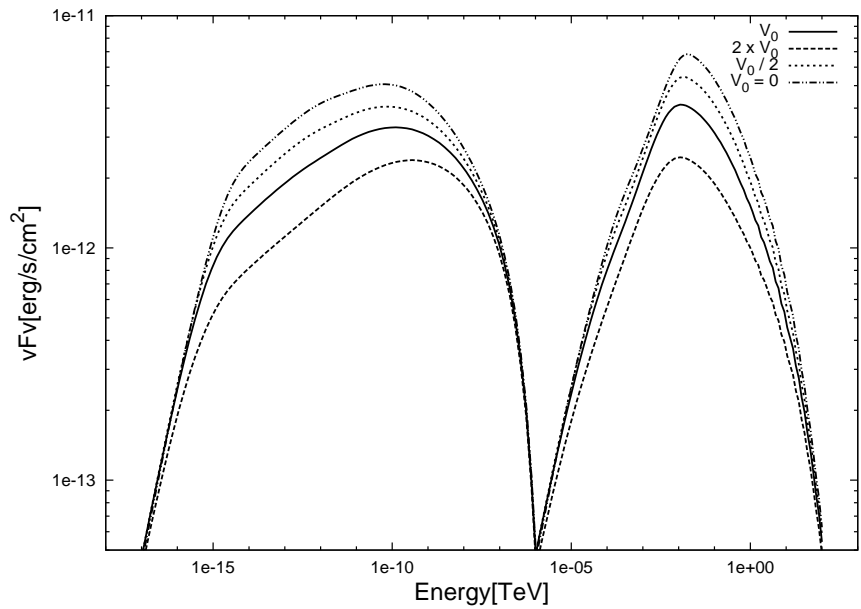}
\caption{\label{fig:SED_Change_V}SED for PWN~G0.9+0.1 with a change in the bulk speed of the particles.}
\end{minipage} 
\end{figure}

The bulk particle motion (particle speed) in the the PWN is modelled by 
\begin{equation}
V(r) = V_0\left(\frac{r}{r_0}\right)^{\alpha_V},
\label{V_profile123}
\end{equation}
and the value for $\alpha_V = 1$ is kept constant in this section, although the value of $V_0$ is changed to $V_0 = 0$, $2V_0$ and $V_0/2$ as can be seen in Figures \ref{fig:Par_Change_V} and \ref{fig:SED_Change_V}. In this section, as in the previous one, we compare our results to those of \cite{Torres2014} and thus we need the same form for the bulk particle motion. To achieve this, we have to use a constant adiabatic timescale, implying in $\alpha_V = 1$. This is a non-physical assumption as mentioned in Section~\ref{sec:Cal_Torres}. However, by using this we find the value for $V_0$ from the adiabatic timescale
\begin{equation}
\tau_{\rm{ad}} = \frac{E}{\dot{E}_{\rm{ad}}},
\label{eq:ad_parmtr}
\end{equation}
where $\dot{E}_{\rm{ad}} = (\nabla \cdot \mathbf{V})E_{\rm{e}}/3$. By using the analytical form of $(\nabla \cdot \mathbf{V})$ in Eq.~\eqref{eq_ap:nablaDOTv} we find that $V_0 = r_0/\tau_{\rm{ad}}$ and for PWN G0.9+0.1. The adiabatic timescale \citep{Torres2014} used was $\sim$ 2 000 yr, giving $V_0 = 5 \times 10^{-5}$ pc/yr for $r_0 = 0.1$ pc and $\tau_{\rm{ad}} = 2~000$ yr.

In Figure~\ref{fig:Par_Change_V} the particle spectrum increases as $V_0$ is lowered. This is due to the fact that for a lower speed, the particles lose less energy due to adiabatic losses as can be seen from the equation after Eq.~\eqref{eq:ad_parmtr}, resulting in more particles at certain energies. The adiabatic energy losses also account for the shift of the peak in the particle spectrum. The radiation spectrum is linked to the particle spectrum and therefore a lower particle spectrum results in a lower radiation spectrum. This effect can be seen in Figure~\ref{fig:SED_Change_V} where the radiation decreases with an increase in the bulk speed of the particles. For high energies SR energy losses dominates, and therefore the radiation spectrum is independent for changes to $V_0$ as seen in Figure~\ref{fig:SED_Change_V} where the solutions converge at high energies for different scenarios of $V_0$.

Changes to $\alpha_V$ will be discussed in Section~\ref{sec:Space} as it is a spatial parameter.

\subsection{Normalisation of the injected particles}
\begin{figure}[h!]
\centering
\begin{minipage}[b]{5in}
\centering
\includegraphics[width=5in]{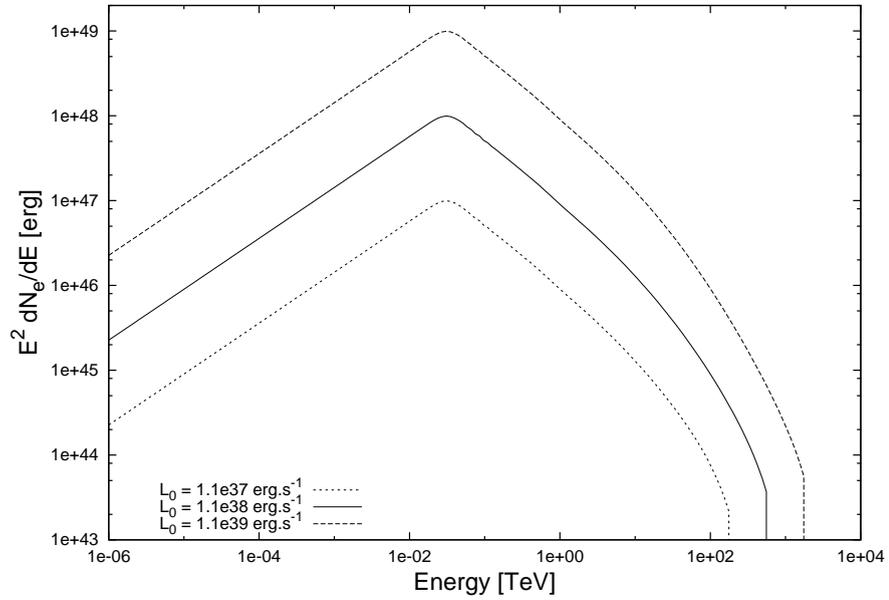}
\caption{\label{fig:Par_Change_L0}Particle spectrum for PWN~G0.9+0.1 with a change in the injection spectrum (change in $L_0$).}
\end{minipage} 
\end{figure}

\begin{figure}[h!]
\centering
\begin{minipage}[b]{5in}
\centering
\includegraphics[width=5in]{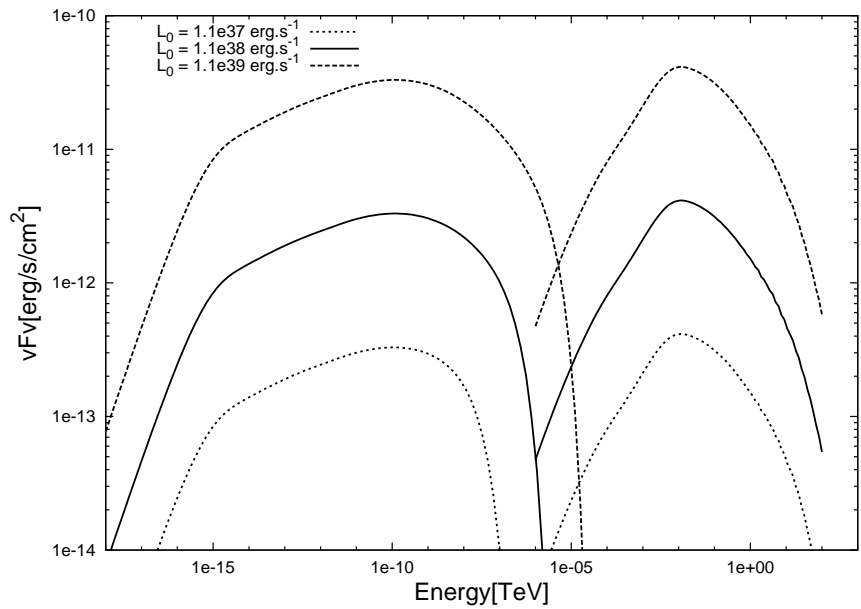}
\caption{\label{fig:SED_Change_L0}SED for PWN~G0.9+0.1 with a change in the injection spectrum (change in $L_0$).}
\end{minipage} 
\end{figure}

The particles in the PWN are injected from the embedded pulsar and the injected spectrum is normalised using the spin-down power of the pulsar. The spin-down power of the pulsar is given by
\begin{equation}
L(t) = L_0 \left(1+\frac{t}{\tau_c}\right)^{-(n+1)/(n-1)},
\label{eq:spinDown123}
\end{equation} 
and the number of injected particles is directly proportional to this spin-down power as discussed in Section~\ref{sec:Injection}. We can thus change $L_0$ to inject more or fewer particles into the PWN. Figures \ref{fig:Par_Change_L0} and \ref{fig:SED_Change_L0} show the effects of this change. If more particles are injected into the PWN, the whole particle spectrum of the PWN will increase and thus also the radiation spectrum and vice versa. In these two figures the normalisation of $L(t)$ is increased and reduced by a factor of ten. This change does not influence the shape of either the particle or the radiation spectrum but simply increases or lowers the amount of particles in the PWN, and thus also the radiation received from the PWN.

The same effect is seen when the value of the conversion efficiency ($\epsilon$) is changed, since $\epsilon$ also changes the normalisation of the injected particles as seen in Eq.~\eqref{normQ}.

\subsection{Characteristic timescale of the embedded pulsar}
\begin{figure}[h!]
\centering
\begin{minipage}[b]{5in}
\centering
\includegraphics[width=5in]{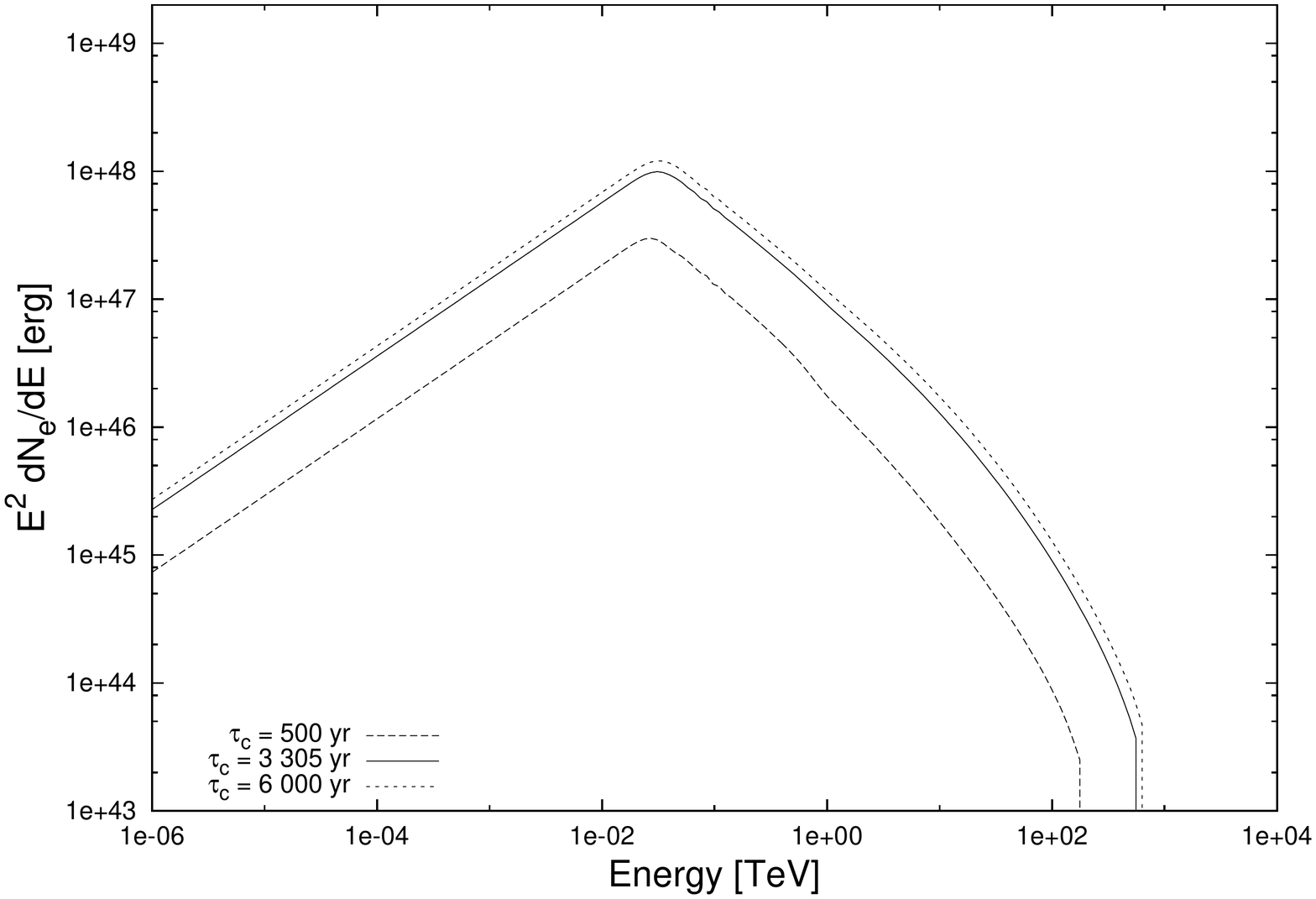}
\caption{\label{fig:Par_Change_tau}Particle spectrum for PWN~G0.9+0.1 with a change in the characteristic timescale of the embedded pulsar (change in $\tau_0$).}
\end{minipage} 
\end{figure}

\begin{figure}[h!]
\centering
\begin{minipage}[b]{5in}
\centering
\includegraphics[width=5in]{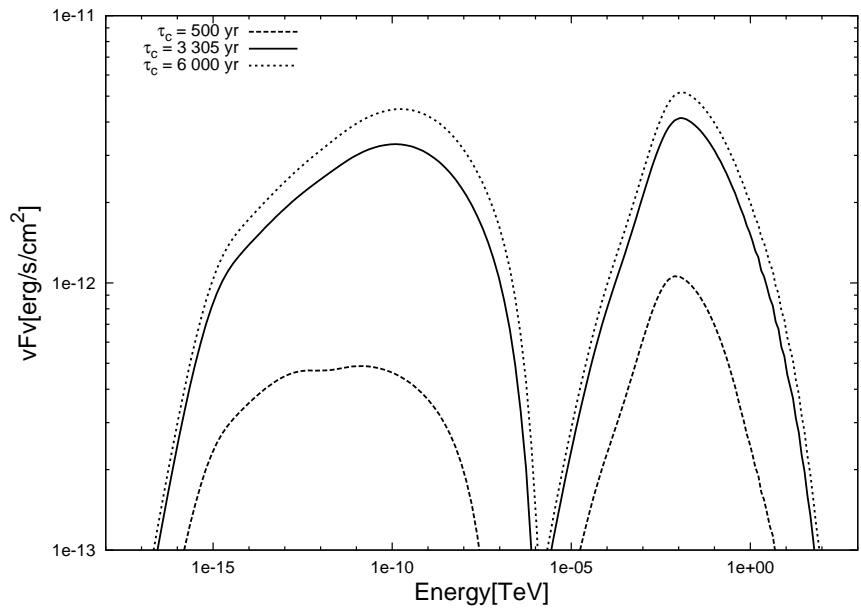}
\caption{\label{fig:SED_Change_tau}SED for PWN~G0.9+0.1 with a change in the characteristic timescale of the embedded pulsar (change in $\tau_0$).}
\end{minipage} 
\end{figure}

Another free parameter is the characteristic spin-down timescale ($\tau_c$) given in Eq.~\eqref{eq:spinDown123} which characterises how fast the pulsar spins down. Figures \ref{fig:Par_Change_tau} and \ref{fig:SED_Change_tau} show what the effects are when changing $\tau_c$. We can see that when the characteristic time is shorter, the pulsar spins down faster, resulting in fewer particles being injected into the PWN and thus the particle and radiation spectrum are both lower. The opposite happens when $\tau_c$ is longer, since more particles are injected into the PWN over time, resulting in a relative increase in the particle and radiation spectrum.

\clearpage
\subsection{Diffusion of particles in the PWN}\label{sec:Var_diff123}
The diffusion coefficient in the model is parametrised as discussed in Section~\ref{sec:Diff}. The diffusion coefficient thus has two free parameters, which can be seen in Eq.~\eqref{eq:kappa123}. Here we consider the parameters $\kappa_0$ and $q$. The value of $E'_0$ is set to 1 TeV. We can now increase or decrease the value of $\kappa_0$ and thus change the normalisation of the diffusion coefficient. We can also change $q$ which has an influence on the energy dependence of the diffusion coefficient ($q=1$ is Bohm diffusion): 
\begin{equation}
\kappa = \kappa_0\left(\frac{E}{E'_0}\right)^{q}.
\label{eq:kappa123}
\end{equation}

\begin{figure}[t]
\centering
\begin{minipage}[b]{5in}
\centering
\includegraphics[width=5in]{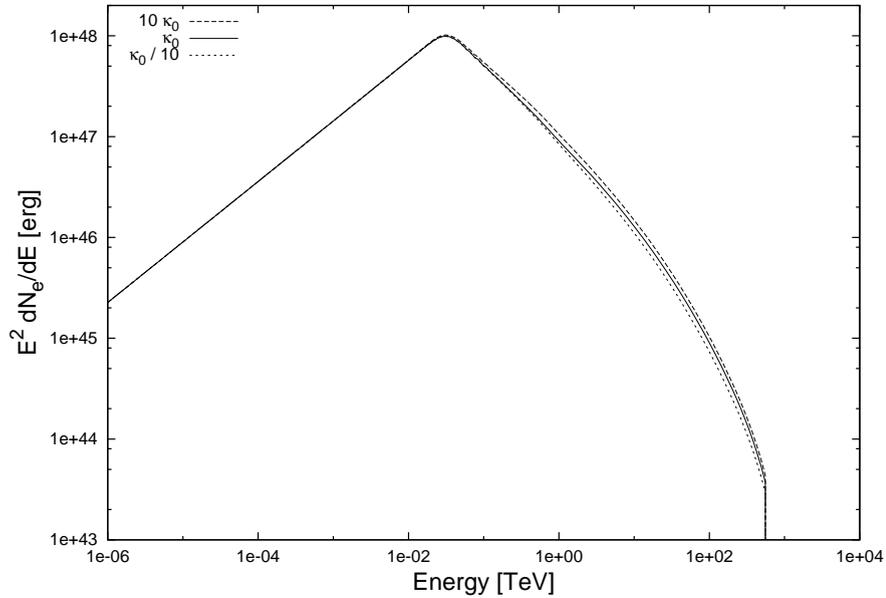}
\caption{\label{fig:Par_Change_kap1}Particle spectrum for PWN~G0.9+0.1 with a change in the normalisation constant of the diffusion coefficient.}
\end{minipage} 
\end{figure}

\begin{figure}[t]
\centering
\begin{minipage}[b]{5in}
\centering
\includegraphics[width=5in]{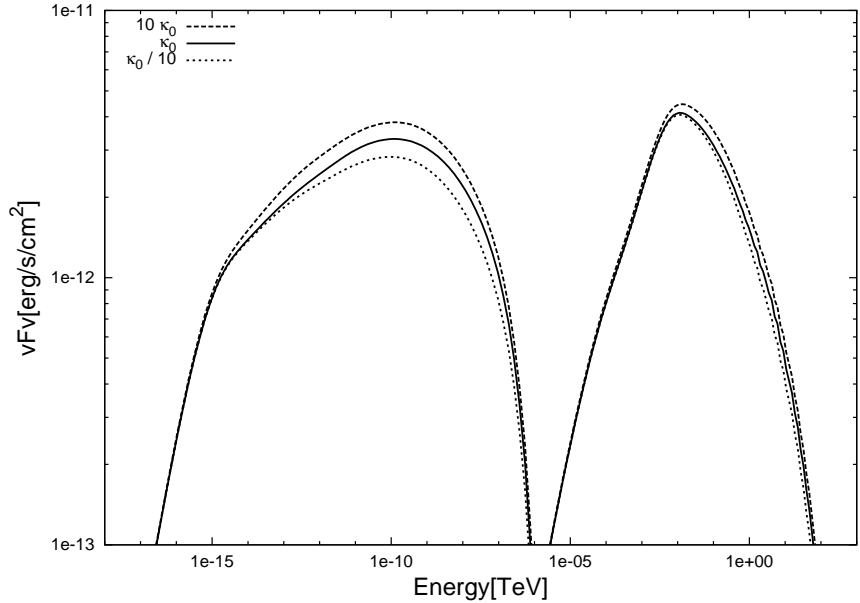}
\caption{\label{fig:SED_Change_kap1}SED for PWN~G0.9+0.1 with a change in the normalisation of the diffusion.}
\end{minipage} 
\end{figure}

First we changed the normalisation constant of the diffusion coefficient by considering $10\kappa_0$ and $\kappa_0/10$. The result of this change can be seen in Figures \ref{fig:Par_Change_kap1} and \ref{fig:SED_Change_kap1}. From Figure~\ref{fig:Par_Change_kap1} we can see that when the normalisation constant of the diffusion coefficient is increased the particle spectrum increases at high energies and stays unchanged at low energies. 

Changes to the normalisation constant of the diffusion coefficient should not change the particle spectrum or the SED, as the energy losses are the same throughout the PWN due to the magnetic field being constant for all zones in this part of the study. Changes to the diffusion coefficient will cause the particles to move to the outer zones faster but not change the shape of the spectrum. The only change is that the dynamical time step in the code is dependent on the diffusion coefficient and thus the effects we see here are numerical effects due to shorter or longer time steps in the code.

%When the normalisation is lowered the particle spectrum also lowers at high energies and stays unchanged for lower energies. The unchanged spectrum at lower energies is due to the fact that at lower energies, the transport of the particles is dominated by the convection (bulk particle motion) and adiabatic energy losses, and not by the diffusion. Therefore changes to the diffusion coefficient do not affect the particle spectrum at lower energies. The at the higher energies, the diffusion starts to play a role in the transport. Here, if the normalisation of the diffusion is increased, the particles will diffuse faster through the PWN and thus have less time to radiate. Therefore, more particles will retain a higher energy. These particles may have escaped the PWN, but we set our outer boundary $r_{\rm{max}}\gg R_{\rm{PWN}}$ as mentioned earlier, therefore the particles never reach the outer boundary. This results in the particle spectrum increasing with increasing $\kappa_0$. This increase in the particle spectrum is seen in the next couple of examples as well and is a somewhat unexpected result. This result would be opposite if the particles effectively escaped from the PWN. This is something that will need to be refined in future studies and possible solutions will be discussed later. The PWN size also increases for a large $\kappa_0$ (see Section~\ref{sec:morpho}). The SED due to the changed normalisation reflects the particle spectrum, where an increased particle spectrum results in the SED also increasing and vice versa.

The same holds for changes to the energy dependence of the diffusion coefficient as it will only change how fast particles of certain energies diffuse, and thus should not change the shape of the spectrum, but will change the time step as mentioned.

%Next the energy dependence of $\kappa$ was changed. Particles with higher energy will diffuse slower when $q$ has a lower value. This results in the particle spectrum becoming softer for higher energy particles as seen in Figure~\ref{fig:Par_Change_kap2}. This is a similar effect to when the normalisation of the diffusion is lowered, which also resulted in the particle spectrum decreasing. Here, as in the previous paragraph, the effect is only visible at the higher energies due to convection dominating the transport at lower energies. Changes in $q$ is an energy-dependent change and thus the effect will be larger for larger particle energies. The radiation caused by the reduced number of high-energy particles will also be reduced. Figure~\ref{fig:SED_Change_kap2} shows this as the radiation the $X$-ray band and the TeV band decrease as $q$ decreases. Again the opposite is expected for effective particle escape.

\begin{figure}[h!]
\centering
\begin{minipage}[b]{5in}
\centering
\includegraphics[width=4.5in]{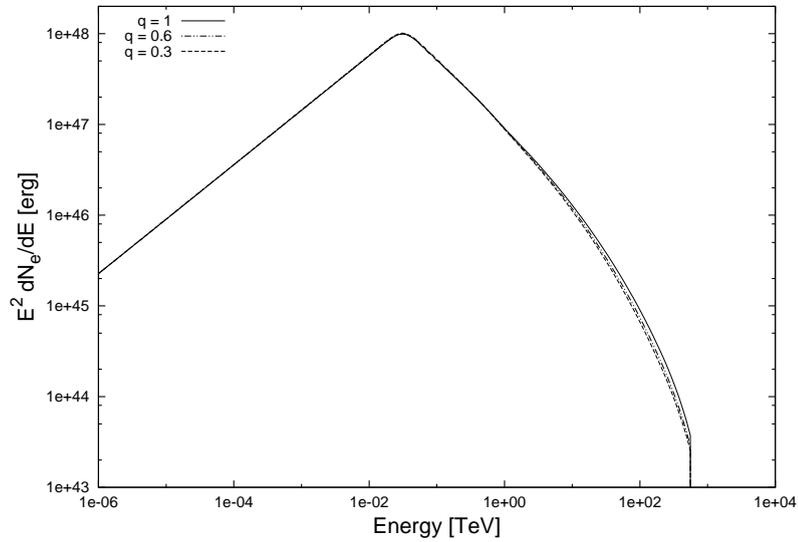}
\caption{\label{fig:Par_Change_kap2}Particle spectrum for PWN~G0.9+0.1 with a change in the energy dependence of the diffusion coefficient.}
\end{minipage} 
\end{figure}

\begin{figure}[h!]
\centering
\begin{minipage}[b]{5in}
\centering
\includegraphics[width=4.5in]{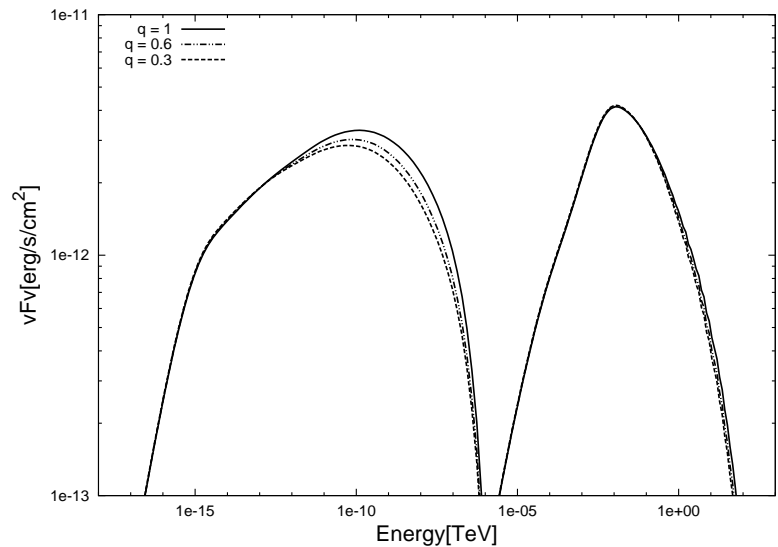}
\caption{\label{fig:SED_Change_kap2}SED for PWN~G0.9+0.1 with a change in the energy dependence of the diffusion coefficient.}
\end{minipage} 
\end{figure}

\subsection{Soft-photon components}
\begin{figure}[h]
\centering
\begin{minipage}[b]{5in}
\centering
\includegraphics[width=4in]{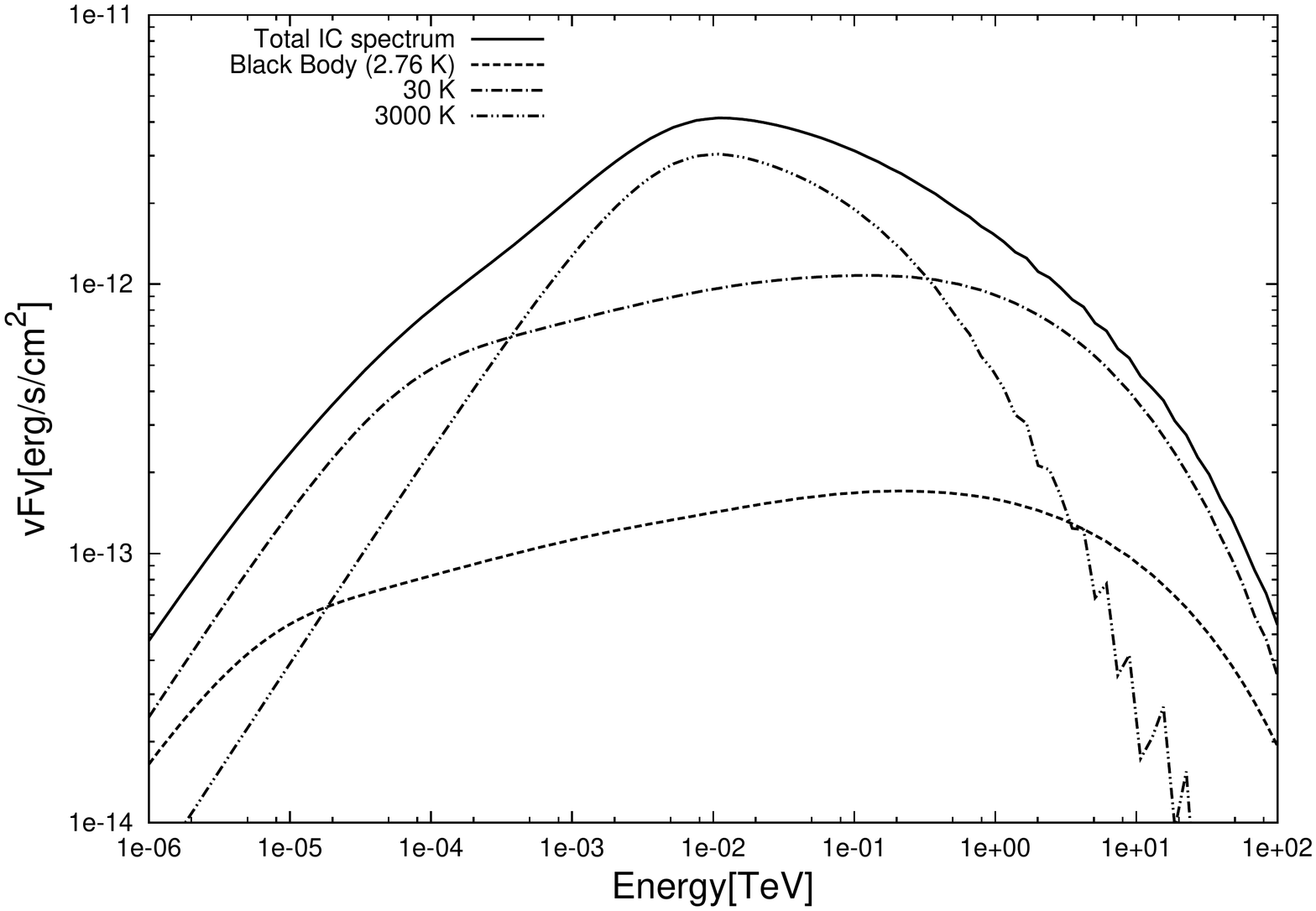}
\caption[IC spectrum for PWN~G0.9+0.1 showing the contribution of different soft-photon components]{\label{fig:SED_soft}IC spectrum for PWN~G0.9+0.1 showing the contribution of different soft-photon components in Table~\ref{tbl:G0.9_Torres}. The solid line is the total radiation, dashed line is the 2.76K CMB component, dashed-dotted line is the 30 K component, and the dashed-dot-dotted line shows the 3 000 K component.}
\end{minipage} 
\end{figure}

Table~\ref{tbl:G0.9_Torres} shows the three different soft-photon components used to model the IC scattering from the PWN. These components can be turned on and off at will and Figure~\ref{fig:SED_soft} shows the contribution of each of these components. 

In Figure~\ref{fig:SED_soft} we can see what contribution each of the three soft-photon target fields makes to the IC radiation received from the PWN. The CMB target field produces a flat spectrum which causes the first small bump on the left hand side of the total IC flux. The starlight at 3~000~K, with an energy density of 25 eV/cm$^3$, has the highest peak and plays the largest role in the overall IC flux. The jaggedness of the IC component due to starlight at high energies is a numerical discretisation effect.

\begin{figure}[h!]
\centering
\begin{minipage}[b]{5in}
\centering
\includegraphics[width=5in]{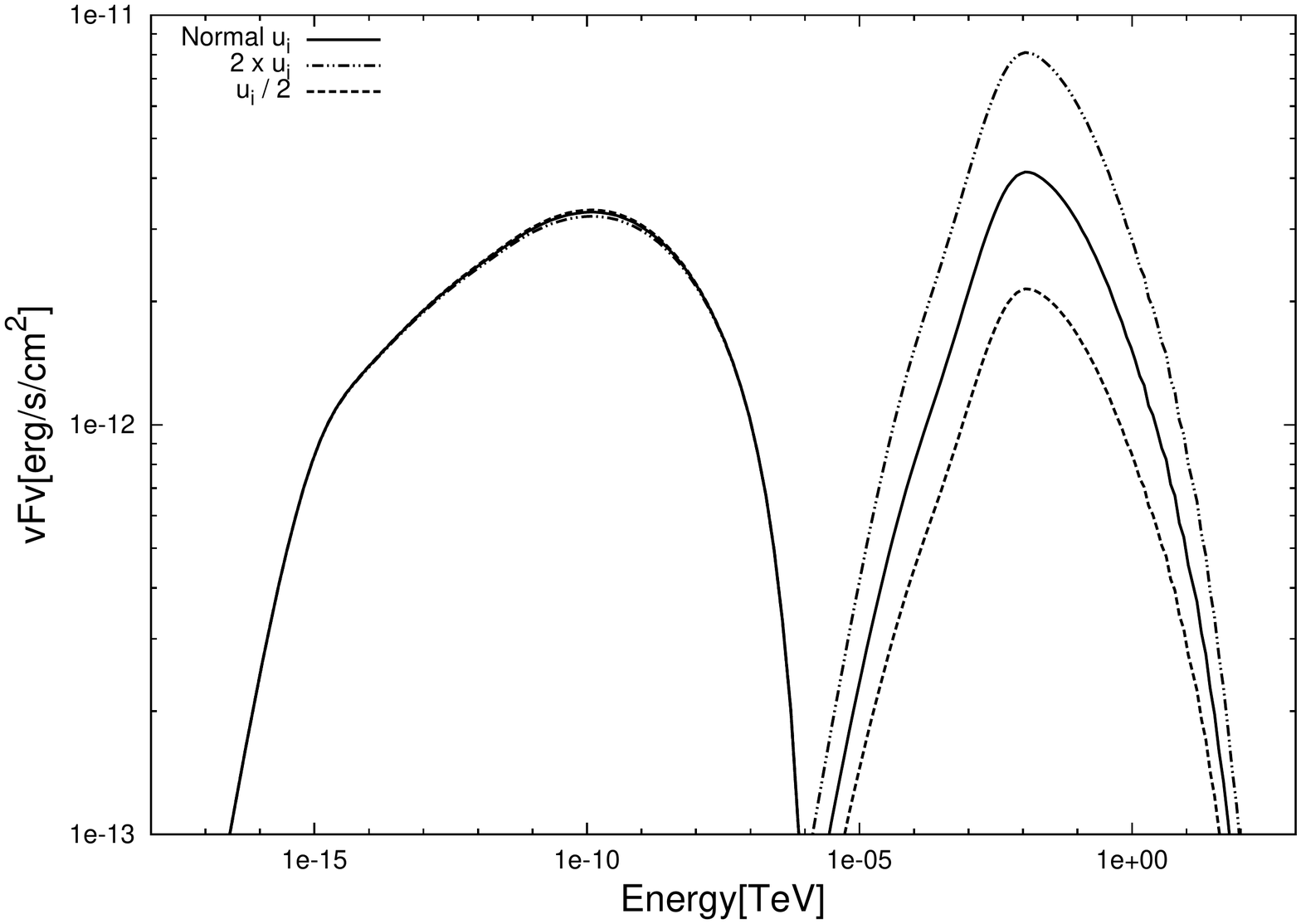}
\caption{\label{fig:Par_Change_u}SED for PWN~G0.9+0.1 with a change in the energy densities of the soft-photon components.}
\end{minipage} 
\end{figure}

\begin{figure}[h!]
\centering
\begin{minipage}[b]{5in}
\centering
\includegraphics[width=5in]{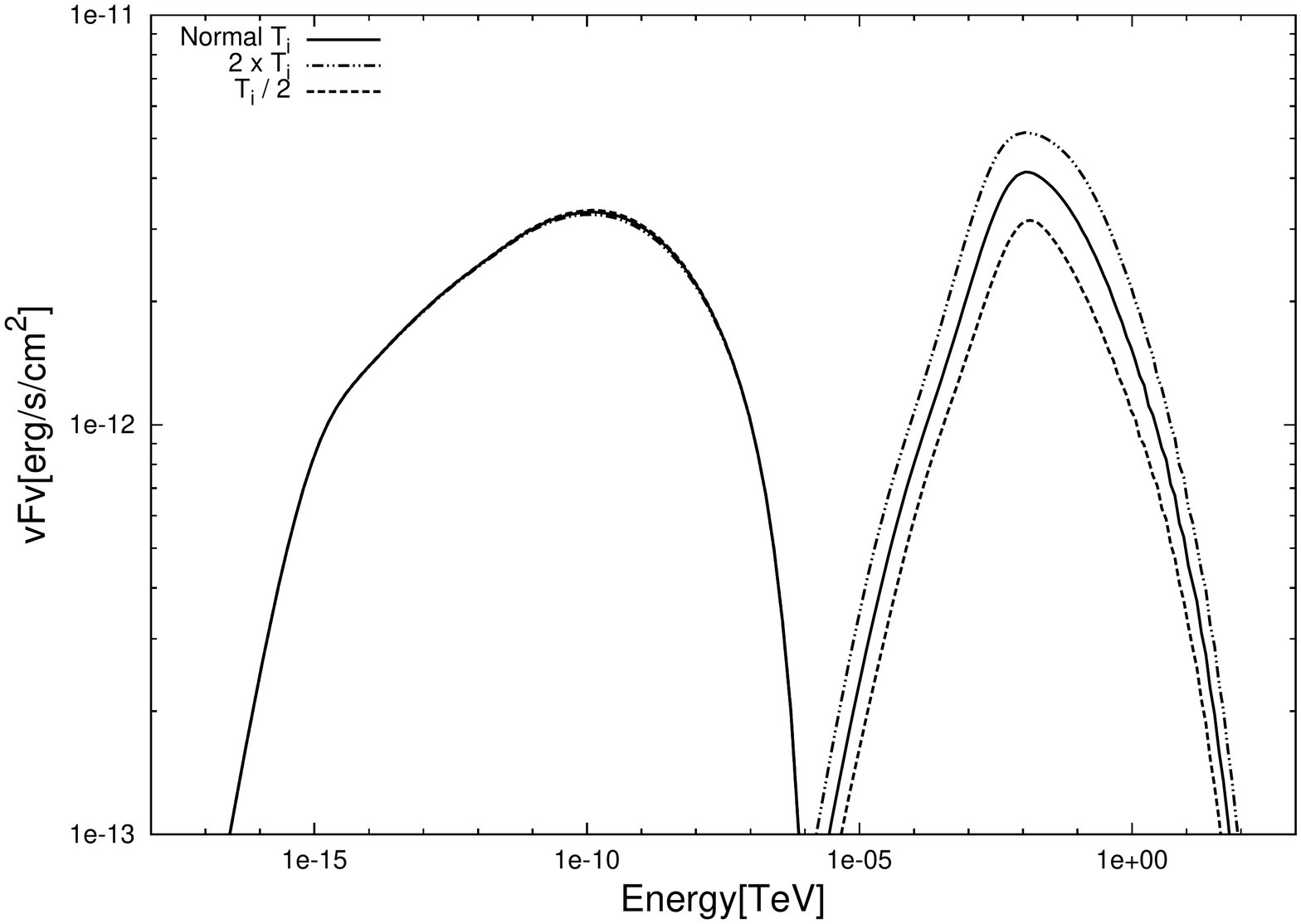}
\caption{\label{fig:SED_Change_T}SED for PWN~G0.9+0.1 with a change in the temperature of the soft-photon components.}
\end{minipage} 
\end{figure}

In Figures \ref{fig:Par_Change_u} and \ref{fig:SED_Change_T} the effect of changes in the energy density and the temperature of the soft-photon components is shown. As seen in Eq.~\eqref{eq:IC_pord_rate}, the IC spectrum is proportional to the soft-photon number density $n(\epsilon)$, which may be written in terms of the total energy density $u_0$ and temperature $T$, so that $n(\epsilon) \propto u_0/T^4$. Thus if the energy density is increased or decreased, the IC radiation will also increase or decrease linearly. This is seen in Figure~\ref{fig:Par_Change_u}. However, when the temperature is increased or decreased for a constant $u_0$, the effects are in the same direction, but smaller as seen from Figure~\ref{fig:SED_Change_T}. This is due to the fact that when the temperature is increased, fewer photons are needed to reach the same energy density $u_0$, leading to a lower normalisation for the cumulative blackbody spectrum. A change in temperature will also have a lesser effect via the blackbody spectral form $\propto (e^{h\nu / kT}-1)^{-1}$.

\subsection{Other parameters}
In the previous sections, the effects of varying some of the most important free parameters were shown. These are, however, not the only free parameters. The braking index $n$ in Eq.~\eqref{eq:spinDown123} is also a free parameter, but as mentioned earlier this is usually set to 3 for dipole rotators. If the braking index is increased, the number of particles injected into the PWN also increases due to the reduced spin-down of the pulsar. Therefore, more particles are injected for longer periods into the PWN. Due to this the particle and radiation spectrum will increase with an increased $n$. The conversion efficiency of spin-down luminosity to particle power, $\epsilon$ in Eq.~\eqref{normQ}, is also a free parameter. If the conversion efficiency is reduced then less of the energy from the pulsar will be converted into particle power and thus a lower particle and radiation spectra will be observed. The last free parameters that will be discussed are the indexes of the injection spectrum. These free parameters, $\alpha_1$ and $\alpha_2$, will influence the slopes and the normalisation of the particle and radiation spectrum. The distance $d$ to the PWN is also a free parameter. The flux from the PWN at Earth scales as $1/d^2$ and the sizes of the spatial bins also linearly dependent on $d$ (influencing the diffusion and convection timescales for each zone) but the latter is a small effect.

\section{Spatially-dependent results from PWN model}
\label{sec:Space}
In the previous sections I showed the total particle spectrum and SED predicted by the code for different parameter choices. This, however, was not the main aim of the code that we have developed, as we are especially interested in the spatial dependence of the radiation from the PWN. Our model is spatially dependent and therefore it is possible to show results regarding the morphology of the PWN. This Section~is therefore dedicated to show the changes in the PWN's morphology when certain parameters are changed. 

\subsection{Effects of changes in the diffusion coefficient and bulk particle motion on the PWN's morphology}\label{sec:morpho}
I calculated the surface brightness for a particular LOS by dividing the emitted photon spectrum from each zone by its subtended solid angle and multiplying this by $4\pi d^2$. I next multiplied this quantity by $E_{\rm{e}}$ and integrated over some energy band. This allowed me to find the size of the PWN as shown in the plots of normalised surface brightness for different energy bands versus radius from the centre of the PWN. These graphs are designed to study the change in the PWN size and therefore they have all been normalised to one. In all the graphs there is a horizontal dotted line indicating where the surface brightness has reduced by two-thirds, yielding the typical PWN size in that energy range.

\begin{figure}[h]
\centering
\begin{minipage}[b]{6in}
\centering
\includegraphics[width=6in]{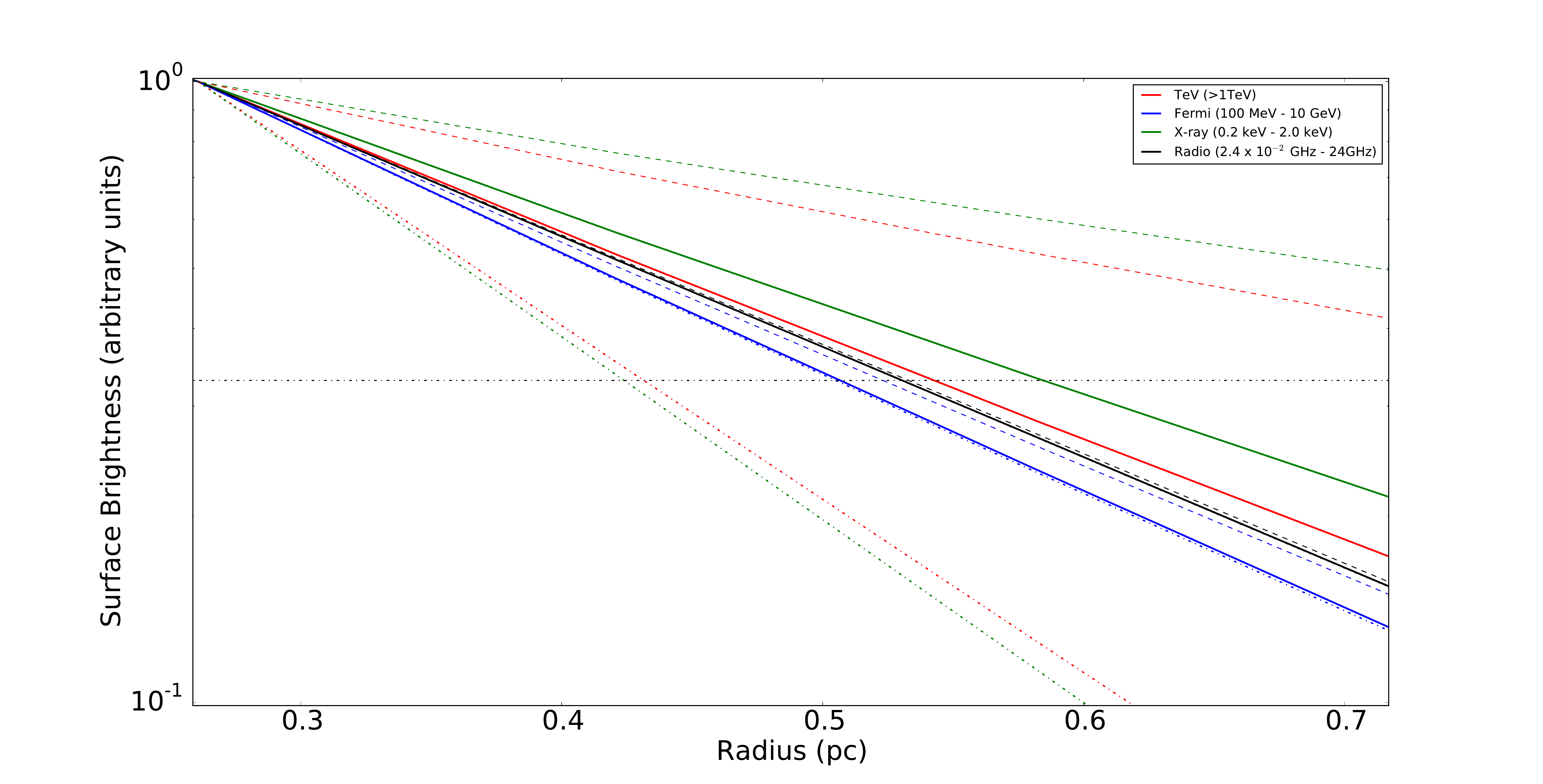}
\caption[Morphology of the PWN for a change in the normalisation of the diffusion coefficient.]{\label{fig:sp_kappa}Morphology of the PWN for a change in the normalisation of the diffusion coefficient. The solid lines indicate $\kappa_0$, the dashed lines indicate 5$\kappa_0$, and the dashed-dotted lines indicate $\kappa_0 /$5.}
\end{minipage} 
\end{figure}

In Figure~\ref{fig:sp_kappa} the value of $\kappa_0$, in Eq.~\eqref{eq:kappa123}, was changed. The effect is shown for four different energy bands, namely, TeV ($>$1 TeV) in red, \textit{Fermi}$-$LAT (100 MeV $-$ 10 GeV) in blue, X-ray (0.2 keV $-$ 2.0 keV) in green, and radio (2.4 $\times 10^{-2}$ GHz $-$ 24 GHz) in black. The solid lines represents Bohm diffusion ($\kappa_0$), the dashed lines are for $5\kappa_0$, and the dashed-dotted lines are for $\kappa_0/5$. From Figure~\ref{fig:sp_kappa} we can see that the size of the PWN increases as the normalisation constant of the diffusion coefficient increases. This is due to the fact that if the diffusion coefficient is larger then the particles move to the outer zones faster, resulting in a larger PWN for certain energy bands. The inverse is also true: the size is smaller if the normalisation constant is reduced.

\begin{figure}[h]
\centering
\begin{minipage}[b]{6in}
\centering
\includegraphics[width=6in]{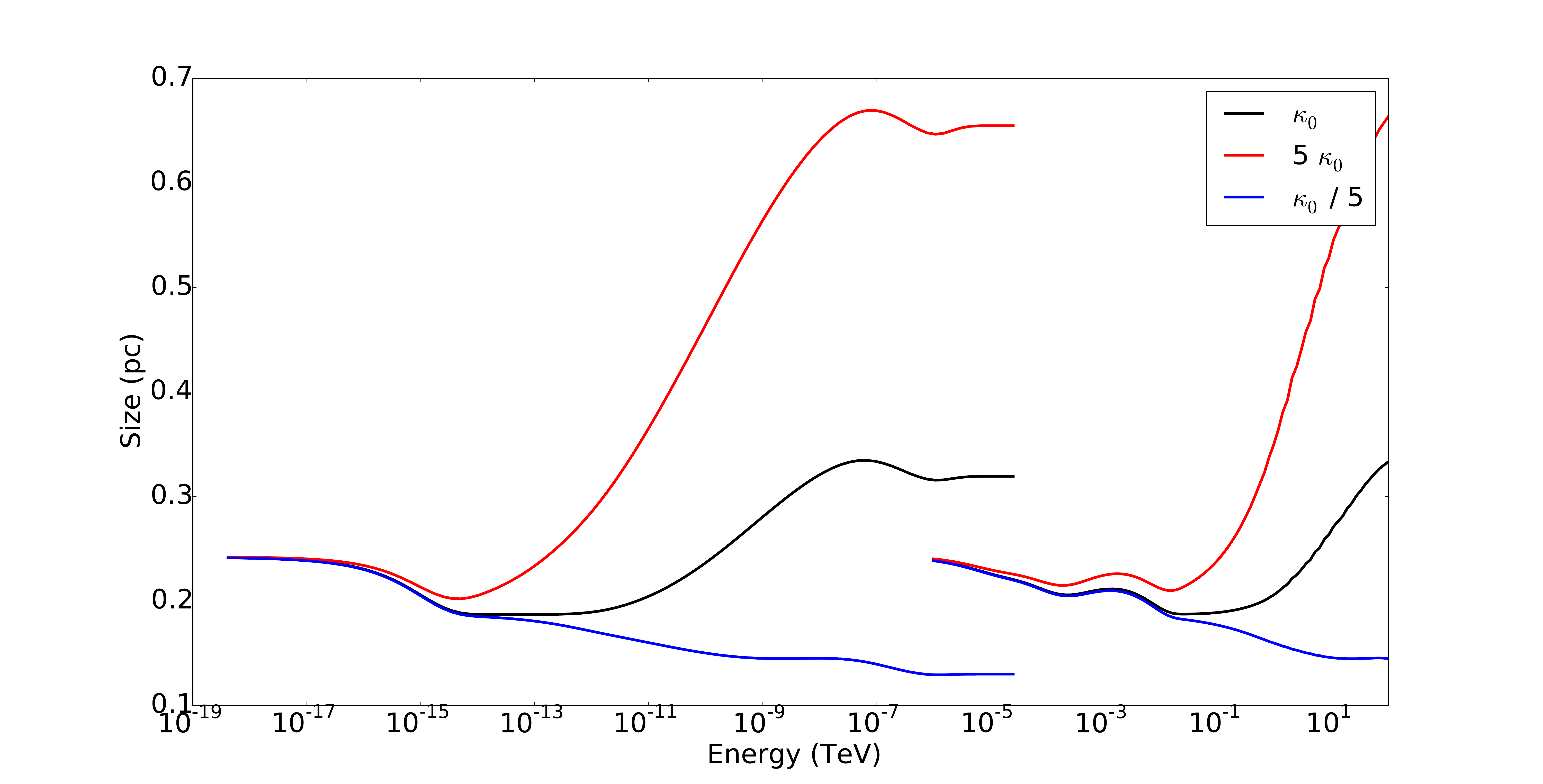}
\caption{\label{fig:SvE_k}Size of the PWN as a function of energy when the normalisation constant of the diffusion coefficient is changed (change in $\kappa_0$).}
\end{minipage} 
\end{figure}
The next result also possible with our code is to see how the size of the PWN changes with energy. Figure~\ref{fig:SvE_k} shows this result for the 3 different scenarios as mentioned with the left showing SR and the right IC. For the first two scenarios, $\kappa_0$ and 5$\kappa_0$, the size of the PWN increases with increased energy. As mentioned in Section~\ref{sec:Var_diff123}, the outer boundary of our model is set much larger than $R_{\rm{PWN}}$, which has the effect that particles do not escape. This effect can be seen here for the first two scenarios, where diffusion plays the largest role in particle transport and causes the high-energy particles to diffuse outward faster than low-energy particles, filling up the outer zones and resulting in a larger size for the PWN at high energies. This effect is larger for high-energy particles due to the energy dependence of the diffusion coefficient. When the third scenario, $\kappa_0/$5, is considered, we see that the opposite happens: the size of the PWN reduces with increasing energy. Here the diffusion coefficient is so small that the energy loss rate due to SR dominates over the diffusion. The particles therefore ``burn off" or expend their energy before they can reach the outer zones (cooling therefore dominates).

\begin{figure}[t]
\centering
\begin{minipage}[b]{6in}
\centering
\includegraphics[width=6in]{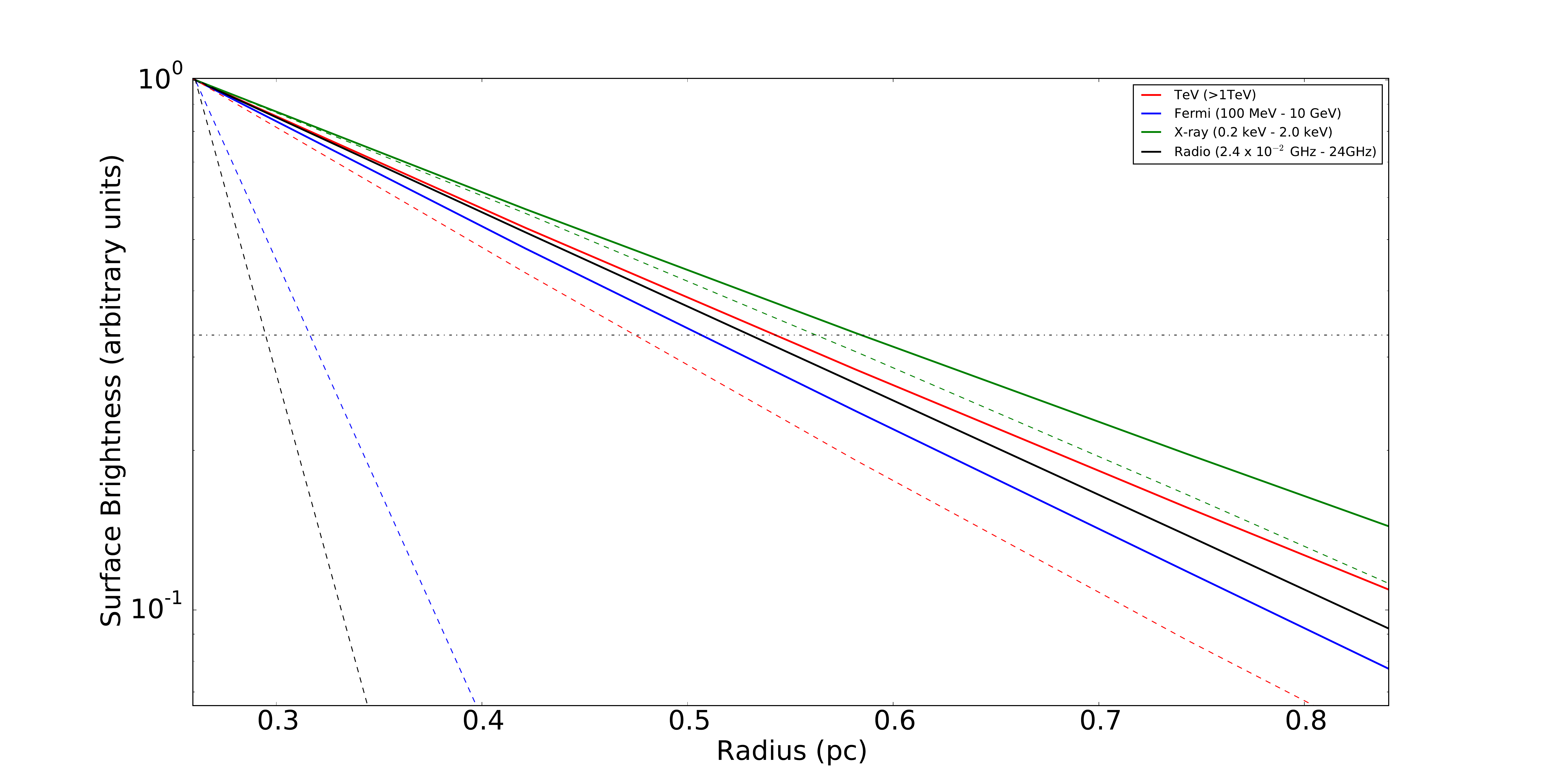}
\caption[Morphology of the PWN for a change in the normalisation of the bulk particle motion.]{\label{fig:sp_V1}Morphology of the PWN for a change in the normalisation of the bulk particle motion. The solid lines indicate $V_0 = 5 \times 10^{-5}$ pc/yr and the dashed lines indicate $V_0 = 0$.}
\end{minipage} 
\end{figure}

Next I studied the effect of varying the bulk motion. In Figure~\ref{fig:sp_V1} the PWN sizes in the different energy bands are shown for two different scenarios. These are for standard particle flow through the PWN as in Eq.~\eqref{V_profile123} (solid lines), and for $V_0 = 0$ which represents no bulk motion at all (dashed lines). It is clear that the size of the PWN decreases for all the energy bands when there is no bulk motion present.

\begin{figure}[b]
\centering
\begin{minipage}[b]{6in}
\centering
\includegraphics[width=5.5in]{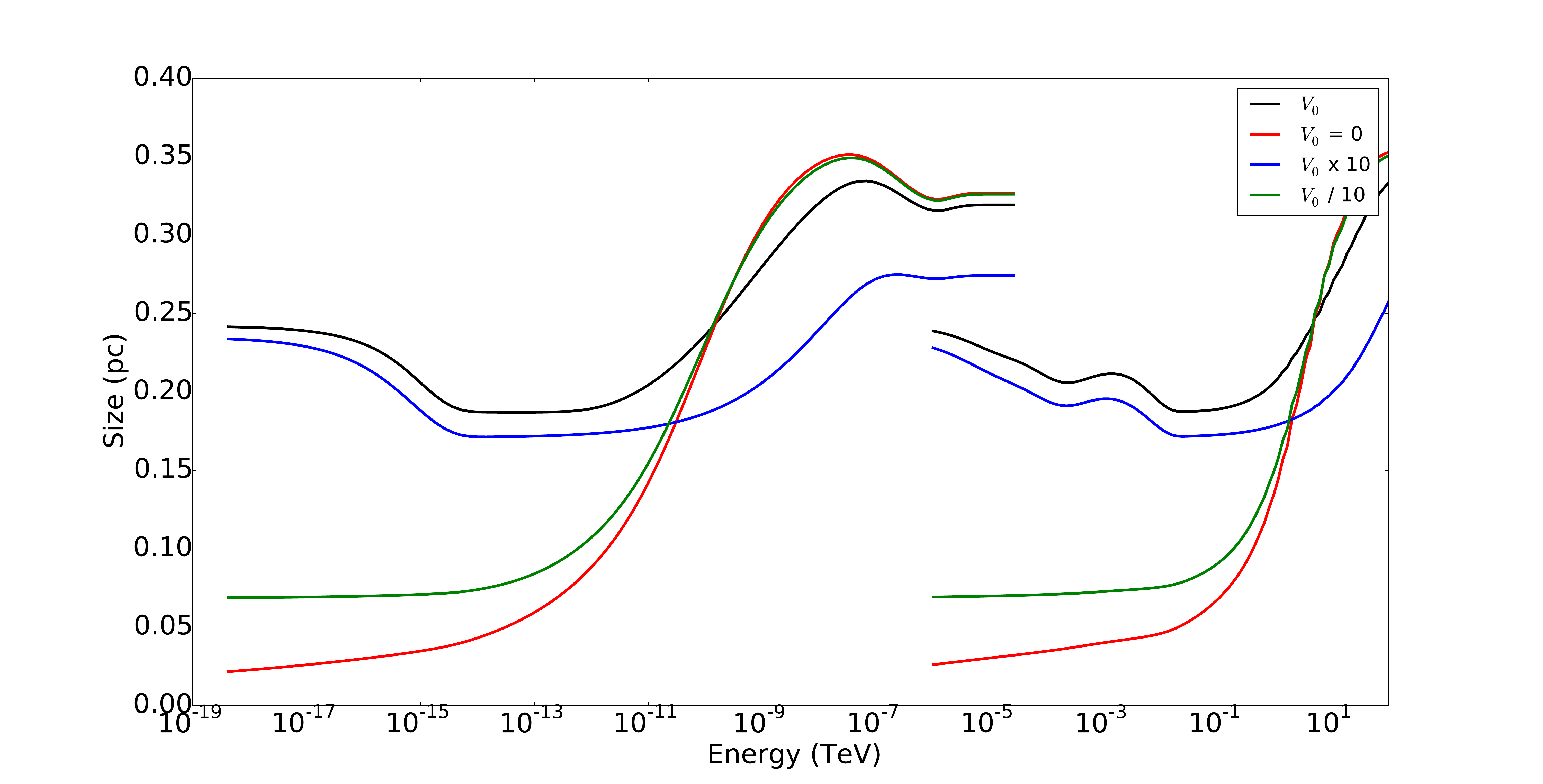}
\caption{\label{fig:SvE_V}Size of the PWN as a function of energy for different normalisations of the bulk particle motion.}
\end{minipage} 
\end{figure}
We now study the size of the PWN as a function of energy for different normalisations of the bulk particle motion. If we first consider the scenario where $V_0 = 0$, we can see that for lower energies the PWN has a smaller size than for higher energies. This is due to the energy dependence of the diffusion coefficient in the PWN. At lower energies the bulk motion dominates and therefore the PWN is smaller due to the slow speed of the particles, and at higher energies the diffusion dominates, causing the particles to move faster towards the outer zones and increasing the size of the PWN. At the highest energies the SR energy losses dominate over all the other effects and cause the particles in the outer zones to lose more energy, i.e., SR cooling reduces the PWN size at the highest energies. The size of the PWN increases monotonically with $V_0$ at low energies, except for $10V_0$. Here the size decreases somewhat as the adiabatic energy losses now dominate the convection, causing the particles to lose energy more rapidly, thereby reducing the size of the PWN.

\subsection{Different cases of $\alpha_V$ and $\alpha_B$: first results}\label{sec:alph_Valph_B}
In Sections \ref{sec:B_field_par} and \ref{sec:V_field_par} I discussed the effects that changes in the normalisation of the magnetic field and the bulk particle speed had on the lepton and radiation spectrum. In Eq.~\eqref{B_Field123} and Eq.~\eqref{V_profile123}, however, we see that the magnetic field may have a spatial and time dependence and the bulk motion only has a spatial dependence. In this section the effects of different spatial dependencies for $B(r,t)$ and $V(r)$ are shown. We note that we have assumed the diffusion coefficient to be spatially independent throughout this work. However, since we are now considering the spatial dependence of the magnetic field in this paragraph, and $\kappa \propto 1/B(r,t)$, this assumption is technically violated here. The effect is small when the divergence of $\vec{\kappa}$ is small, which we will assume to be the case in this section. This spatial dependence of the diffusion coefficient can be implemented in future by adding another convective term to the transport equation. 

In Eq.~\eqref{eq:a_v+a_b=-1} I showed that the following relationship holds: $\alpha_V + \alpha_B = -1$. For this section the time dependence of the magnetic field is kept unchanged, with $\beta_B =-1.3$. Four different situations are shown. The first is $\alpha_B = 0$ and $\alpha_V = 1$, which is consistent with \cite{Torres2014}. Next three extreme cases are shown that comply with the relationship in Eq.~\eqref{eq:a_v+a_b=-1} as shown in figure legends that follow with the magnetic field kept constant in the first zone.

\begin{figure}[b]
\centering
\begin{minipage}[b]{5in}
\centering
\includegraphics[width=5in]{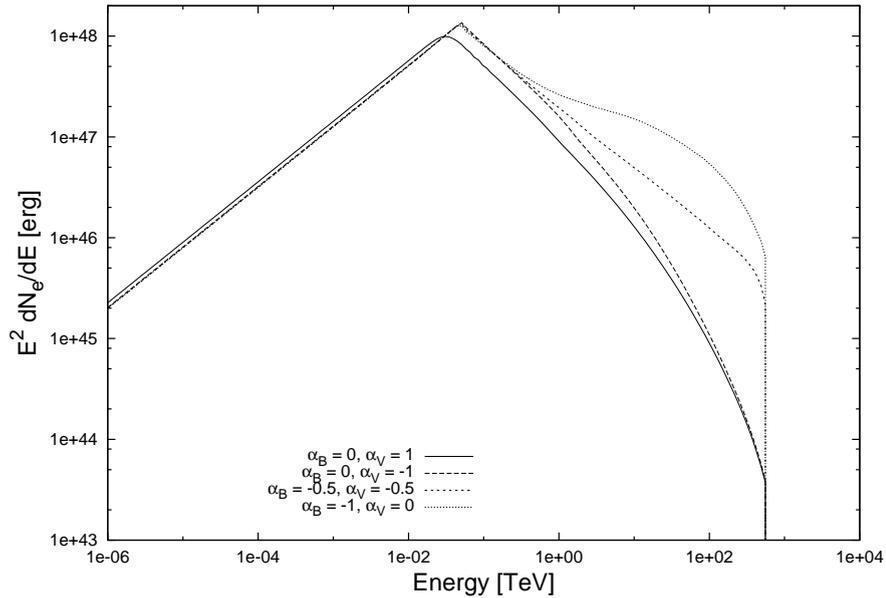}
\caption{\label{fig:Par_Change_alp}Particle spectrum for PWN~G0.9+0.1 with a change in the parametrised magnetic field and bulk particle motion.}
\end{minipage} 
\end{figure}

\begin{figure}[t]
\centering
\begin{minipage}[b]{5in}
\centering
\includegraphics[width=5in]{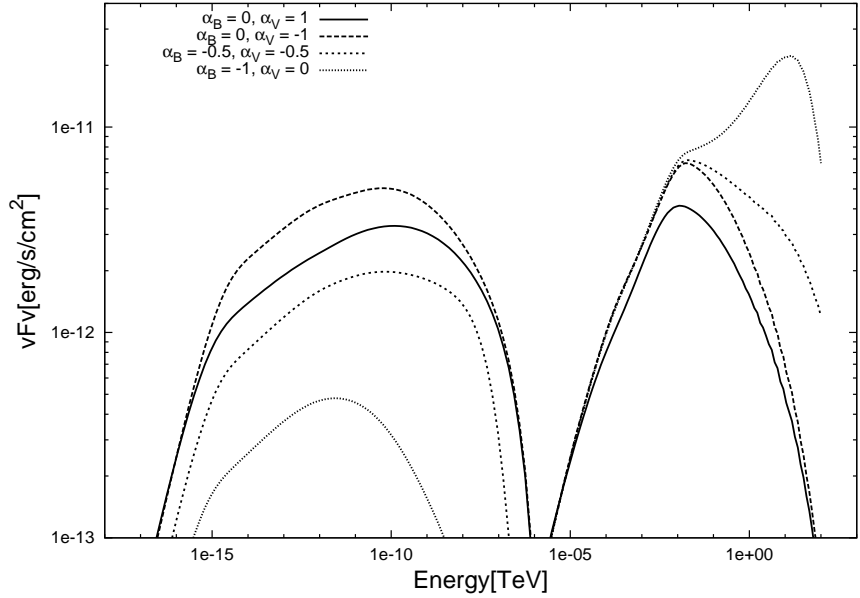}
\caption{\label{fig:SED_Change_alp}SED for PWN~G0.9+0.1 with a change in the parametrised magnetic field and bulk particle motion.}
\end{minipage} 
\end{figure}

In Figure~\ref{fig:Par_Change_alp} the particle spectrum is shown for four different situations, with the solid line showing the result for $\alpha_B = 0$ and $\alpha_V = 1$ as is assumed by \cite{Torres2014}. In this case the magnetic field is constant for the entire PWN, but the bulk speed increases with $r$. The particles move extremely fast as they propagate farther from the centre of the PWN. They therefore lose more energy due to adiabatic energy losses relative to the other cases. Thus the solid line is lower than the other situations and the peak of the spectrum is also shifted to the left.

Next I considered the following three situations: $\alpha_B = 0$ and $\alpha_V = -1$, $\alpha_B = -0.5$ and $\alpha_V = -0.5$, and $\alpha_B = -1$ and $\alpha_V = 0$. These three situations all comply with Eq.~\eqref{eq:a_v+a_b=-1} and we can see from both Figures~\ref{fig:Par_Change_alp} and \ref{fig:SED_Change_alp} that changes to the magnetic field have a more profound impact on the particle spectrum and SED than changes in the speed. If the spatial dependence of the magnetic field changes from 0 to -0.5 and -1, the magnetic field is first constant over all space and then decreases as $r^{-0.5}$ and finally it reduces rapidly as $r^{-1}$. The effect of this can be seen in the particle spectrum as the number of high-energy particles increases for a decreased magnetic field as mentioned in Section~\ref{sec:B_field_par}. This effect is emphasised in the situation where $\alpha_B = -1$ (where the magnetic field reduces as $r^{-1}$) resulting in a very small magnetic field at the outer edges of the PWN. This can also be seen in the radiation spectrum in Figure~\ref{fig:SED_Change_alp} where a decreased magnetic field results in reduced radiation in the SR band, as discussed previously, and the increased radiation in the IC band is due to more particles being present at those energies. This increase in the high-energy particles is quite large, though (possibly indicating a violation of our assumption that the divergence of $\vec{\kappa}$ is small in this case). We note that our model currently does not take into account the fact that the cutoff energy due to particle escape ($E_{\rm{max}}$) should also be a function of the magnetic field. This is because in reality $\sigma \propto B^2$ (we have assumed $\sigma$ to be constant), and therefore $E_{\rm{max}} \sim \sqrt{B^2/(1+B^2)}$, which will have the effect that if the magnetic field is reduced, $\sigma$ and therefore $E_{\rm{max}}$ will decrease. This may cause the high-energy particles to be cut off at lower and lower energies as the magnetic field decreases due to more efficient particle escape, and therefore the build up of high-energy particles may be partially removed (we say `partially' since the Larmor radius of the most energetic particles in the outer zones is still smaller than the PWN size by a factor of a few, inhibiting efficient escape of particles from the PWN). The question of particle escape may also be addressed by refining our outer boundary condition. This is something that will be addressed in the future.

Above I showed the effects on the particle spectrum and also the SED for four different scenarios of the free parameters parametrising the spatial dependence of the magnetic field in Eq.~\eqref{B_Field123} and the bulk particle motion in Eq.~\eqref{V_profile123}. Now I show the size of the PWN as a function of energy for the same four scenarios. 

\begin{figure}[b]
\centering
\begin{minipage}[b]{6in}
\centering
\includegraphics[width=6in]{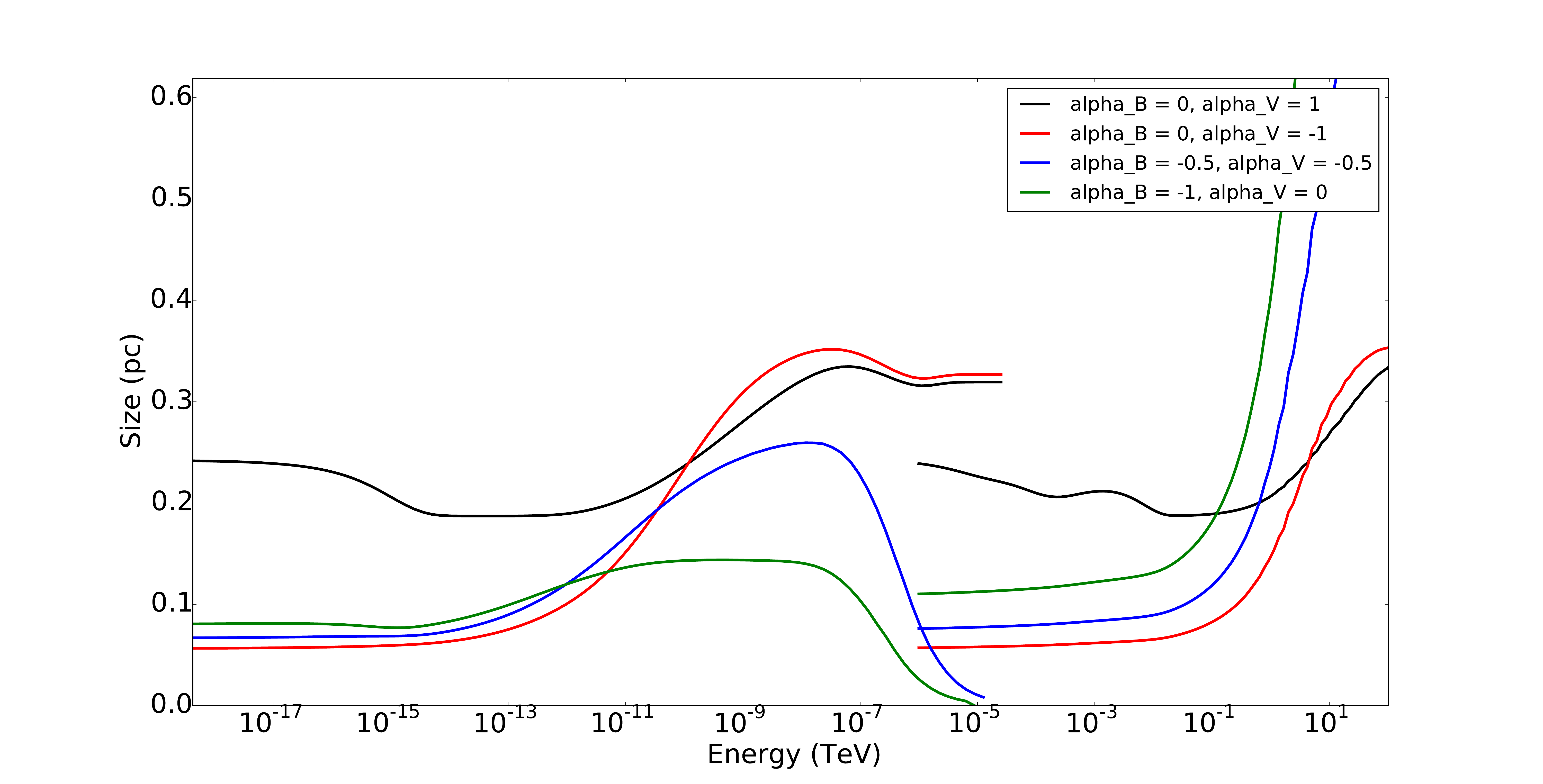}
\caption{\label{fig:SvE_alpB}Size of the PWN as a function of energy for changes in $\alpha_B$ and $\alpha_V$.}
\end{minipage} 
\end{figure}

From Figure~\ref{fig:SvE_alpB} we can see that in scenario one (black line, $\alpha_B = 0$ and $\alpha_V = 1$) the PWN size for low energies is always larger than for all the other scenarios. This is due to the speed being directly proportional to $r$ in this case, resulting in the particles moving faster as they move further out from the centre of the PWN. This will result in the outer zones filling up with particles, while not escaping. This may point to the fact that our outer boundary was chosen to be much larger than the radius of the PWN ($r_{\rm{max}} \gg R_{\rm{PWN}}$). For scenario two (red line, $\alpha_B = 0$ and $\alpha_V = -1$), the size of the PWN  at low energies follows the same pattern as for the high-energy photons, since the energy-dependent diffusion now dominates convection. At lower energies, we see that PWN is smaller than in scenario one, as the speed is now proportional to $r^{-1}$, which results in a slower bulk motion and thus fewer low-energy particles move to the outer zones.

In scenario three, blue line ($\alpha_B = -0.5$ and $\alpha_V = -0.5$), and four, green line ($\alpha_B = -1$ and $\alpha_V = 0$) the magnetic field has a spatial dependence. This causes the magnetic field to reduce as one moves farther away from the centre of the PWN. This reduced magnetic field will lead to increased diffusion as mentioned in the first part of this section. For these two scenarios the dependence of the bulk motion on radius is weaker and therefore diffusion dominates the particle transport. Once again we can see the energy dependence of the diffusion, since the PWN is initially smaller and then increases as we go to higher energies. As can be seen at very high energies, the PWN size becomes very large, which is not the case for the SR component. The first is due to the pile up of high-energy particles (leading to substantially increased IC emission, Figure~\ref{fig:SED_Change_alp}), while the second is due to the fact that SR is severely inhibited for the very low magnetic field. We lastly note that a larger bulk speed leads to a relatively larger PWN size at low energies where convection dominates (see especially the black lines).

\chapter{Summary, conclusion and future work} % Write in your own chapter title
\label{ch:concl}
\lhead{Chapter 4. \emph{Summary, conclusion and future work}} % Write in your own chapter title to set the page header
This study focused on modelling the evolution of PWNe, with the main aim being to create a spatially-dependent temporal code to model the morphology of PWNe. While we performed a multi-wavelength study, our focus was on the high-energy gamma-ray band, as I am part of the H.E.S.S.\ Collaboration and also SA-GAMMA, which focuses on high-energy astrophysics. We solved a Fokker-Planck-type transport equation to model the particle evolution inside a PWN, injecting a broken power-law particle spectrum, and allowing the particle spectrum to evolve over time, taking into account energy losses due to SR, IC scattering, and adiabatic cooling of the PWN due to expansion. The transport of particles also took into account particle diffusion and convection in the form of a bulk particle motion. The main results found from the code will be summarised bellow.

\section{The spatial-temporal-energetic PWN model}\label{sec:sum_Model}
I made the simplified assumption that the geometrical structure of a PWN may be modelled as a sphere into which particles are injected by a pulsar and allowed to diffuse, convect, and undergo energy losses in the ambient nebula. I assumed spherical symmetry so that the only changes in the particle spectrum would be in the radial direction. The model was set up with the pulsar in the centre, and dividing the surrounding PWN into concentric zones (shells). The boundary of the innermost zone was assumed to coincide with the termination shock of the pulsar wind where the particles were injected into the PWN.

The transport of the particles was modelled using a Fokker-Planck-type transport equation similar to the Parker equation \citep{Parker1965} as discussed in Chapter \ref{ch:Model}. The transport equation was rewritten in a form that was more suitable for our type of study: (this transformation from momentum to energy space was discussed in Appendix~\ref{app:transport})
\begin{equation}
\frac{\partial U_E}{\partial t} = -\mathbf{V} \cdot (\nabla U_E) +  \kappa \nabla^2 U_E + \frac{1}{3}(\nabla \cdot \mathbf{V})\left( \left[\frac{\partial U_E}{\partial \ln E} \right] - 2U_E \right) +  \frac{\partial }{\partial E}(\dot{E}_{\rm{rad}}U_E) +  Q'(\mathbf{r},E,t).
\label{eq:sum_transportFIN}
\end{equation}
This equation consists of an injection of particles $Q(\mathbf{r},E,t)$ or source term, energy losses due to radiation $\frac{\partial }{\partial E}(\dot{E}_{\rm{rad}}U_E)$, adiabatic energy losses due to cooling $\frac{1}{3}(\nabla \cdot \mathbf{V})\left( \left[\frac{\partial U_E}{\partial \ln E} \right] - 2U_E \right)$, spatially-independent diffusion $\kappa \nabla^2 U_E$, and convection of particles $-\mathbf{V} \cdot (\nabla U_E)$. The radiative energy losses were due to two processes, SR and IC scattering and the diffusion of the particles was considered to be Bohm-like. The solution of this detailed equation represented a major step forward when compared to previous studies, since other authors frequently solved a linearised and sometimes more basic version of this equation (neglecting spatial dependence of $N{\rm{e}}$ or some of these terms) using average timescales to characterise each of these processes.

I first considered solving the Fokker-Planck-type transport equation using an Euler method, but it soon became clear that this method was not stable for this type of differential equation. After considering different numerical models we decided to use a DuFort-Frankel numerical scheme that is stable for parabolic differential equations as long as the timestep is short enough. The discretised form of the Fokker-Planck-type transport equation can be found in Eq.~\eqref{eq:fin_EQ}. The boundary conditions for our model were as follows: the PWN was initially assumed to be devoid of particles, with the spatial boundary conditions being assumed to be reflective at the inner boundary, as particles were injected at the termination shock and could not diffuse inward. This was done by assuming zero flow of particles at the inner boundary as given by Eq.~\eqref{eq:reflective_boundary}. At the outer boundary particles were allowed to escape the PWN by setting the particle spectrum to zero. We later discovered that the way we handled the outer boundary may need to be refined and I will discuss this in the next paragraphs. Injection of particles took place at the first zone. Importantly, we found that our predicted particle and radiation spectra converged when using a suitable number of radial and energy bins and small enough time step.

We could next calculate the SED from the PWN once the particle spectrum was known for each spatial zone. The SED consisted of two components, the SR and IC spectrum. We assumed three blackbody soft-photon background fields. These were the CMB, Galactic infrared photons, and optical starlight. Once the SED for each zone was known, we could integrate over space to find the total SED as viewed from Earth.

The main aim of the code we developed was to study the PWN morphology through calculating the surface brightness. We used the radiation at different distances from the centre of the PWN and performed a line-of-sight integration to project the total flux onto a flat surface on the plane of the sky. We could then find the radiation profile and estimate the size of the PWN as function of energy.

\section{Calibration and results}
In the previous section I summarised the development and implementation of our model. Once the model was finalised, I calibrated the code by comparing it with results from two independent codes, using PWN G0.9+0.1 as calibration source. First the model was compared to that of \cite{VdeJager2007}. They used a one-zone model and treated the particle transport in a very simplified way, incorporating only the SR energy loss timescale. Our new model reproduces the results from \cite{VdeJager2007} quite well, after our respective parametrisations of the magnetic field were conformed and we removed the effects of convection from our model. In addition, I decided to use a more recent model \citep{Torres2014} as a second calibration. In their paper they modelled several sources, including PWN G0.9+0.1. They view the PWN as a single sphere, similar to \cite{VdeJager2007}, but model the transport of the particles by considering the balance of energy losses, injection, and escape. The way they parametrise their magnetic field is different from our implementation, but for young PWNe, the time dependence of the magnetic field was modelled in the same way. Our model fitted the results from \cite{Torres2014} very well, and this was a second confirmation that our model is well calibrated. I modelled three other sources, G21.5-0.9, G54.1+0.3, and HESS J1356$-$645. For the first two sources, our model also calibrated well with \cite{Torres2014}, but for HESS J1356-645 we could not reproduce their results exactly. This is due to the fact that HESS J1356$-$645 is an older PWN, pointing to the fact that our model is currently only suitable for modelling young PWNe.

Next, I performed a parameter study to see what effects changes in all the free parameters had on the particle spectrum and the SED. First, I looked at how the PWN changed as it ages. I saw that the PWN accumulated particles over time and at some stage reached a maximum number of particles. After this the particle spectrum decreased due to the pulsar spinning down and fewer particles were injected into the PWN. For an age of 15~000 yr, I saw that the embedded pulsar had spun down and significant energy loss and particle escape had taken place, resulting in a very low particle spectrum and SED. For this case there was a build up of particles at high energies due to the decreased magnetic field, which in turn increased the diffusion and suppressed SR losses. This effect of the particle build up will be mentioned again and is something that needs future refinement. Next changes to the normalisation of the magnetic field and also the bulk particle motion were considered. When changing the magnetic field, one could see the inverse magnetic dependence of the diffusion coefficient, as a lower magnetic field resulted in reduced SR losses and an increased particle intensity. This effect is especially visible at high energies since SR losses scale with $E^2$. The SR radiation is also proportional to $B^2$, and this was seen in the SED so that an increased magnetic field increased the SR radiation. When the normalisation of the bulk particle motion was changed, it impacted the energy loss rate due to adiabatic cooling. Thus an increased bulk speed caused the entire particle spectrum to lose more energy, shifting down and to the left.

I considered changes to the injected particles spectrum. I first changed the normalisation of the injection spectrum and also changed the characteristic pulsar spin-down timescale. Both of these effects influenced the amount of particles injected into the PWN. When the normalisation of the injected particles was increased, more particles was injected into the PWN, resulting in a higher particle spectrum and SED. Similarly, when the characteristic timescale of the embedded pulsar was increased, it took longer for the pulsar to spin down, resulting in more particles being injected into the PWN and vice versa.

We also studied the effects changing to the diffusion coefficient on the particle spectrum and SED. Here we saw that changes to the diffusion should not change the shape of the particle spectrum or the SED and the changes we saw there where due to changes in the diffusion coefficient that changes the length of the time step in the code. Finally, I showed the contribution that each of the soft-photon components had on the total IC spectrum and also showed that if the energy density or temperature of the soft-photon components were increased, the radiation in the IC spectrum also increased and vice versa.

\section{Spatially-dependent results}
The main aim of the development of this code was for it to be able to give results regarding the morphology of the PWN. Following the parameter study I investigated the effects of changes in the particle bulk motion and the diffusion coefficient on the PWN size. This was done by plotting the normalised surface brightness as a function of radius for different energy bands, as well as plotting the size of the PWN as a function of energy. I saw that if the normalisation of the diffusion coefficient was increased, the size of the PWN also increased. This is an expected result, as faster diffusion results in the particles moving faster towards the outer zones. However, we saw that the size of the PWN increased as the energy increased for a large diffusion coefficient. By increasing the normalisation of the bulk particle speed, we observed the same effect as for an increased diffusion coefficient. Particles also reached the outer boundaries faster, increasing the size of the PWN. This example illustrates the potential of the model to constrain certain parameters: since we can predict the energy dependence of the PWN size, we can constrain quantities such as diffusion and convection (these should be relatively small if cooling is to dominate so that the size will decrease with energy). In other words, since we are now able to concurrently fit energy and radially dependent data (spectra and emission profiles), we can potentially derive stronger constraints on key quantities characterising the PWN.

We lastly showed the effect of changes to the spatial parametrisation of the magnetic field and particle bulk motion. We investigated four different scenarios. The first scenario mimicked the way adiabatic losses were treated in \cite{Torres2014} who assumed a constant magnetic field in space. Their usage of a constant adiabatic timescale implied a bulk particle speed that increased with distance from the centre of the PWN. This is an unphysical situation as mentioned earlier. The other three scenarios were discussed in detail in Section~\ref{sec:alph_Valph_B} and the most important point to make here is the fact that when the magnetic field decreased with $r$ and became very small, the diffusion coefficient increased substantially, causing  a build up of particles at high energies. Here the effect of choosing $r_{\rm{max}}\gg R_{PWN}$ was most clear and we realised that we had treated $\sigma$ and $B$ as independent variables. A refined treatment in future should mitigate this problem. Furthermore, we also assumed that the diffusion coefficient was spatially independent. However, a spatially-dependent magnetic field implies a spatially dependent diffusion coefficient, pointing to further revision on our part (i.e., adding a convection-like term to the transport equation). 

\section{Future work}
The spatial-temporal-energetic model we presented is a first approach to modelling PWNe for multiple spatial bins, thus there are many of improvements that can be made to the code, for example:
\begin{itemize}
\item The code currently has a problem with a build up of particles at high energies when the magnetic field decreases rapidly with radius. This is partially due to the fact that we chose $r_{\rm{max}}\gg R_{PWN}$. We will revise this boundary condition in future. One way in which this could be refined is by using an MHD code to model the morphology of the PWN in more detail and to find a more accurate value for the time-dependent radius of the PWN. This will allow us to use this radius as the outer boundary which will enable the particles to escape more efficiently from the PWN. Furthermore, treating $\sigma$ as being dependent on the magnetic field will also aid by lowering the maximum energy of particles that are contained within the PWN.
\item Throughout this study we have assumed that the diffusion coefficient was spatially independent. However, by considering the spatial dependence of the magnetic field, and the fact that $\kappa \propto 1/B(r,t)$, this assumption may technically be violated in some cases. The code should be generalised in future to handle a spatially-dependent diffusion coefficient by adding another convective term to the transport equation.
\end{itemize}
Our model currently produces morphological information for PWNe. This is a advantage that few other PWN models possess. This opens up a wide field of new research possibilities:
\begin{itemize}
\item In future a population study should be done to investigate currently know trends, e.g., the X-ray luminosity that correlates with the pulsar spin-down luminosity. We should also probe unknown trends, e.g., investigate whether there is a correlation between the TeV surface brightness of the PWN and the spin-down luminosity of the pulsar.
\item The future of high-energy astrophysics is bright with the development of new gamma-ray telescopes, like H.E.S.S.\ II and CTA. Both these telescopes will reveal more sources as they have lower energy thresholds and increased sensitivities. CTA will also yield more information regarding the morphology of currently known PWNe due to improved angular resolution. This will necessitate the development, application, and refinement of spatially-dependent codes as more morphological aspects will need modelling. 
\end{itemize}
Some follow-up projects or refinements to the model that could be considered are the following:
\begin{itemize}
\item Many observed PWNe are not spherically symmetric. Some older PWNe are offset from the pulsar, revealing a bullet shape as mentioned in Chapter~\ref{ch:ThBack}. This is due to either an inhomogeneity in the ISM in which the PWN expands, causing an asymmetric reverse shock and thus an offset PWN, or to the pulsar receiving some kick velocity at the birth of the PWN, thus moving away from the centre. The radiation peaks at the pulsar position, thus also causing the bullet shape. These effects could be added to the model to simulate a more realistic situation.
\item Currently our code has a stationary outer boundary. This should be revised, since the PWN expands as it ages.
\item The code currently only applicable to young PWN. This should be addressed so that all ages of PWNe can be modelled, e.g., by including a more complex parametrisation of the magnetic field and adding the effect of an asymmetric reverse shock to the code.
\item The code currently assumes spherical symmetry. This can be revised by expanding the model to 2 or 3 spatial dimensions. One could also add anisotropic effects such as considering distinct equatorial and polar outflows (injection) of particles.
\item One can obtain more realistic spatial and time dependencies of the magnetic field and bulk flow speeds using an MHD code. This can then be implemented into our PWN code, yielding refined SR and adiabatic losses and convection. 
\end{itemize}

%% ----------------------------------------------------------------
% Now begin the Appendices, including them as separate files

\addtocontents{toc}{\vspace{2em}} % Add a gap in the Contents, for aesthetics

\appendix % Cue to tell LaTeX that the following 'chapters' are Appendices

% Appendix A

\chapter{Mathematical derivations}
\label{AppendixA}
\lhead{Appendix A. \sc{Mathematical Derivations}}

In this appendix, I collect some mathematical results that were not included in the main part of the thesis to improve the readability of the main text.

\section{Logarithmic bins}\label{appen:delSR}
In Section~\ref{sec:Mdimen} the geometry of the model is discussed and the fact that some bin sizes are increased logarithmically is mentioned. This is in contrast to the linear case where the bins are all the same size. The way this is handled is shown here by looking at the lepton energies. If $E_i$ are the discretised energies, then
\begin{equation}
  E_i = E_{\rm{min}}e^{i \delta}, i=0..M-1
\label{eq:discretised}
\end{equation}
and
\begin{equation}
  E_{\rm{max}} = E_{\rm{min}}e^{(M-1) \delta},
\label{eq:discretised2}
\end{equation}
with $E_{\rm{min}}$ and $E_{\rm{max}}$ the minimum and maximum lepton energies respectively, $\delta$ a step value for the lepton energies, and $M$ the total number of bins allocated to the lepton energy vector. Equation \eqref{eq:discretised2} is used to calculate the size of $\delta$ by noting that 
\begin{equation}
  e^{(M-1)\delta} = \frac{E_{\rm{max}}}{E_{\rm{min}}}
\end{equation}
thus
\begin{equation}
  \delta = \frac{1}{M-1}\ln\left(\frac{E_{\rm{max}}}{E_{\rm{min}}}\right).
\label{eq:deltaSR!}
\end{equation}
The bin widths are not constant, but can be calculated as follows,
\begin{equation}
\begin{split}
  \left(\triangle E\right)_i &= E_{i+1}-E_i\\
  &=E_{\rm{min}}e^{(i+1)\delta} - E_{\rm{min}}e^{i\delta}\\ 
  &=E_{\rm{min}} \left[e^{i\delta}\left(e^{\delta}-1\right)\right].
\end{split}
\end{equation}
In our case the value for $\delta$ will always be much smaller than one, $\delta\ll 1$, and thus by using a Taylor expansion we can write $e^{\delta} \approx 1+\delta$. Thus
\begin{equation}
\begin{split}
  \left(\triangle E\right)_i & \approx E_{\rm{min}}\left[e^{i\delta}\left(1+\delta-1\right)\right]\\
  &=E_{\rm{min}}e^{i\delta}\delta\\
  &=\delta E_i.
\end{split}
\end{equation}

\section{Normalisation of the particle injection spectrum}\label{appen:normQ}
In Eq.~\eqref{brokenpowerlaw} I showed that the particle (lepton) injection spectrum at the termination shock in the PWN is modelled by a broken power law, with $Q_0$ the normalisation constant. I also showed that by using the spin-down luminosity $L(t) = L_0\left(1+t/\tau_0\right)^{-2}$ of the pulsar, with  $\tau_0$ the characteristic spin-down timescale of the pulsar and $L_0$ the initial spin-down luminosity, one can write
\begin{equation}
  \epsilon L(t) = \int_{E_{\rm{min}}}^{E_{\rm{b}}}QE_{\rm{e}}dE_{\rm{e}} + \int_{E_{\rm{b}}}^{E_{\rm{max}}}QE_{\rm{e}}dE_{\rm{e}},
  \label{normQAp}
\end{equation}
with $\epsilon$ the conversion efficiency of the time-dependent spin-down luminosity $L(t)$ to power in the particle spectrum. This equation can be used to normalise $Q_0$.  This is done by discretising Eq.~(\ref{normQAp}) as follows
\begin{equation}
  \epsilon L(t) = \delta\frac{Q_0}{E_{\rm{b}}^{\alpha_1}}\sum_{i=0}^{i=i_{\rm{b}}}E_i^{\alpha_1+2} + \delta\frac{Q_0}{E_{\rm{b}}^{\alpha_2}}\sum_{i=i_{\rm{b}}+1}^{i=M-1}E_i^{\alpha_2+2},
  \label{ap1}
\end{equation}
where $E_{\rm{b}}$ is the energy where the break in the lepton spectrum occurs and $i_b$ is the corresponding bin index for the break energy. Eq.~(\ref{ap1}) can now be manipulated to give
\begin{equation}
  Q_0(t) = \frac{\epsilon L(t)}{\delta}\left(E_b^{-\alpha_1}\sum_{i=0}^{i=i_b}E_i^{\alpha_1+2} + E_b^{-\alpha_2}\sum_{i=i_b+1}^{i=M-1}E_i^{\alpha_2+2} \right)^{-1}.
\label{ap:Q_0}
\end{equation}

\section{The Fokker-Planck-type transport equation}\label{app:transport}
\subsection{General transport equation}
The general transport equation in terms of momentum is given by \citep{Moraal2013}
\begin{equation}
\frac{\partial f}{\partial t} = -\nabla \cdot \mathbf{S} + \frac{1}{p^2}\frac{\partial }{\partial p}\left(p^2\left\langle\dot{p}\right\rangle_{\rm{tot}} f\right) + Q(\mathbf{r},p,t)
\label{eq:trans_Mich}
\end{equation}
with
\begin{equation}
\begin{split}
  \nabla \cdot \mathbf{S} &= \nabla \cdot \left(\mathbf{V}f - \mathbf{\underline{K}} \nabla f\right)\\
  & = \nabla \cdot \left(\mathbf{V}f\right) - \nabla \cdot (\mathbf{\underline{K}} \nabla f) \\
  & = \mathbf{V} \cdot (\nabla f) + f(\nabla \cdot \mathbf{V}) - \nabla \cdot (\mathbf{\underline{K}} \nabla f)
\end{split}
\label{eq:nablaS}
\end{equation}
where the symbols have been defined in Section~\ref{sec:Injection}. By substituting Eq.~\eqref{eq:nablaS} into Eq.~\eqref{eq:trans_Mich}, and using the total energy loss rate as 
\begin{equation}
\left\langle\dot{p}\right\rangle_{\rm{tot}} = \dot{p}_{\rm{rad}} + \dot{p}_{\rm{ad}},
\end{equation}
where $\dot{p}_{\rm{rad}}$ is the energy loss rate due to radiation, and $\dot{p}_{\rm{ad}} = \frac{1}{3}(\nabla \cdot \mathbf{V})p$ the adiabatic energy rate of change (usually a loss rate during the expansion phase), we find
\begin{equation}
\small
\begin{split}
  \frac{\partial f}{\partial t} &= -\left[ \mathbf{V} \cdot (\nabla f) + f(\nabla \cdot \mathbf{V}) - \nabla \cdot (\mathbf{\underline{K}} \nabla f)\right] + \frac{1}{p^2}\frac{\partial }{\partial p}\left(p^3\frac{1}{3}(\nabla \cdot \mathbf{V}) f\right) + \frac{1}{p^2}\frac{\partial }{\partial p}\left(p^2\left\langle\dot{p}\right\rangle_{\rm{rad}} f\right) + Q(\mathbf{r},p,t) \\
  & = -\mathbf{V} \cdot (\nabla f) - f(\nabla \cdot \mathbf{V}) + \nabla \cdot (\mathbf{\underline{K}} \nabla f) +  \frac{1}{p^2}\left(\frac{1}{3}(\nabla \cdot \mathbf{V})\left[3p^2f+p^3   \frac{\partial f}{\partial p} \right]    \right) + \frac{1}{p^2}\frac{\partial }{\partial p}\left(p^2\left\langle\dot{p}\right\rangle_{\rm{rad}} f\right) + Q(\mathbf{r},p,t) \\
  & = -\mathbf{V} \cdot (\nabla f) - f(\nabla \cdot \mathbf{V}) + \nabla \cdot (\mathbf{\underline{K}} \nabla f) +  f(\nabla \cdot \mathbf{V}) + \left(p \frac{\partial f}{\partial p} \right)\left( \frac{1}{3}(\nabla \cdot \mathbf{V}) \right) + \frac{1}{p^2}\frac{\partial }{\partial p}\left(p^2\left\langle\dot{p}\right\rangle_{\rm{rad}} f\right) + Q(\mathbf{r},p,t) \\
  & = -\mathbf{V} \cdot (\nabla f) + \nabla \cdot (\mathbf{\underline{K}} \nabla f) + \frac{1}{3}(\nabla \cdot \mathbf{V})\frac{\partial f}{\partial \ln p} + \frac{1}{p^2}\frac{\partial }{\partial p}\left(p^2\left\langle\dot{p}\right\rangle_{\rm{rad}} f\right) + Q(\mathbf{r},p,t).
\end{split}
\label{eq:trans_Moraal}
\end{equation}
Note the cancellation of the $f(\nabla \cdot \mathbf{V})$ terms. Eq.~\eqref{eq:trans_Moraal} corresponds to Eq.~(18) of \cite{Moraal2013}. Furthermore, by using $U_p(\mathbf{r},p,t) = 4\pi p^2 f(\mathbf{r},p,t)$ also given in \cite{Moraal2013}, and $\mathbf{\underline{K}} = \kappa(p)$, we can rewrite Eq.~\eqref{eq:trans_Moraal} as follows:
\begin{equation}
\small
\begin{split}
  \frac{\partial f}{\partial t} =& -\mathbf{V} \cdot (\nabla f) + \nabla \cdot (\mathbf{\underline{K}} \nabla f) + \frac{1}{3}(\nabla \cdot \mathbf{V})\frac{\partial f}{\partial \ln p} + \frac{1}{p^2}\frac{\partial }{\partial p}\left(p^2\left\langle\dot{p}\right\rangle_{\rm{rad}} f\right) + Q(\mathbf{r},p,t) \\
  \frac{1}{4\pi p^2}\frac{\partial U_p}{\partial t} =& -\frac{1}{4\pi p^2}\mathbf{V} \cdot (\nabla U_p) + \frac{1}{4\pi p^2}(\kappa \nabla^2 U_p) + \frac{1}{3}(\nabla \cdot \mathbf{V}) \frac{1}{4\pi p^2}\left( \frac{\partial U_p}{\partial \ln p} - 2U_p \right) + \\
  &\frac{1}{p^2}\frac{\partial }{\partial p}\left(\left\langle\dot{p}\right\rangle_{\rm{rad}} \left[ \frac{U_p}{4\pi } \right]\right) + Q(\mathbf{r},p,t) \\
  \frac{\partial U_p}{\partial t} =& -\mathbf{V} \cdot (\nabla U_p) + (\kappa \nabla^2 U_p) + \frac{1}{3}(\nabla \cdot \mathbf{V})\left( \frac{\partial U_p}{\partial \ln p} - 2U_p \right) +  \frac{\partial }{\partial p}\left(\left\langle\dot{p}\right\rangle_{\rm{rad}} U_p \right) + Q'(\mathbf{r},p,t) 
\end{split}
\label{eq:trans_Carlo}
\end{equation}
by using $\frac{\partial f}{\partial \ln p} = p\left( \frac{1}{4\pi p^2}\frac{\partial U_p}{\partial p} - \frac{2}{4\pi p^3}U_p \right)$, and setting $Q'(\mathbf{r},p,t) = 4\pi p^2 Q(\mathbf{r},p,t)$.

\subsection{Writing the transport equation in terms of energy}
In this section I will rewrite the transport equation in terms of energy by using the relation $E^2 = p^2c^2 + E_0^2$ (I will use the symbol $E$ instead of $E_{\rm{e}}$ for the particle energy). In this part the source term $ Q'(\mathbf{r},p,t) $ will be neglected and added later. Thus we start with Eq.~\eqref{eq:trans_Carlo}
\begin{equation}
\begin{split}
  \frac{\partial U_p}{\partial t} &= -\mathbf{V} \cdot (\nabla U_p) + (\kappa \nabla^2 U_p) + \frac{1}{3}(\nabla \cdot \mathbf{V})\left( \frac{\partial U_p}{\partial \ln p} - 2U_p \right) + \frac{\partial }{\partial p}\left(\left\langle\dot{p}\right\rangle_{\rm{rad}}U_p\right)
\end{split}
\label{eq:trans_Carlo2}
\end{equation}
and use the following:
\begin{equation}
\begin{split}
  p &= \sqrt{\frac{(E^2 - E_0^2)}{c^2}} \\
  \therefore \frac{d p}{d E} &= \frac{1}{2}\left(\frac{(E^2 - E_0^2)}{c^2}\right)^{-\frac{1}{2}} \frac{2E}{c^2}\\
  &= \frac{E}{pc^2}.
\end{split}
\label{eq:dedp}
\end{equation}
But
\begin{equation}
U_p dp = U_E dE,
\end{equation}
with $U_E$ the number of particles per unit volume and energy. Therefore:
\begin{equation}
\begin{split}
  U_p &= U_E \frac{\sqrt{(E^2- E_0^2)/c^2} }{E}c^2\\
  &=cU_E\frac{\sqrt{(E^2- E_0^2)} }{E}.
\end{split}
\label{eq:dedp2}
\end{equation}
In Eq.~\eqref{eq:trans_Carlo2} we have to calculate $\partial U_p/\partial \ln p$ and I will first show how this is done.
\begin{equation}
\begin{split}
  \frac{\partial U_p}{\partial \ln p} & = \frac{\partial U_p}{\partial \ln E}\frac{\partial \ln E}{\partial \ln p}\\
  & = \frac{\partial U_p}{\partial \ln E}\frac{p}{E}\frac{dE}{dp} \\
  & = \frac{\partial U_p}{\partial \ln E}     \left(\frac{pc}{E}\right)^2\\
  & = \frac{\partial }{\partial \ln E}\left( cU_E \frac{\sqrt{(E^2-E_0^2)}}{E}\right)     \left(\frac{pc}{E}\right)^2\\
  & = \left(\frac{p^2c^3}{E}\right)\frac{\partial }{\partial E}\left( U_E \frac{\sqrt{(E^2-E_0^2)}}{E}\right)     \\ 
  & = \left(\frac{p^2c^3}{E}\right)\left[\frac{\partial U_E}{\partial E}\frac{\sqrt{(E^2-E_0^2)}}{E} + U_E\left( \frac{1}{\sqrt{(E^2-E_0^2)}} - \frac{\sqrt{(E^2-E_0^2)}}{E^2} \right) \right]   \\
  & = \left(\frac{p^2c^3}{E}\right)\left[\frac{\partial U_E}{\partial E}\frac{\sqrt{(E^2-E_0^2)}}{E} + U_E\left( \frac{E^2 - (E^2 - E_0^2)}{\sqrt{(E^2-E_0^2)}E^2}\right) \right]   \\
  & = \left(\frac{p^2c^3}{E}\right)\left[\frac{\partial U_E}{\partial E}\frac{\sqrt{(E^2-E_0^2)}}{E} + U_E\left( \frac{E_0^2}{\sqrt{(E^2-E_0^2)}E^2}\right) \right]   \\
  & = c\left(\frac{p^2c^2}{E^2}\right)\left[\frac{\partial U_E}{\partial \ln E}\frac{\sqrt{(E^2-E_0^2)}}{E} \right] + c\left(\frac{p^2c^2}{E^2}\right)\left[U_E\left( \frac{E_0^2}{\sqrt{(E^2-E_0^2)}E}\right) \right].
\end{split}
\label{eq:dudlnp}
\end{equation}
We need the rest of the terms also in terms of $U_E$, thus
\begin{equation}
\begin{split}
 \frac{\partial U_p}{\partial t} &= \frac{\partial }{\partial t}\left(U_E\frac{\sqrt{(E^2-E_0^2)/c^2}}{E} c^2\right)\\
 &=c\frac{\sqrt{(E^2-E_0^2)}}{E}\frac{\partial U_E}{\partial t} ,
\end{split}
\end{equation}
and
\begin{equation}
\begin{split}
  \mathbf{V} \cdot (\nabla U_p) &= \mathbf{V} \cdot \left(\nabla cU_E\frac{\sqrt{(E^2- E_0^2)} }{E}\right)\\
  &= c\mathbf{V} \cdot \left(\nabla U_E\frac{\sqrt{(E^2- E_0^2)} }{E}\right),
\end{split}
\end{equation}
and
\begin{equation}
\begin{split}
\kappa \nabla^2 U_p &= \left(\kappa \nabla^2 cU_E\frac{\sqrt{(E^2- E_0^2)} }{E}\right)\\
&= c\left(\kappa \nabla^2 U_E\frac{\sqrt{(E^2- E_0^2)} }{E}\right),
\end{split}
\end{equation}
and
\begin{equation}
\begin{split}
&\frac{\partial }{\partial p}\left(\left\langle\dot{p}\right\rangle_{\rm{rad}} U_p\right) = \frac{\partial}{\partial E}\left( \left[\frac{\partial p}{\partial E}\frac{\partial E}{\partial t}\right] U_E \frac{\partial E}{\partial p} \right) \frac{\partial E}{\partial p} \\
&=\frac{\partial }{\partial E}(\dot{E}_{\rm{rad}}U_E)\frac{pc^2}{E}.
\end{split}
\end{equation}
We now have all the terms for $U_p$ in terms of $U_E$, and thus Eq.~\eqref{eq:trans_Carlo2} becomes
\begin{equation}
\begin{split}
  c\frac{\sqrt{(E^2-E_0^2)}}{E}\frac{\partial U_E}{\partial t} =& -c\mathbf{V} \cdot \left(\nabla U_E\frac{\sqrt{(E^2- E_0^2)} }{E}\right)\\ 
  +&  c\left(\kappa \nabla^2 U_E\frac{\sqrt{(E^2- E_0^2)} }{E}\right)\\
  +& \frac{1}{3}(\nabla \cdot \mathbf{V})\left\{ c\left(\frac{p^2c^2}{E^2}\right)\left[\frac{\partial U_E}{\partial \ln E}\frac{\sqrt{(E^2-E_0^2)}}{E} \right] + \right. \\
  &\left.c\left(\frac{p^2c^2}{E^2}\right)\left[U_E\left( \frac{E_0^2}{\sqrt{(E^2-E_0^2)}E}\right) \right] - 2cU_E\frac{\sqrt{(E^2- E_0^2)} }{E} \right\} \\
  +& \frac{\partial }{\partial E}(\dot{E}_{\rm{rad}}U_E)\frac{pc^2}{E}.
\end{split}
\label{eq:fin_before_assum}
\end{equation}
It is possible to simplify Eq.~\eqref{eq:fin_before_assum} when considering relativistic particles (e.g., $E_{\rm{e}}\sim 10^{11} - 10^{14}~\rm{erg}$, $\gamma \sim 10^5 - 10^8$).  We can then assume that the $E\gg E_0$, so that the particle energy $E\simeq pc$, therefore $\sqrt{(E^2-E_0^2)}/{E} \simeq 1$, $\gamma^{-1} \simeq 0$ and Eq.~\eqref{eq:fin_before_assum} reduces to
\begin{equation}
\begin{split}
  c\frac{\partial U_E}{\partial t} &= -c\mathbf{V} \cdot (\nabla U_E)\\ 
  &+  c \kappa \nabla^2 U_E\\
  &+ \frac{1}{3}(\nabla \cdot \mathbf{V})\left( c\left[\frac{\partial U_E}{\partial \ln E} \right] + c\left[U_E\left( \frac{1}{\gamma^2}\right) \right] - 2cU_E \right) \\
  &+ c\frac{\partial }{\partial E}(\dot{E}_{\rm{rad}}U_E).
\end{split}
\label{eq:fin_after_assum1}
\end{equation}
We finally arrive at (reinserting $Q$ in terms of $E$ now) 
\begin{equation}
\begin{split}
  \frac{\partial U_E}{\partial t} &= -\mathbf{V} \cdot (\nabla U_E) +  \kappa \nabla^2 U_E + \frac{1}{3}(\nabla \cdot \mathbf{V})\left( \left[\frac{\partial U_E}{\partial \ln E} \right] - 2U_E \right) +  \frac{\partial }{\partial E}(\dot{E}_{\rm{rad}}U_E) +  Q(\mathbf{r},E,t).
\end{split}
\label{eq:fin_transport(E)}
\end{equation}
For the rest of the thesis, we use the symbol $N_{\rm{e}}$ to indicate $U_{E}$ with the units $\rm{cm}^{-3} \rm{erg}^{-1}$.

\subsection{Discretisation of the Fokker-Planck-type transport equation}\label{appen:disctr}
\begin{figure}[h]
\centering
\begin{minipage}[b]{3in}
\includegraphics[width=3in]{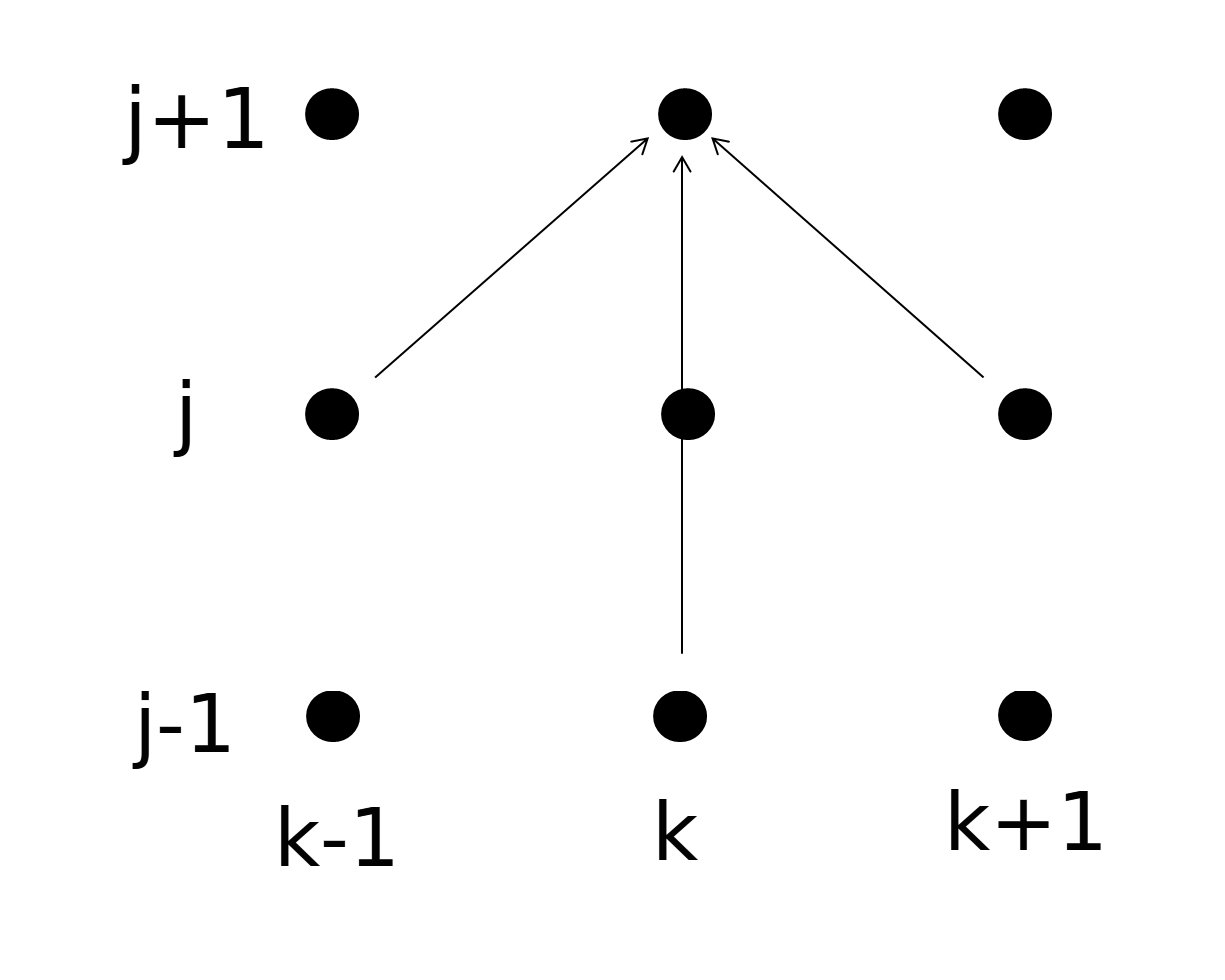}
\caption{\label{fig:DuFort}DuFort-Frankel numerical scheme.}
\end{minipage} 
\end{figure}

In this section I will show how the following Fokker-Planck-type transport equation is descretised (see Eq.~[\ref{eq:fin_transport(E)}]): 
\begin{equation}
\begin{split}
  \frac{\partial N_{\rm{e}}}{\partial t} &= -\mathbf{V} \cdot (\nabla N_{\rm{e}}) +  \kappa \nabla^2 N_{\rm{e}} + \frac{1}{3}(\nabla \cdot \mathbf{V})\left( \left[\frac{\partial N_{\rm{e}}}{\partial \ln E} \right] - 2N_{\rm{e}} \right) +  \frac{\partial }{\partial E}(\dot{E}_{\rm{rad}}N_{\rm{e}}) +  Q(\mathbf{r},E,t).
\end{split}
\label{eq:fin_transport(E)_2}
\end{equation}
 Before we can start with the discretisation, we first consider the term $\frac{1}{3}(\nabla \cdot \mathbf{V})\left( \left[\frac{\partial N_{\rm{e}}}{\partial \ln E} \right] - 2N_{\rm{e}} \right)$ and write it in the following form:
\begin{equation}
\begin{split}
&\frac{1}{3}(\nabla \cdot \mathbf{V})\left( \left[\frac{\partial N_{\rm{e}}}{\partial \ln E} \right] - 2N_{\rm{e}} \right) \\
& = \frac{1}{3}(\nabla \cdot \mathbf{V}) \frac{\partial N_{\rm{e}}}{\partial \ln E} - \frac{2}{3}(\nabla \cdot \mathbf{V})N_{\rm{e}}\\
& = \frac{1}{3}(\nabla \cdot \mathbf{V})E \frac{\partial N_{\rm{e}}}{\partial E} + \frac{1}{3}(\nabla \cdot \mathbf{V})N_{\rm{e}} - (\nabla \cdot \mathbf{V})N_{\rm{e}}\\
& = \frac{\partial}{\partial E}\left(\frac{1}{3}(\nabla \cdot \mathbf{V})EN_{\rm{e}} \right) - (\nabla \cdot \mathbf{V})N_{\rm{e}} \\
& = \frac{\partial}{\partial E}\left(\dot{E}_{\rm{ad}}N_{\rm{e}}\right) - (\nabla \cdot \mathbf{V})N_{\rm{e}},
\end{split}
\label{ap_eq:qwerty}
\end{equation}
where $\dot{E}_{\rm{ad}}$ is the energy change due to adiabatic heating or cooling. With the adiabatic energy change now in this form, we can add it to the radiation energy losses to give us a term for the total energy change $\dot{E}_{\rm{tot}} = \dot{E}_{\rm{rad}} + \dot{E}_{\rm{ad}}$. Thus the transport equation becomes:
\begin{equation}
\begin{split}
  \frac{\partial N_{\rm{e}}}{\partial t} &= -\mathbf{V} \cdot (\nabla N_{\rm{e}}) +  \kappa \nabla^2 N_{\rm{e}} + \frac{\partial }{\partial E}(\dot{E}_{\rm{tot}}N_{\rm{e}}) - (\nabla \cdot \mathbf{V})N_{\rm{e}} +  Q(\mathbf{r},E,t).
\end{split}
\label{eq:fin_transport(E)_3}
\end{equation}
As a first approach to discretise Eq.~\eqref{eq:fin_transport(E)_3} an Euler method was used, but it soon became clear that this method was not stable.  The next step was to discretise the equation by using a DuFort-Frankel scheme. This scheme is used to solve parabolic differential equations, i.e., equations of the form
\begin{equation}
  \frac{\partial u}{\partial t} = \eta_0\frac{\partial^2 u}{\partial x^2}
\label{ap:duf1}
\end{equation}
can be discretised as 
\begin{equation}
  \frac{u_{k}^{(j+1)}-u_{k}^{(j-1)}}{2\triangle t}=\eta_{0}\frac{u_{k+1}^{(j)}-(u_{k}^{(j-1)}+u_{k}^{(j+1)})+u_{k-1}^{(j)}}{\triangle x^{2}}.
\label{ap:duf2}
\end{equation}
This scheme is stable for small time steps, where $j$ is the time step and $k$ the spatial step. A visual representation of this can be seen in Figure~\ref{fig:DuFort}.

To discretise Eq.~(\ref{eq:fin_transport(E)_3}) we first have to do a Taylor expansion of some function $f(x)$ to find the correct way to discretise the energy loss term, $\frac{\partial}{\partial E}(\dot{E}_{\rm{tot}}N_{\rm{e}})$ due to the different logarithmic bin sizes in energy. For two different bin sizes $h_1$ and $h_2$ we have
\begin{equation}
\begin{split}
f(x+h_1) = f(x) + h_1f^\prime(x) + \frac{h_1^2}{2}f^{\prime\prime}(x)\\
f(x-h_2) = f(x) - h_2f^\prime(x) + \frac{h_2^2}{2}f^{\prime\prime}(x).
\end{split}
\end{equation}
By multiplying by $h_2^2$ and $h_1^2$, we find
\begin{equation}
\begin{split}
h_2^2 f(x+h_1) = h_2^2f(x) + h_2^2h_1f^\prime(x) + \frac{h_2^2h_1^2}{2}f^{\prime\prime}(x)\\
h_1^2 f(x-h_2) = h_1^2f(x) - h_1^2h_2f^\prime(x) + \frac{h_1^2h_2^2}{2}f^{\prime\prime}(x).
\end{split}
\end{equation}
By then subtracting the second equation from the first we find
\begin{equation}
\begin{split}
h_2^2& f(x+h_1) - h_1^2 f(x-h_2) = \left(h_2^2 - h_1^2\right)f(x) + \left(h_2^2h_1+h_1^2h_2\right)f^\prime(x)\\
&\therefore f^\prime(x) = \frac{h_{2}^{2}f(x+h_{1})-h_{1}^{2}f(x-h_{2})+(h_{1}^{2}-h_{2}^{2})f(x)}{h_{1}h_{2}(h_{2}+h_{1})}\\
&\therefore f^\prime(x) = \frac{1}{h_2+h_1}\left[\frac{h_2}{h_1}f(x+h_1) - \frac{h_1}{h_2}f(x-h_2) + \left(\frac{h_1}{h_2}-\frac{h_2}{h_1}\right)f(x)\right].
\end{split}
\label{ap:h}
\end{equation}
For the case when $h_1 = h_2$ this reduces back to the usual expression for $f^\prime(x)$. In using Eq.~(\ref{ap:h}), $\frac{\partial}{\partial E}(\dot{E}_{\rm{tot}}N_{\rm{e}})$ becomes
\begin{equation}
\small
\begin{split}
\frac{\partial}{\partial E}(\dot{E}_{\rm{tot}}N_{\rm{e}}) = \frac{1}{dE_{i+1,j,k}+dE_{i,j,k}}\left[r_{\rm{a}}(\dot{E}_{\rm{tot}}N_{\rm{e}})_{i+1,j,k} - \frac{1}{r_{\rm{a}}}(\dot{E}_{\rm{tot}}N_{\rm{e}})_{i-1,j,k} + \left(\frac{1}{r_{\rm{a}}}-r_{\rm{a}}\right)(\dot{E}_{\rm{tot}}N_{\rm{e}})_{i,j,k}  \right]
\end{split}
\end{equation}
where $r_{\rm{a}} = h_2/h_1 = dE_{i+1}/dE_{i}$, $dE_{i,j,k}$ is the energy bin size and $i,j,k$ are the indices for energy, time, and space respectively. For simplicity, I am going to rename the energy term as $\frac{\partial}{\partial E}(\dot{E}_{\rm{tot}}N_{\rm{e}}) = X$, as the discretisation for this term in Eq.~(\ref{eq:fin_transport(E)_2}) is done for now and will be finalised in Eq.~(\ref{ap:finEQ}).

Next we consider the diffusion coefficient $\kappa$. We assume that it is not spatially dependent and therefore by assuming spherical symmetry we can rewrite $\kappa \nabla^2 N_{\rm{e}})$ as 
\begin{equation}
\begin{split}
\kappa \nabla^2 N_{\rm{e}}& = \kappa \frac{1}{r^2}\frac{\partial}{\partial r}\left(r^2\frac{\partial N_{\rm{e}}}{\partial r}\right)\\
& = \frac{2\kappa}{r} \frac{\partial N_{\rm{e}}}{\partial r} + \kappa \frac{\partial^2 N_{\rm{e}}}{\partial r^2}.
\end{split}
\end{equation}

The transport equation that has to be discretised, by adding the injection $Q$ back, is thus
\begin{equation}
\begin{split}
\frac{\partial N_{\rm{e}}}{\partial t} = Q + X + \frac{2\kappa}{r} \frac{\partial N_{\rm{e}}}{\partial r} + \kappa \frac{\partial^2 N_{\rm{e}}}{\partial r^2} - \mathbf{V} \cdot (\nabla N_{\rm{e}}) - (\nabla \cdot \mathbf{V})N_{\rm{e}}.
\end{split}
\end{equation}
It can now be fully discretised by using the DuFort-Frankel scheme as given in Eq.~(\ref{ap:duf2}).
\begin{equation}
\begin{split}
  \frac{(N_{\rm{e}})_{i,j+1,k}-(N_{\rm{e}})_{i,j-1,k}}{2\triangle t} = & ~Q_{i,j,1} + X\\
  & + \frac{2\kappa}{r}\frac{(N_{\rm{e}})_{i,j,k+1} - (N_{\rm{e}})_{i,j,k-1}}{2\triangle r}\\
  & + \kappa\frac{(N_{\rm{e}})_{i,j,k+1} - \left[(N_{\rm{e}})_{i,j-1,k}+ (N_{\rm{e}})_{i,j+1,k}\right]+(N_{\rm{e}})_{i,j,k-1}}{\triangle r^{2}}\\
  & - \frac{V_{i,j,k}}{2\triangle r}\left( \left( N_{\rm{e}} \right)_{i,j,k+1} - (N_{\rm{e}})_{i,j,k-1} \right)\\
  & - (\nabla \cdot \mathbf{V})_{i,j,k}(N_{\rm{e}})_{i,j,k}.
\end{split}
\label{ap_eq:asd}
\end{equation}
Note that the injection term $Q_{i,j,1}$ is only non-zero in the first spatial zone, since it is considered a boundary condition as discussed in Section~\ref{sec:boundary_conditions}. The term $(\nabla \cdot \mathbf{V})_{i,j,k}$ is calculated analytically as 
\begin{equation}
(\nabla \cdot \mathbf{V})_{i,j,k} = \left( \alpha_V + 2 \right) \left( \frac{V_{i,j,k}}{r_{k}} \right) 
\label{eq_ap:nablaDOTv}
\end{equation}
by using the parametrised form of the velocity given in Eq.~\eqref{V_profile}.

It is possible to simplify Eq.~\eqref{ap_eq:asd} to be more useful as shown in the next sets of equations:
\begin{equation}
\begin{split}
(N_{\rm{e}})_{i,j+1,k}-(N_{\rm{e}})_{i,j-1,k} = & ~2Q_{i,j,1}\triangle t + 2X\triangle t\\
& + \frac{2\kappa \triangle t}{r\triangle r}\left(\left(N_{\rm{e}}\right)_{i,j,k+1} - (N_{\rm{e}})_{i,j,k-1}\right)\\
& + \frac{2\kappa \triangle t}{\triangle r^2}\left(\left(N_{\rm{e}}\right)_{i,j,k+1} - \left[\left(N_{\rm{e}}\right)_{i,j-1,k}+ (N_{\rm{e}})_{i,j+1,k}\right]+(N_{\rm{e}})_{i,j,k-1}\right)\\
& - V_{i,j,k}\left( \frac{\triangle t}{\triangle r} \right)\left( \left(N_{\rm{e}}\right)_{i,j,k+1} - (N_{\rm{e}})_{i,j,k-1}\right)\\
& - 2(\nabla \cdot \mathbf{V})_{i,j,k}(N_{\rm{e}})_{i,j,k} \triangle t.
\end{split}
\label{ap:ne}
\end{equation}
By replacing $\frac{\partial}{\partial E}(\dot{E}_{\rm{tot}}N_{\rm{e}}) = X$, grouping similar terms in Eq.~\eqref{ap:ne} and setting $\beta = \frac{2\kappa \triangle t}{(\triangle r)^2}$, $\gamma = \frac{2\kappa \triangle t}{r\triangle r}$, and $\eta = \frac{V_{i,j,k} \triangle t}{\triangle r}$ and writing $\dot{E}_{\rm{tot}}\triangle t = dE_{\rm{loss}}$, we find that 
\begin{equation}
\begin{split}
(1+\beta)(N_{\rm{e}})_{i,j+1,k} &= (1-\beta)(N_{\rm{e}})_{i,j-1,k}\\
&+ (\beta + \gamma - \eta)(N_{\rm{e}})_{i,j,k+1}\\
&+ (\beta - \gamma + \eta)(N_{\rm{e}})_{i,j,k-1}\\
&+ \left[ \frac{2 \triangle t}{dE_{i+1,j,k}+dE_{i,j,k}} \right]\left[ r_{\rm{a}} \dot{E}_{i+1,j,k}(N_{\rm{e}})_{i+1,j,k} - \frac{1}{r_{\rm{a}}}\dot{E}_{i-1,j,k}(N_{\rm{e}})_{i-1,j,k} \right]\\
&+ \left[ \frac{2 \triangle t}{dE_{i+1,j,k}+dE_{i,j,k}} \right]\left[ \frac{1}{r_{\rm{a}}} - r_{\rm{a}} \right]\left[ \dot{E}_{i,j,k} \left( \frac{(N_{\rm{e}})_{i,j+1,k} + (N_{\rm{e}})_{i,j-1,k}}{2} \right) \right]\\
&- 2(\nabla \cdot \mathbf{V})_{i,j,k}(N_{\rm{e}})_{i,j,k} \triangle t\\
&+ 2Q_{i,j,1}\triangle t.
\end{split}
\label{ap_eq:asd2}
\end{equation}
Note that the term $\left( \frac{(N_{\rm{e}})_{i,j+1,k} + (N_{\rm{e}})_{i,j-1,k}}{2} \right)$ is the average of $N_{\rm{e}}$ over two time steps. Equation \eqref{ap_eq:asd2} can now be finalised by setting 
\begin{equation}
z = \left[ \frac{1}{dE_{i+1,j,k}+dE_{i,j,k}} \right]\left[ \frac{1}{r_{\rm{a}}} - r_{\rm{a}} \right]\left(dE_{\rm{loss}}\right)_{i,j,k}
\end{equation}
and thus the final equation that can now be implemented in the code is
\begin{equation}
\begin{split}
(1-z+\beta)(N_{\rm{e}})_{i,j+1,k} = &2Q_{i,j,1} \triangle t\\
& + (1+z-\beta)(N_{\rm{e}})_{i,j-1,k} \\
& + (\beta+\gamma-\eta)(N_{\rm{e}})_{i,j,k+1} \\
& + (\beta-\gamma+\eta)(N_{\rm{e}})_{i,j,k-1} \\
& -2(\nabla \cdot \mathbf{V})_{i,j,k}\triangle t (N_{\rm{e}})_{i,j,k}\\
& + \frac{2}{(d E_{i+1,j,k}+dE_{i,j,k})}\\
& \left(r_{\rm{a}} \left(dE_{\rm{loss}}\right)_{i+1,j,k}(N_{\rm{e}})_{i+1,j,k} - \frac{1}{r_{\rm{a}}}\left(dE_{\rm{loss}}\right)_{i-1,j,k}(N_{\rm{e}})_{i-1,j,k}\right).
\end{split}
 \label{ap:finEQ}
\end{equation}

	% Appendix Title

%\input{./Appendices/AppendixB} % Appendix Title

%\input{./Appendices/AppendixC} % Appendix Title

\addtocontents{toc}{\vspace{2em}}  % Add a gap in the Contents, for aesthetics
\backmatter

%% ----------------------------------------------------------------
\label{Bibliography}
\lhead{\emph{Bibliography}}  % Change the left side page header to "Bibliography"
\bibliographystyle{apj}  % Use the "unsrtnat" BibTeX style for formatting the Bibliography
\bibliography{Bibliography}  % The references (bibliography) information are stored in the file named "Bibliography.bib"

\end{document}